\documentclass[preprint]{aastex}

\received{ }
\accepted{ }

\slugcomment{Accepted for publication in the June 1, 2001 edition of {\it The Astrophysical Journal}} 

\shorttitle{Spectropolarimetry of SN 1999em}
\shortauthors{Leonard et al.}

\newcommand{\kms}{km s$^{-1}$}
\newcommand{\Bband}{B}
\newcommand{\Vband}{V}
\newcommand{\BminusV}{({\Bband}{\rm -}{\Vband})}

\newcommand{\halpha}{H$\alpha$}

\begin{document}

\title{Is it Round?  Spectropolarimetry of the Type II-P Supernova 1999em}

\vspace{2cm}

\author{Douglas C. Leonard, Alexei V. Filippenko, and David R. Ardila}
\affil{Department of Astronomy, University of California, Berkeley,
California 94720-3411}
\email{dleonard@astro.berkeley.edu, alex@astro.berkeley.edu,dardila@astro.berkeley.edu}

\and

\author{Michael S. Brotherton} 
\affil{National Optical Astronomy Observatories\footnote{The National
Optical Astronomy Observatories are operated by the Association of Universities
for Research in Astronomy, Inc., under cooperative agreement with the National
Science Foundation.}, 950 N. Cherry Ave., P. O. Box 26732, Tucson, AZ 85726}
\email{mbrother@ohmah.tuc.noao.edu}

\vspace{1cm}

\begin{abstract}

We present the first multi-epoch spectropolarimetry of a type II plateau
supernova (SN~II-P), with optical observations of SN~1999em on days 7, 40, 49,
159, and 163 after discovery.  These data are used to probe the geometry of the
electron-scattering atmosphere before, during, and after the plateau phase,
which ended roughly 90 days after discovery.  Weak continuum polarization with
an unchanging polarization angle ($\theta \approx 160^{\circ}$) is detected at
all epochs, with $p \approx 0.2\%$ on day 7, $p \approx 0.3\%$ on days 40 and
49, and $p \approx 0.5\%$ in the final observations.  Distinct polarization
modulations across strong line features are present on days 40, 49, 159, and
163.  Uncorrected for interstellar polarization (which is believed to be quite
small), polarization peaks are associated with strong P-Cygni absorption
troughs and nearly complete depolarization is seen across the H$\alpha$
emission profile.  The temporal evolution of the continuum polarization and
sharp changes across lines indicate polarization intrinsic to SN~1999em.  When
modeled in terms of the oblate, electron-scattering atmospheres of H\"{o}flich,
the observed polarization suggests an asphericity of at least $7\%$ during the
period studied.  The temporal polarization increase may indicate greater
asphericity deeper into the ejecta.  We discuss the implications of asphericity
on the use of type II-P supernovae as primary extragalactic distance indicators
through the Expanding Photosphere Method (EPM).  If asphericity produces
directionally dependant flux and peculiar galaxy motions are characterized by
$\sigma_{v_{rec}} = 300$ \kms, it is shown that the agreement between previous
EPM measurements of SNe~II and distances to the host galaxies predicted by a
linear Hubble law restrict mean SN~II asphericity to values less than $30\%$
($3\sigma$) during the photospheric phase.

\end{abstract}

\medskip

\section{INTRODUCTION}
\label{sec3:1}

The assumption of spherical symmetry in core-collapse supernova (SN) explosions
is under assault.  On the observational front, high-velocity ``bullets'' of
matter in SN~remnants (e.g., Taylor, Manchester, \& Lyne 1993), the Galactic
distribution (Lyne 1998) and high velocities of pulsars (Strom et al. 1995;
Cordes \& Chernoff 1998), the aspherical morphology of many young SN remnants
(Manchester 1987; see, however, Gaensler 1998), and the asymmetric distribution
of material inferred from direct speckle imaging of young supernovae (SNe;
e.g., SN~1987A, Papaliolios et al. 1989; see, however, H\"{o}flich 1990)
collectively argue for asymmetry in the explosion mechanism and/or distribution
of SN ejecta.  Moreover, recent advances in the understanding of the
hydrodynamics and distribution of material in the preexplosion core (Bazan \&
Arnett 1994; Lai \& Goldreich 2000) coupled with results obtained through
multidimensional numerical explosion models (Burrows, Hayes, \& Fryxell 1995)
imply that asphericity may be a generic feature of the explosion process
(Burrows 2000).  It has even recently been proposed that some core-collapse SNe
produce gamma-ray bursts (e.g., Bloom et al. 1999) through the action of a jet
of material aimed fortuitously at the observer, the result of a ``bipolar'',
jet-induced, SN explosion (Khokhlov et al. 1999; MacFadyan \& Woosley 1999;
Wheeler et al. 2000).  Such ideas shake the theoretical foundation of the
conventional spherical neutrino-driven SN explosion (Colgate \& White 1966),
and beg for a direct observational probe of early-time SN geometry.

Shapiro \& Sutherland (1982) first proposed that polarimetry could provide a
powerful tool with which to study of young SN geometry.  The most basic
question is whether the SN is round.  Because electron scattering dominates the
opacity of SNe during the photospheric phase (Branch 1980), any continuum
radiation that emerges at an angle to the surface normal is linearly polarized.
For a spherical source that is unresolved (true for all young, recent SNe) the
electric vectors cancel exactly, resulting in zero net polarization.
Aspherical sources, however, will yield residual polarization due to incomplete
cancellation; see Figure 1 of Leonard, Filippenko, \& Matheson (2000b). In
general, linear polarizations of $\sim1\%$ are expected for moderate
($\sim20\%$) SN asphericity.  The exact polarization amount varies with the
degree of asphericity, as well as with the viewing angle and the extension and
density profile of the electron-scattering atmosphere (H\"{o}flich 1991a [H91];
H\"{o}flich 1995).

SNe come in many varieties, with the main classification based on the presence
(type II) or absence (type I) of hydrogen in the spectrum near maximum light
(see Filippenko 1997 for a recent review). Except for SNe~Ia, whose progenitors
are carbon-oxygen white dwarfs that accrete matter from a companion and undergo
thermonuclear runaway, all other SN types (II-P, II-L, IIn, IIb, Ib, Ic) are
believed to result from the explosion of evolved, massive stars (initial mass
$\gtrsim 8-10\ M_{\odot}$) that have reached the ends of their nuclear-burning
lives.  These core-collapse events present a smorgasbord of spectral and
photometric properties, leading to a proliferation of subclassifications.  A
consensus is emerging, however, that much of this diversity is due to the state
of the progenitor's hydrogen envelope (helium as well in some cases) and
immediate circumstellar environment (CSE) at the time of the explosion.

\begin{figure}
\begin{center}
\rotatebox{270}{
 \scalebox{0.45}{
	\plotone{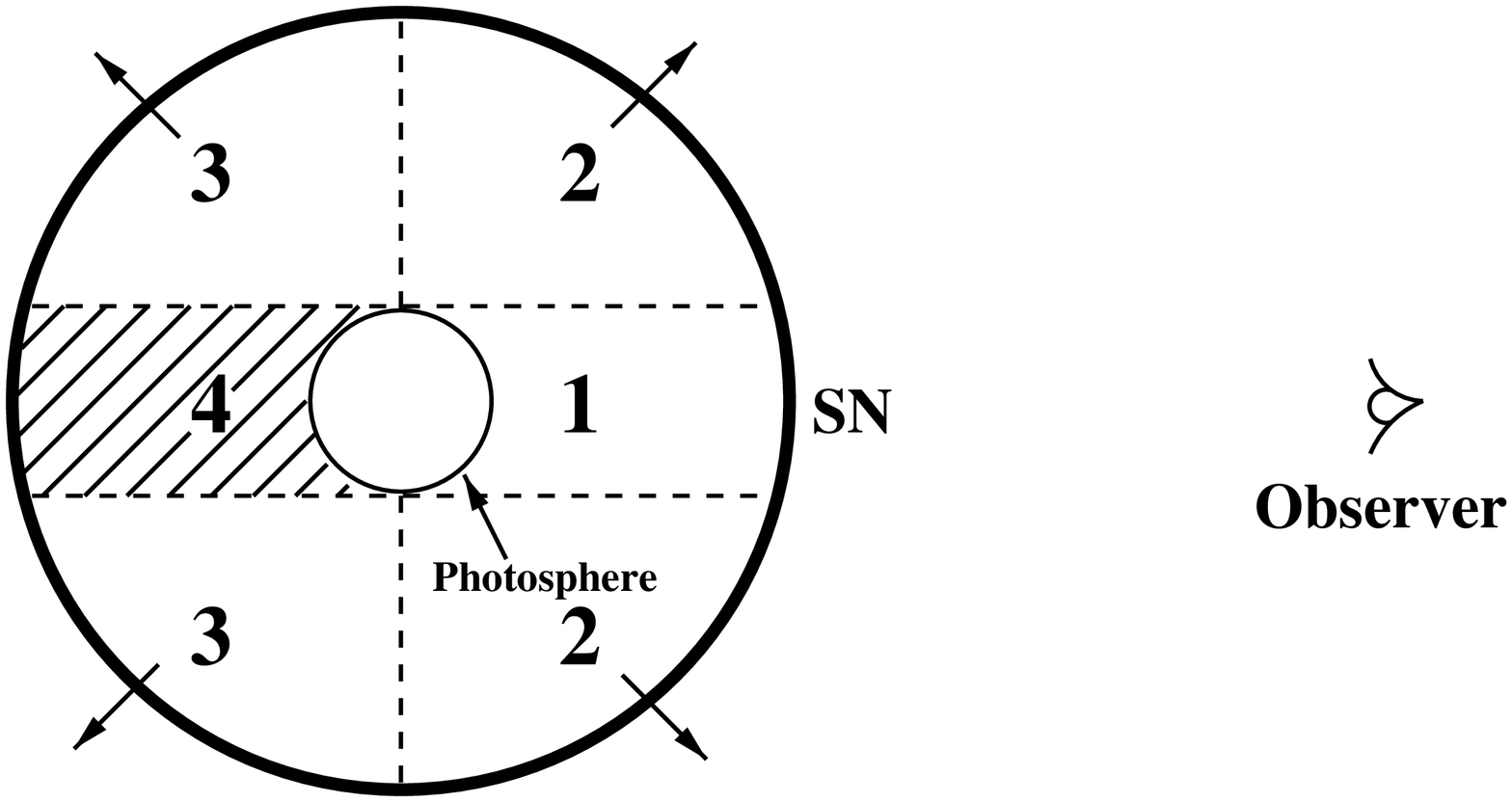}
		}
		}
\end{center}
\caption[Schematic representation of an expanding SN atmosphere during the
photospheric phase.]
{Schematic representation of an expanding SN atmosphere during the
photospheric phase.  Specific regions referred to in the text are indicated.  }
\label{fig1:3}
\end{figure}

``Normal'' core-collapse events are thought to result from isolated stars with
thick hydrogen envelopes (generally of several solar masses) intact at the time
of the explosion.  Their light curves show a distinct plateau (hence the
moniker ``SN II-P''), resulting from an enduring period (sometimes as long as
150 days) of nearly constant luminosity as the hydrogen recombination wave
recedes through the envelope and slowly releases the trapped, shock-deposited
energy.  The spectral and photometric evolution of SNe~II-P has been
successfully reproduced by computer simulations of exploding red supergiant
stars (e.g., Eastman, Schmidt, \& Kirshner 1996; Weaver \& Woosley 1993;
Imshennik \& Nadyozhin 1974; Eastman et al. 1994; Popov 1993).  The basic model
includes a source of thermal photons that electron-scatter their way out of the
envelope (Branch 1980), with line features imprinted onto the continuum by
resonance scattering in the envelope (see, e.g., Branch et al.  2000).  The
line features in photospheric-phase SNe are characterized by ``P-Cygni''
profiles, with blueshifted absorption troughs with emission centered near the
rest wavelength.  These characteristics are common for lines formed in
expanding atmospheres.  The general situation is illustrated by
Figure~\ref{fig1:3}.  The absorption component results from continuum photons
being scattered out of the line-of-sight (l-o-s) by gas approaching the
observer in region 1.  Photons scattered into the l-o-s by gas in regions 2 and
3 produce the emission component centered near the line's rest wavelength,
$\lambda_{\circ}$.

\clearpage

The observed properties of SNe~II-P are consistent with those expected for a
massive star exploding into the near-vacuum of the ambient interstellar medium.
Some massive stars, however, cast off much of their hydrogen envelope prior to
exploding, perhaps through stellar winds or due to the influence of a
companion.  It is the explosions of these stars that are believed to produce
SNe~II-L (``L'' for their linearly declining light curves, lacking a plateau)
or SNe~IIn (``n'' for the presence of relatively narrow emission lines lacking
the customary P-Cygni absorption components [Schlegel 1990]).  Much of the
spectral peculiarity of these events likely results from the interaction
between the SN and the dense CSE (Chugai 1997; Branch et al. 2000) in which
the star explodes.  Progenitors that have lost their {\em entire} hydrogen
envelope well before the explosion produce SNe~Ib (Type I due to the absence of
hydrogen in their spectra), and those that have been stripped of both hydrogen
and helium produce SNe~Ic.  Stars that explode with just a low-mass outer layer
of hydrogen remaining are placed into the transitional category ``SN~IIb''
since the spectrum shows hydrogen at early times but then becomes
helium-dominated later on (e.g., SN~1993J; Matheson et al. 2000).  It is
possible, then, to roughly rank core-collapse events in order of increasing
hydrogen envelope mass at the time of explosion: Ic, Ib, IIb, IIn, II-L and,
finally, II-P.  The exact sequence, especially the relative positions of IIb,
IIn, and II-L, is still quite uncertain.

The geometry of SNe~II-P are of particular interest due to their cosmological
utility as extragalactic distance indicators through the Expanding Photosphere
Method (EPM; Kirshner \& Kwan 1974; Schmidt et al. 1994a [S94]), a variant of
the Baade method used to measure distances to variable stars (Baade 1926).
This technique is especially attractive since it is a {\it primary} distance
indicator, unlike the ``standard candle'' method used to determine distances to
SNe~Ia.  It is therefore not subject to all the uncertainties that plague the
various rungs of the extragalactic distance ladder.  Thus far, EPM has been
applied to 18 SNe~II spanning distances from 0.049 Mpc (SN~1987A) to 180 Mpc
(SN~1992am; Schmidt et al. 1994b).  Using SNe~II alone, a Cepheid-independent
value of $H_0 = 73 \pm 7 {\rm\ km\ s^{-1}\ Mpc^{-1}} $ has been derived (S94).
However, unlike the empirically-based method used to measure distances to
SNe~Ia, distances derived to SNe~II-P rely on the assumption of a spherically
symmetric flux distribution during the early stages of development (i.e., the
plateau).  In this way, spectropolarimetry of SNe~II-P provides a critical test
of the cosmological utility of these core-collapse events.

Due largely to the difficulty of obtaining the required signal-to-noise (S/N)
ratio, detailed spectropolarimetric studies exist for only three events,
SN~1987A (Jeffery 1991a and references therein), SN~1993J (Trammell, Hines, \&
Wheeler 1993; H\"{o}flich et al. 1996; Tran et al. 1997), and SN~1998S (Leonard
et al. 2000a; Wang et al. 2000), none of which were of the classic II-P
variety.  SN~1987A resulted from the explosion of a {\it blue} supergiant, and
had a peak intrinsic polarization of $p \approx 0.7\%$ during the first 100
days.  SN~1993J, a Type IIb event, was polarized at $p \approx 1.5\%$ at early
times.  SN~1998S, a peculiar Type IIn event, was polarized up to $3\%$ at early
times.  The polarization seen in these three objects suggests that
core-collapse events, at least those of the unusual varieties represented by
them, may indeed be aspherical at early times.  It is also curious that
SN~1993J and SN~1998S, which had both lost much of their hydrogen envelopes
prior to exploding, were significantly more polarized than SN~1987A, which
likely had retained most of its envelope.  This trend is also seen in the
sample of broadband polarimetric studies (or low S/N ratio spectropolarimetry)
that have been carried out on SNe (Wang et al. 1996; see Wheeler [2000] for a
comprehensive list of polarimetric observations of SNe), and tentatively
suggests that the degree of polarization increases with decreasing envelope
mass (Wheeler 2000).  In fact, a polarization of nearly $7\%$ has recently been
inferred for a SN~Ic (Wang et al. 2000a).

It seems that the deeper we peer into the SN atmosphere, the more evidence
there is for asphericity.  This leads naturally to the conclusion that it is
the collapse or explosion mechanism itself that is strongly asymmetric, and
that evidence for this asymmetry is damped by the addition of outer envelope
material (e.g., Chevalier \& Soker 1989).  Since a SN~II-P gradually reveals
its inner layers as the recombination front recedes through the hydrogen
envelope during the plateau phase, a single object observed at multiple epochs
during (and after) the plateau era provides an important test of this
hypothesis: explosion asymmetry should result in increasing evidence for
asphericity with time.  With the larger aperture telescopes capable of
generating high S/N ratio spectropolarimetry in reasonable observing times now
available (e.g., Keck, VLT), the time is ripe for a detailed investigation into
early-time SN II-P geometry.

\begin{figure}
\begin{center}
\rotatebox{270}{
 \scalebox{0.95}{
	\plotone{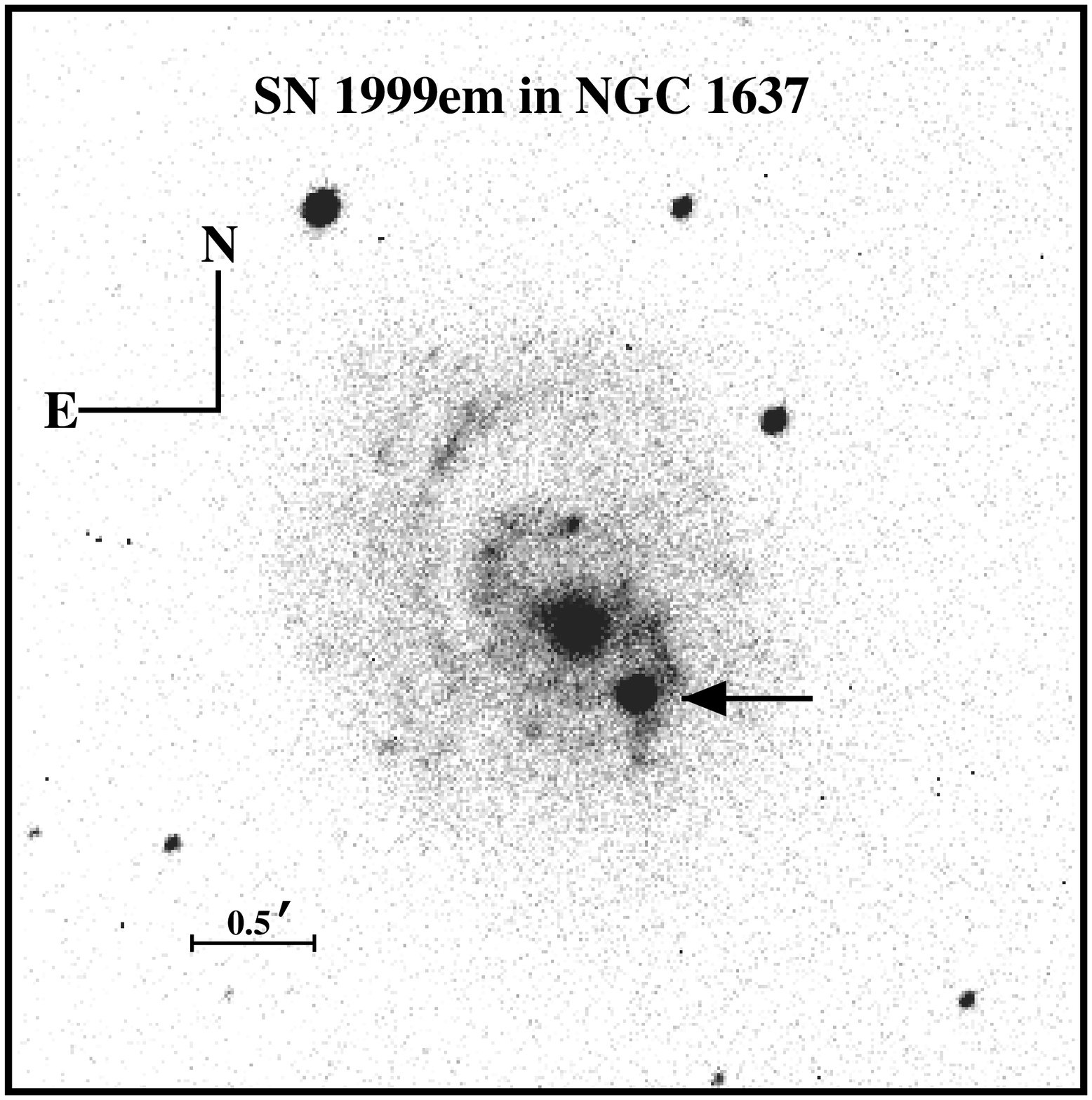}
		}
		}
\end{center}
\caption[$B$-band image of SN~1999em in NGC~1637, 2 days after discovery]
{$B$-band image of NGC~1637 taken on 1999 October 31 with the Katzman Automatic
Imaging Telescope (Treffers et al. 1997; A. V. Filippenko et al., in
preparation).  SN~1999em ({\em arrow}) is measured to be $15.\arcsec1$
west and $17.\arcsec2$ south of the galaxy nucleus (cf., Jha et al. 1999).}
\label{fig3:1}
\end{figure}

Supernova (SN) 1999em was discovered on 1999 October 29 UT\footnote{UT dates
are used throughout this paper.} by Li (1999) at an unfiltered magnitude of
$m \approx 13.5$ mag in the nearly face-on ($i \approx 31^{\circ}$, from
LEDA\footnote{http://www-obs.univ-lyon1.fr/leda/home\_leda.html.}) SBc galaxy
NGC~1637 (Fig.~\ref{fig3:1}).  It was quickly identified as a Type II event,
with prominent P-Cygni features of hydrogen Balmer and \ion{He}{1}
$\lambda$5876 and a blue continuum (Jha et al. 1999).  Immediately after
discovery a spectral, spectropolarimetric, and photometric campaign was
initiated at Lick Observatory.  Nearly daily optical photometry and
spectroscopy of the young SN were obtained with the Katzman Automatic Imaging
Telescope (KAIT) and the Lick 1-m Nickel reflector, respectively, providing
unprecedented temporal coverage of this Type~II event during the photospheric
phase.  Spectropolarimetry was obtained using the Lick 3-m and Keck-I 10-m
telescopes. 

We focus here on the spectropolarimetric observations of SN~1999em, and defer a
more detailed discussion of the photometric and spectral data to a separate
paper (Leonard et al. 2001), in which the EPM distance to SN~1999em ($d \approx
6$ Mpc) is derived.  For our present purposes, we note that both its
light-curve shape and spectral properties firmly establish SN~1999em as a Type
II-P event, with a plateau lasting roughly 90 days from discovery.  In
addition, the EPM analysis determines the time of explosion to be $t_\circ
\approx 5$ days before discovery, consistent with the absence of the SN in a
CCD image of NGC~1637 taken 9 days before discovery (limiting magnitude about
19; Li 1999).  We note that at an estimated distance of $\sim 6$ Mpc,
SN~1999em reached an extinction-corrected ($E [B - V] = 0.05$ mag [Baron et
al. 2000]) peak $B$-band magnitude of $M_B = -15.3 \pm 0.2$ mag, which is
somewhat fainter than average for SNe II-P (Patat et al. 1994).  Its color
evolution may also have been somewhat unusual (Leonard et al. 2001).  SNe~II-P
are known to be a heterogeneous class of objects, and these differences set a
cautionary note on generalizing about the Type II-P class as a whole based on
our findings for this one object.

We obtained spectropolarimetry of SN~1999em on five occasions, with
observations on days 7, 40, 49, 159, and 163 after discovery.  We outline the
observations and basic reduction techniques in \S~\ref{sec3:2} and present the
data in \S~\ref{sec3:3}.  Section 4 discusses the spectropolarimetry, the total
flux spectra, and the implications of asphericity on the use of SNe~II-P as
extragalactic distance indicators through EPM. Conclusions are summarized in
\S~\ref{sec3:5}.  Details of the data reduction process and a discussion of
the interpretation of low-polarization spectropolarimetry similar to what is
found for SN~1999em are given in Appendix~A.  Some
preliminary results of this study were announced by Leonard, Filippenko, \&
Chornock (1999) and summarized by Leonard, Filippenko, \& Matheson (2000b).

\section{Observations and Reductions} 
\label{sec3:2}

We obtained spectropolarimetry of SN~1999em in 1999 on November 5, December 8,
and December 17 (7, 40, and 49 days after discovery, respectively) with the
Kast double spectrograph (Miller \& Stone 1993) with polarimeter at the
Cassegrain focus of the Shane 3-m telescope at Lick Observatory. Similarly, on
2000 April 5 and 9 (159 and 163 days after discovery, respectively), we used
the Low-Resolution Imaging Spectrometer (Oke et al. 1995) in polarimetry mode
(Cohen 1996\footnote{Instrument manual available at
http://www2.keck.hawaii.edu:3636/.}) at the Cassegrain focus of the Keck-I 10-m
telescope.  The slit position angle (P.A.)  was generally set near the
parallactic angle (Filippenko 1982) for the midpoint of each set of
observations in order to minimize differential light loss.  A journal of
observations is given in Table~1.

\addcontentsline{lot}{table}{\protect\numberline{3.1}Journal of
Spectropolarimetric Observations of SN~1999em}
\begin{figure}
\vskip -1.3in
\hskip -3.5in
\rotatebox{90}{
\scalebox{1.4}{
\plotone{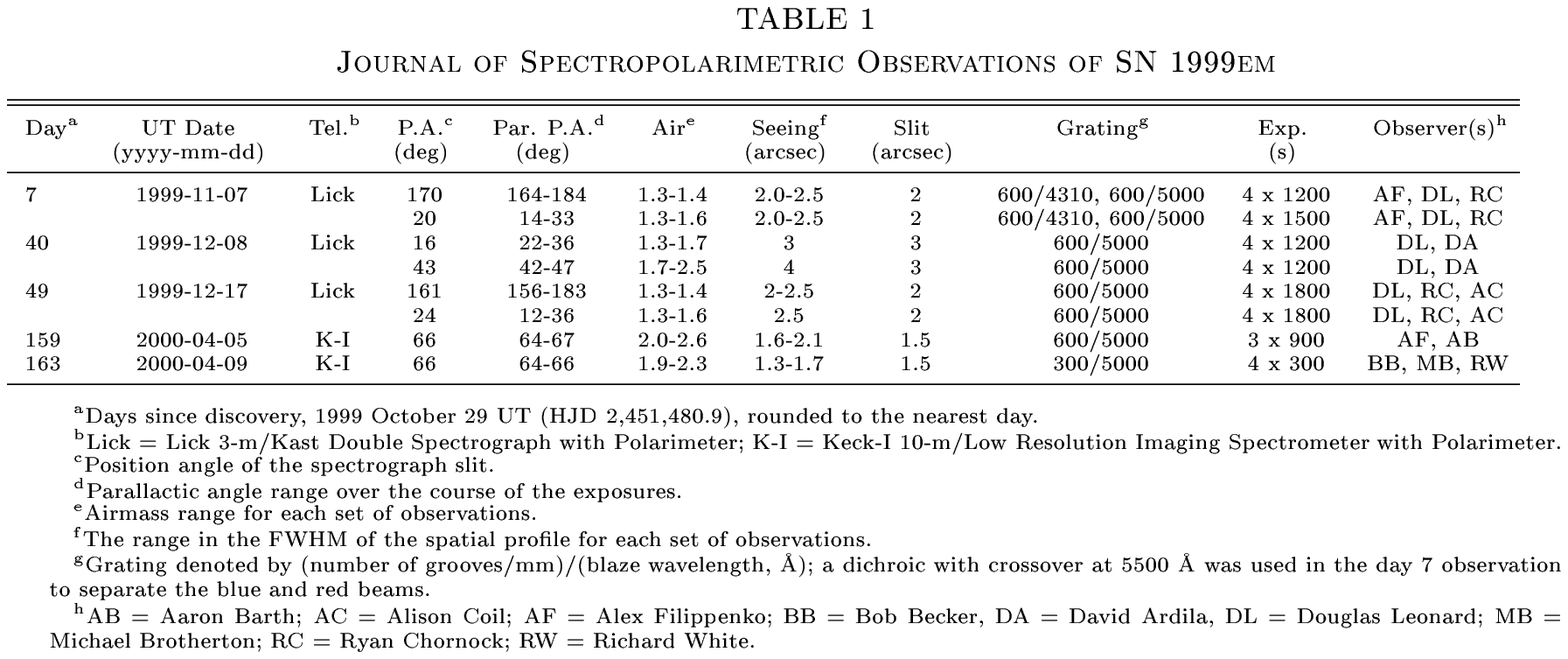}
} }
\end{figure}

One-dimensional sky-subtracted spectra were extracted optimally (Horne 1986),
with a width of from 3 to 5 times the full width at half maximum (FWHM) of the
object's spatial profile (i.e., the ``seeing,'' although guiding errors and
dome turbulence can also contribute).  This extraction width was found to
represent a good compromise between the increased signal-to-noise (S/N) ratio
obtained with wide extractions and the reduction of galaxy light contamination
achieved using a narrow aperture.\footnote{It is also advantageous to use a
large extraction window for spectropolarimetry since spurious polarization
features can be introduced by interpolating the counts in fractional pixels at
the edges of the extraction aperture.  Experiments with different extraction
widths reveal that the amplitude of such features is substantial (up to 0.5\%
in a Stokes parameter) when a width of $1-2$ times the seeing is used, but
becomes negligible ($< 0.01\%$) when the extraction window exceeds 3 times the
seeing.}  We subtracted a linear interpolation of the median values of
background windows on either side of the object from the object's spectrum;
since the polarimetry optics restrict the useful (unvignetted) slit length to
only $40^{\prime\prime}$ at Lick and $20^{\prime\prime}$ at Keck, background
regions were necessarily quite narrow and close to the object.  Each spectrum
was then wavelength and flux calibrated, corrected for continuum atmospheric
extinction and telluric absorption bands (Wade \& Horne 1988), and rebinned to
a 2~\AA\ bin$^{-1}$ linear scale.  Details of the extractions are given in
Table~2.

\addcontentsline{lot}{table}{\protect\numberline{3.2}Reduction Information for
SN~1999em Spectropolarimetry}
\begin{figure}
\vskip -2.1in
\hskip -1.3in
\rotatebox{0}{
\scalebox{1.3}{
\plotone{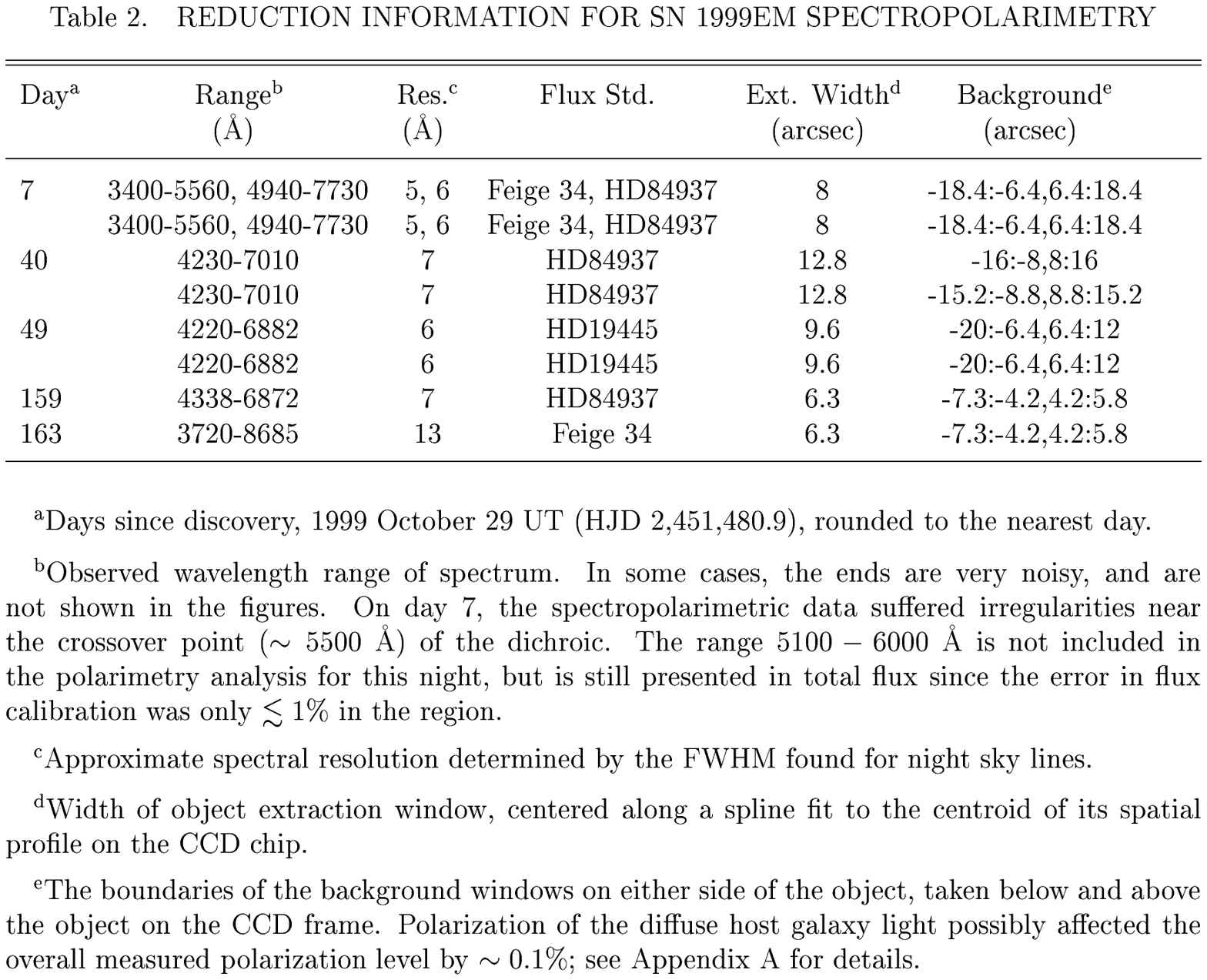}
} }
\end{figure}

We performed polarimetric analysis according to the methods outlined by Miller,
Robinson, \& Goodrich (1988) and Cohen et al. (1997).  The P.A. offset between
the half-wave plate and the sky coordinate system was determined for each
observing run using polarized standard stars from the lists of Hiltner (1956),
Mathewson \& Ford (1970), Hsu \& Breger (1982), and Schmidt, Elston, \& Lupie
(1992b).  The absolute P.A. adopted for the standards is given in Table~3, along
with the polarization angle measured for other polarized standards observed,
but not used to determine the offset.  When multiple polarization standards
were observed on the same night, they always agreed to within $\pm 0.5^\circ$.

\addcontentsline{lot}{table}{\protect\numberline{3.3}Polarized Standard Stars
Observed during Nights of SN~1999em Observations}
\begin{figure}
\vskip -2.1in
\hskip -1.6in
\rotatebox{0}{
\scalebox{1.3}{
\plotone{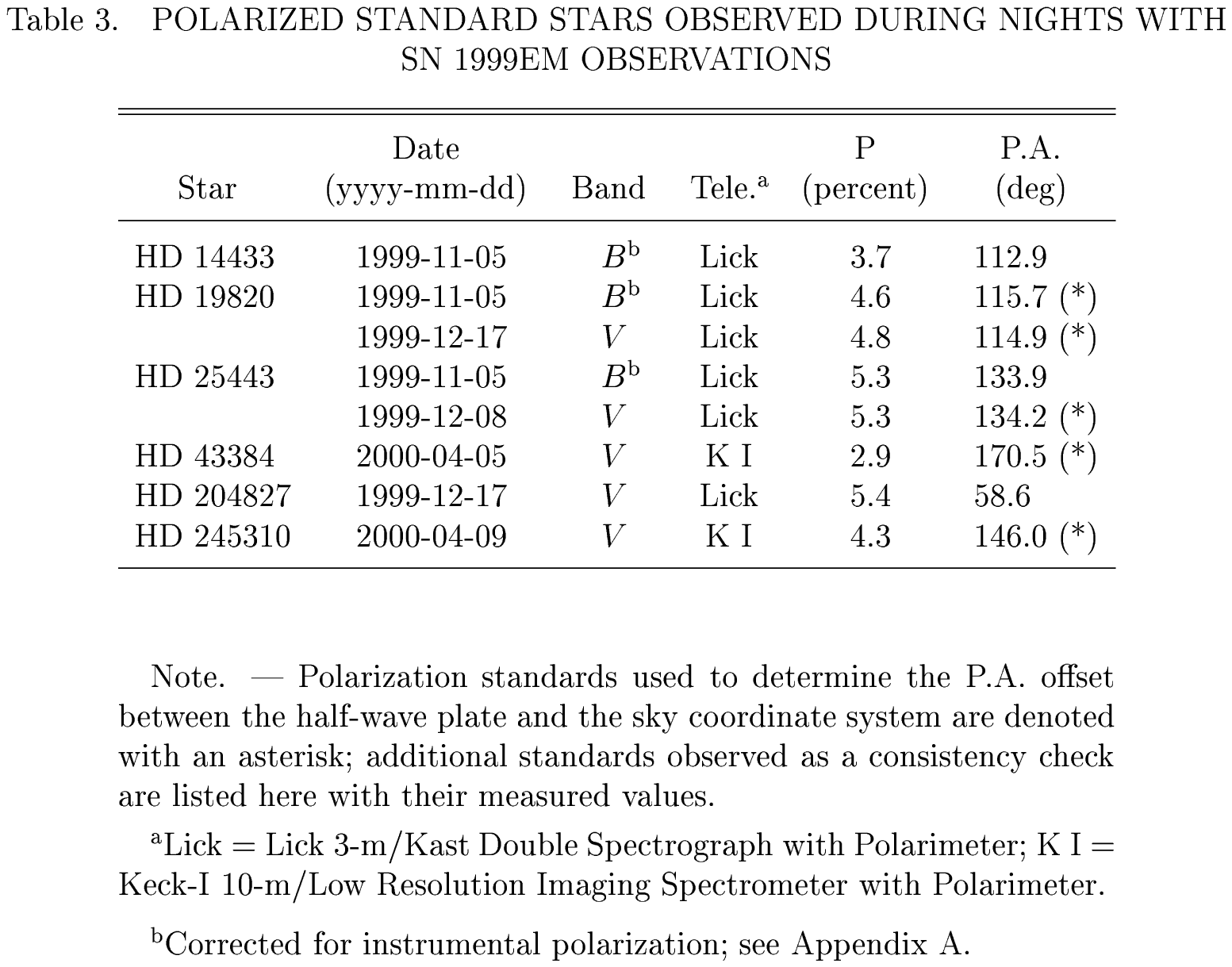}
} }
\end{figure}

There were a few peculiarities encountered in the reductions.  Briefly, on day
7, we used a dichroic filter that split the spectrum at about 5500~\AA\ and
sent the light to separate gratings that dispersed the spectra in the blue and
red arms of the spectrograph.  This yielded spectral coverage from 3400~\AA\ to
7700~\AA, but the region around the dichroic cutoff ($5100 - 6000$~\AA)
suffered from irregularities at the $1\%$ level that did not calibrate out of
the polarimetry; thus, we discarded the range $5100 - 6000$~\AA\ in the
polarimetric analysis.  In addition, instrumental polarization ranging up to
$\sim 0.4\%$ was detected at blue wavelengths ($\lambda \lesssim 5000$~\AA) in
this setup and had to be removed using the observation of the null standard on
this night.  On days 40 and 49 we observed without the dichroic and obtained
continuous coverage over $4300 - 7000$~\AA\ using only the red arm of the
spectrograph.  Observations of the null standards HR 7783 and HR 8086 on day 49
showed a flat, well-behaved polarization response.  A small polarization ($<
0.2\%$) was measured to both nulls.  However, observations of these ``nulls''
on nights when additional null standards were observed suggest that much of
this polarization is likely intrinsic to the stars themselves (i.e., the
interstellar medium along the line-of-sight [l-o-s]) and not of instrumental
origin.  Therefore, we did not apply any instrumental correction to the data
from these nights.  Observations of null standards at Keck were always found to
be null to within 0.1\%, suggesting negligible instrumental polarization in the
day 159 and 163 data.  On day 159, a CCD readout failure occurred during the
observation of the final waveplate position ($67.5^{\circ}$), leaving us with
only 3 of the desired 4 observations in the set.  While it was still possible
to derive the Stokes parameters, the data had potential systematic errors.
Fortunately, SN~1999em was observed again just four days later at Keck, largely
confirming the results of the earlier observation.  A complete discussion of
these issues, along with an investigation into sources of systematic
uncertainty in measurements of SN~polarization, is given in Appendix~A.

The displayed polarization $p$ is the ``rotated Stokes parameter'' (RSP),
calculated by rotating the normalized Stokes parameters $q$ and $u$ by a fit to
the continuum P.A. curve so that all of the continuum polarization falls in a
single Stokes parameter (here, rotated $q$).  This is preferable to calculating
either the ``traditional'' $p$ ($p_{trad}\equiv \sqrt{q^2 + u^2}$), or a
``debiased'' $p$ ($p_{deb} \equiv \pm \sqrt{\mid q^2 + u^2 - (\sigma^2_q +
\sigma^2_u)\mid}$, where $p_{deb}$ assumes the sign of [$q^2 + u^2 - \sigma^2_q
- \sigma^2_u$]; see Stockman \& Angel 1978), since the former is biased high
and is positive-definite while the latter gives poor results for data with low
$p/\sigma_p$ (e.g., Miller et al. 1988).  RSP provides a good measure of $p$
when the P.A. is a slowly varying function of wavelength.  However, when there
are sharp P.A. rotations across line features, RSP always underestimates the
true polarization, and it is necessary to examine both RSP and URSP (rotated
$u$) to determine the line's polarization.  To measure absorption-line and
emission-line polarizations, we first binned the flux data into 10~\AA\ (rest)
bins (about 1.5 resolution elements wide for the standard Lick and Keck
setups), and then derived $q$ and $u$ from the binned data.  Although binning
always reduces peak polarization, it is more reliable because of the reduced
noise, which is especially problematic here since polarization peaks are often
associated with line troughs.  However, since features may be quite sharp, it
often happens that the polarization change is detectable in only a few
(rebinned) pixels.  In such cases, we carefully examine the polarization
properties of the original spectrum (always binned 2~\AA\ bin$^{-1}$) to see if
the polarization feature is seen consistently across many pixels or is due to a
single deviant pixel, before concluding that a true polarization change exists.
We determined observed $B$-band and $V$-band polarization by calculating the
``debiased'', flux-weighted averages of $q$ and $u$ over the intervals
$3950-4900$~\AA\ and $5050-5950$~\AA, respectively, converting to $p$ and
$\theta$ only at the final step for quoting results.  A more extensive
description of the adopted analysis techniques is given in Appendix~A.

\section{Results}
\label{sec3:3}

Figures~\ref{fig3:2}, \ref{fig3:3}, \ref{fig3:4}, and \ref{fig3:5} show the
observed polarization data for SN~1999em on days 7, 40, 49, and 161 (a
combination of the observations on day 159 and 163) after discovery.  The data
from days 7, 40, and 49 are the weighted averages of two successive
observations taken on those nights, and the ``day 161'' data are the weighted
average of data obtained on days 159 and 163; see Appendix~A for details.  A
montage of all the epochs studied is shown in Figure~\ref{fig3:10}.  Measured
values for the observed $V$-band polarization are given in Table~4, along
with line trough polarization for \ion{Fe}{2} $\lambda 5169$ and \ion{Na}{1} D
$\lambda\lambda 5890, 5896$, and the approximate polarization change measured
across the H$\alpha$ profile.

The day 7 flux spectrum shows features that are typical for a very young
SN~II-P: a blue continuum with P-Cygni lines of hydrogen Balmer and He I
$\lambda 5876$. The blue edges of absorption troughs indicate velocities in
excess of $10,000 {\rm\ km\ s}^{-1}$\ for the fastest moving gas. The spectra
from days 40 and 49 are typical for a SN~II-P during the plateau phase,
characterized by a thermal continuum with P-Cygni hydrogen Balmer, \ion{Fe}{2}
$\lambda\lambda 4924, 5018, 5169 $, \ion{Sc}{2} $\lambda\lambda 5526, 5657,
6245 $, and \ion{Na}{1} D $\lambda\lambda 5890, 5896$.  The plateau phase for
SN~1999em lasted until about 90 days after discovery, suggesting that a massive
hydrogen envelope remained at the time of explosion (Leonard et al. 2001).  The
day 161 spectrum is emission-line dominated, with prominent H$\alpha$ and
[\ion{Ca}{2}] $\lambda\lambda 7291, 7324$, implying that the transition to the
nebular phase was underway; P-Cygni absorptions still indicate significant
continuum light, however, and the fully nebular phase is not expected until
after about day 300 (e.g., Jeffery 1991a).

\addcontentsline{lot}{table}{\protect\numberline{3.4}SN~1999em Polarization Data}
\begin{figure}
\vskip -0.1in
\hskip -2.2in
\rotatebox{90}{
\scalebox{1.2}{
\plotone{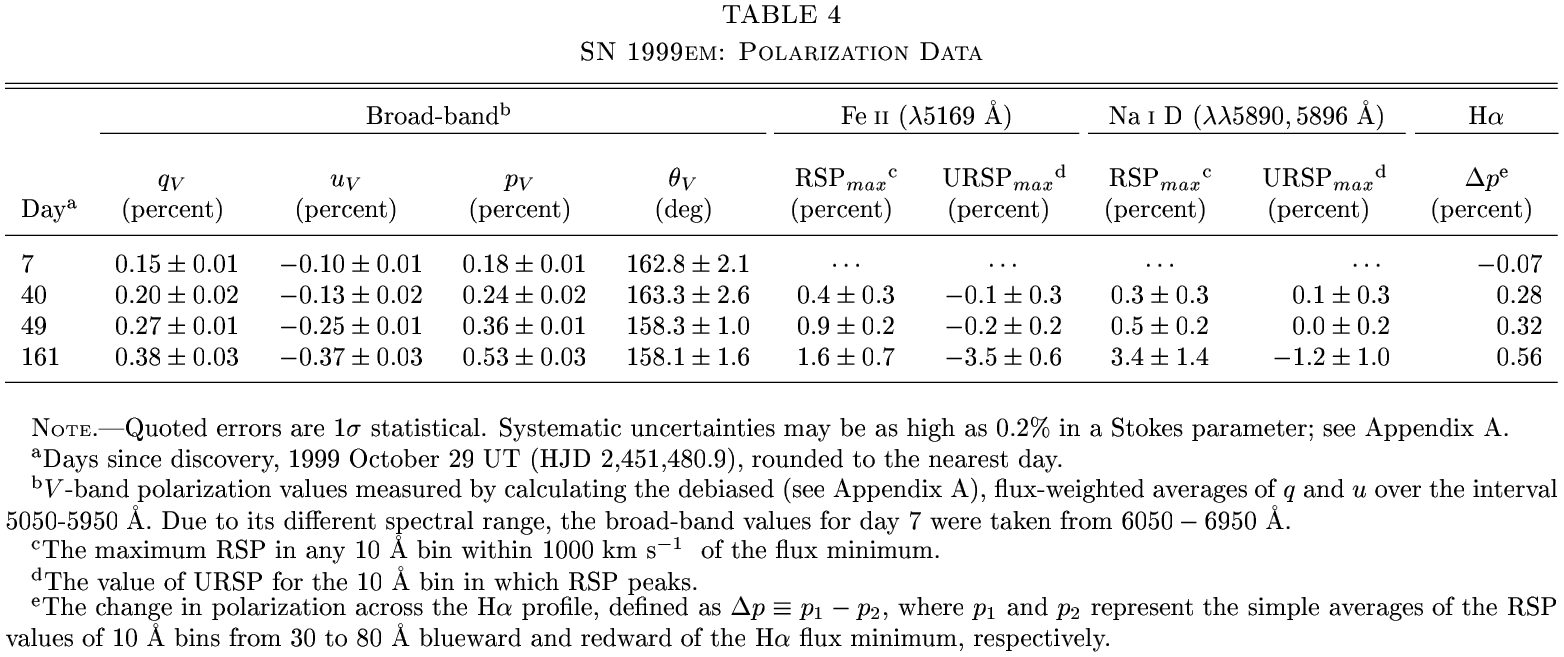}
} }
\end{figure}

\begin{figure}
\begin{center}
 \scalebox{0.55}{
	\plotone{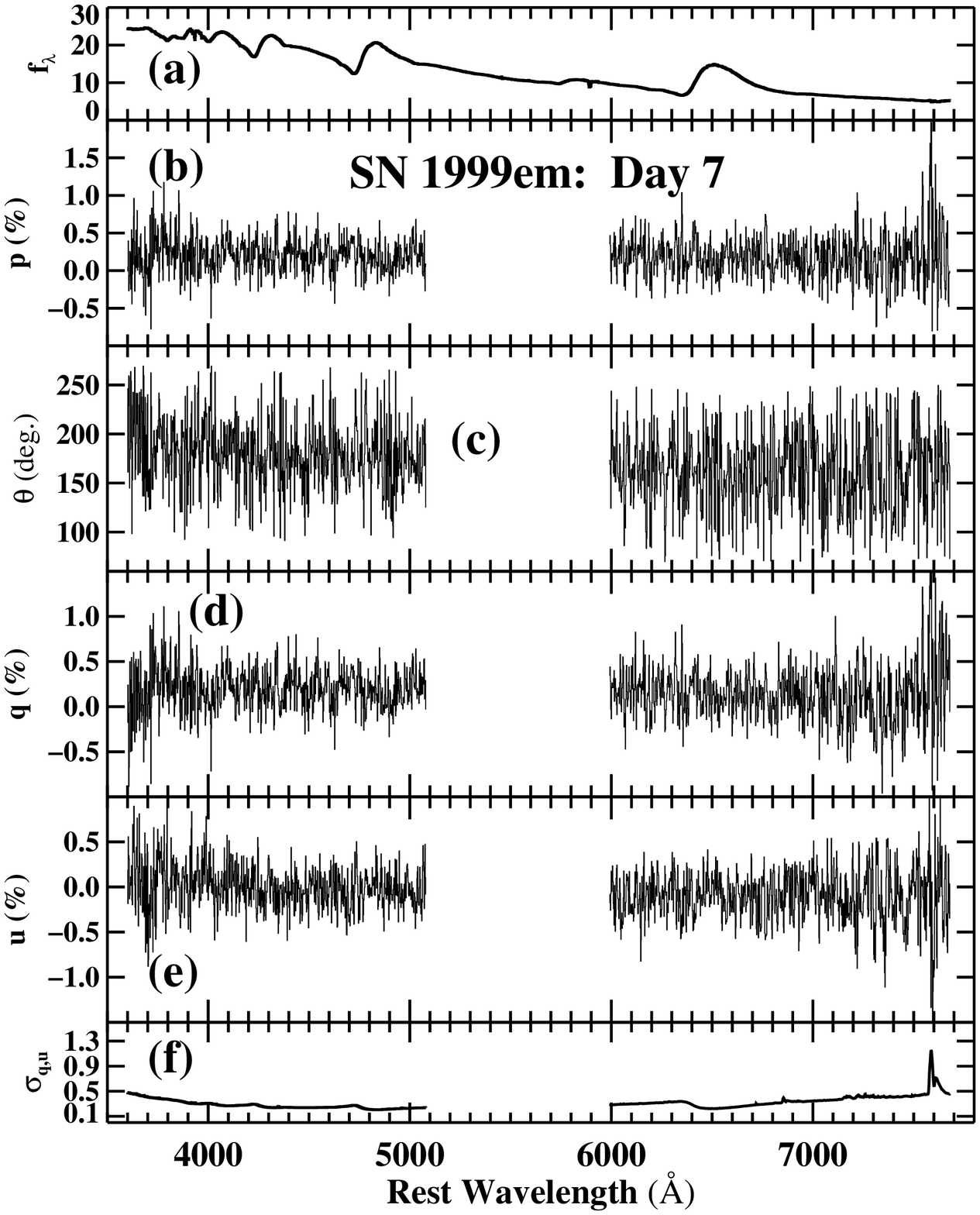}
		}
\end{center}
\caption[Polarization data for SN~1999em 7 days after discovery] {Polarization
data for SN~1999em, obtained 1999 November 5, only 7 days after discovery.  In
this, and all subsequent figures, the NED recession velocity of 717 km s$^{-1}$
has been removed. Instrumental polarization at blue wavelengths has been
removed from the data on this night only (see Appendix A).  ({\it a}) Total
flux, in units of $10^{-15}$ ergs s$^{-1}$ cm$^{-2}$ \AA$^{-1}$.  ({\it b})
Observed degree of polarization.  ({\it c}) Polarization angle in the plane of
the sky. ({\it d, e}) The normalized $q$ and $u$ Stokes parameters. ({\it f})
Average of the (nearly identical) $1\sigma$ statistical uncertainties in the
Stokes $q$ and $u$ parameters for the displayed binning of 2~\AA\ bin$^{-1}$.
The polarization shown in this and all plots is actually the ``rotated Stokes
parameter'' (RSP; see Appendix B).  To facilitate comparison, the ordinate
ranges of panels ({\it b}), ({\it c}), ({\it d}), ({\it e}), and ({\it f}) are
the same in Figures~\ref{fig3:2}, \ref{fig3:3}, and ~\ref{fig3:4}. }

\label{fig3:2}
\end{figure}

\begin{figure}
\begin{center}
 \scalebox{0.90}{
	\plotone{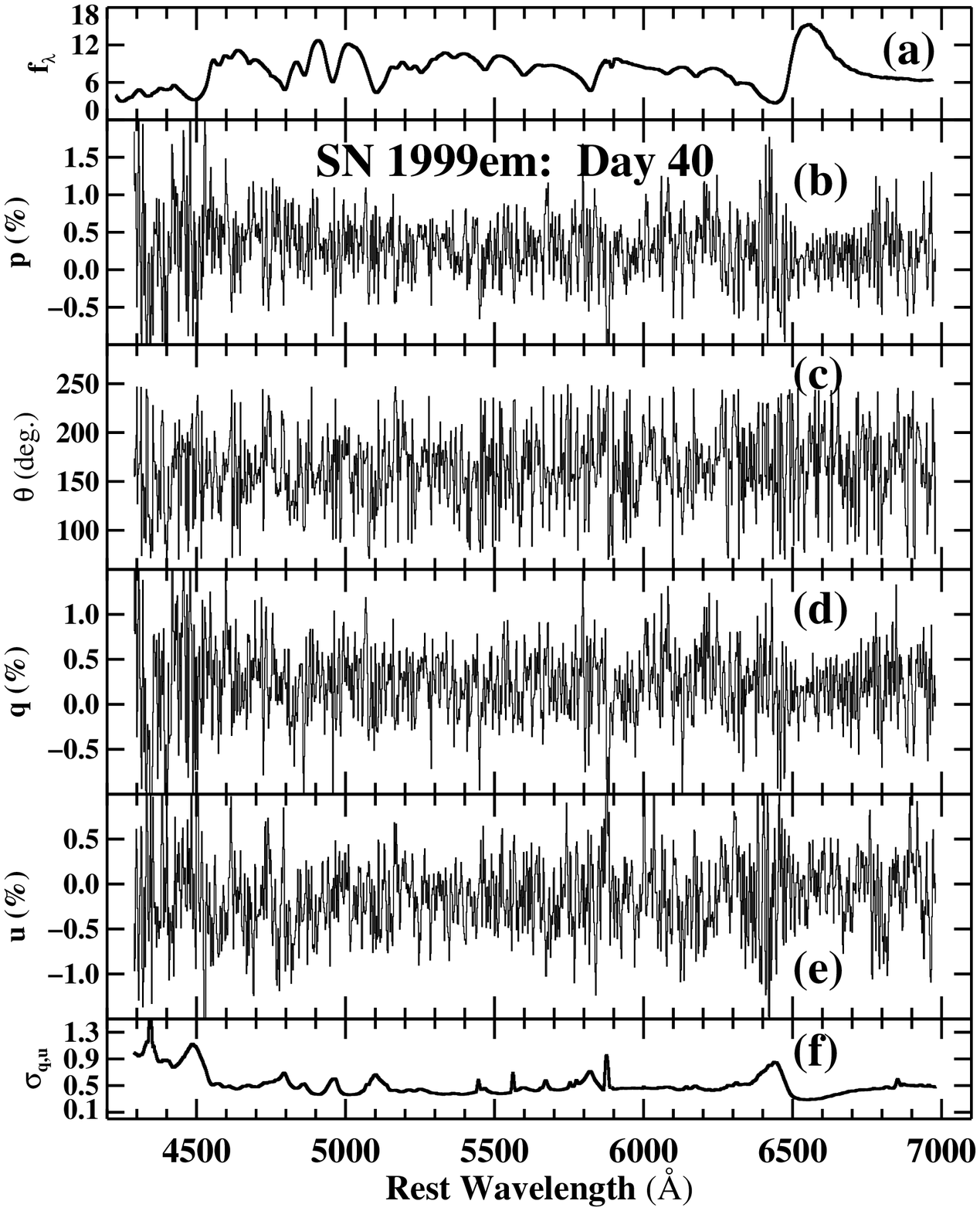}
		}
\end{center}
\caption[Polarization data for SN~1999em 40 days after discovery] {As in
Figure~\ref{fig3:2}, but for data obtained on 1999 December 8.}
\label{fig3:3}
\end{figure}

\begin{figure}
\begin{center}
 \scalebox{0.9}{
	\plotone{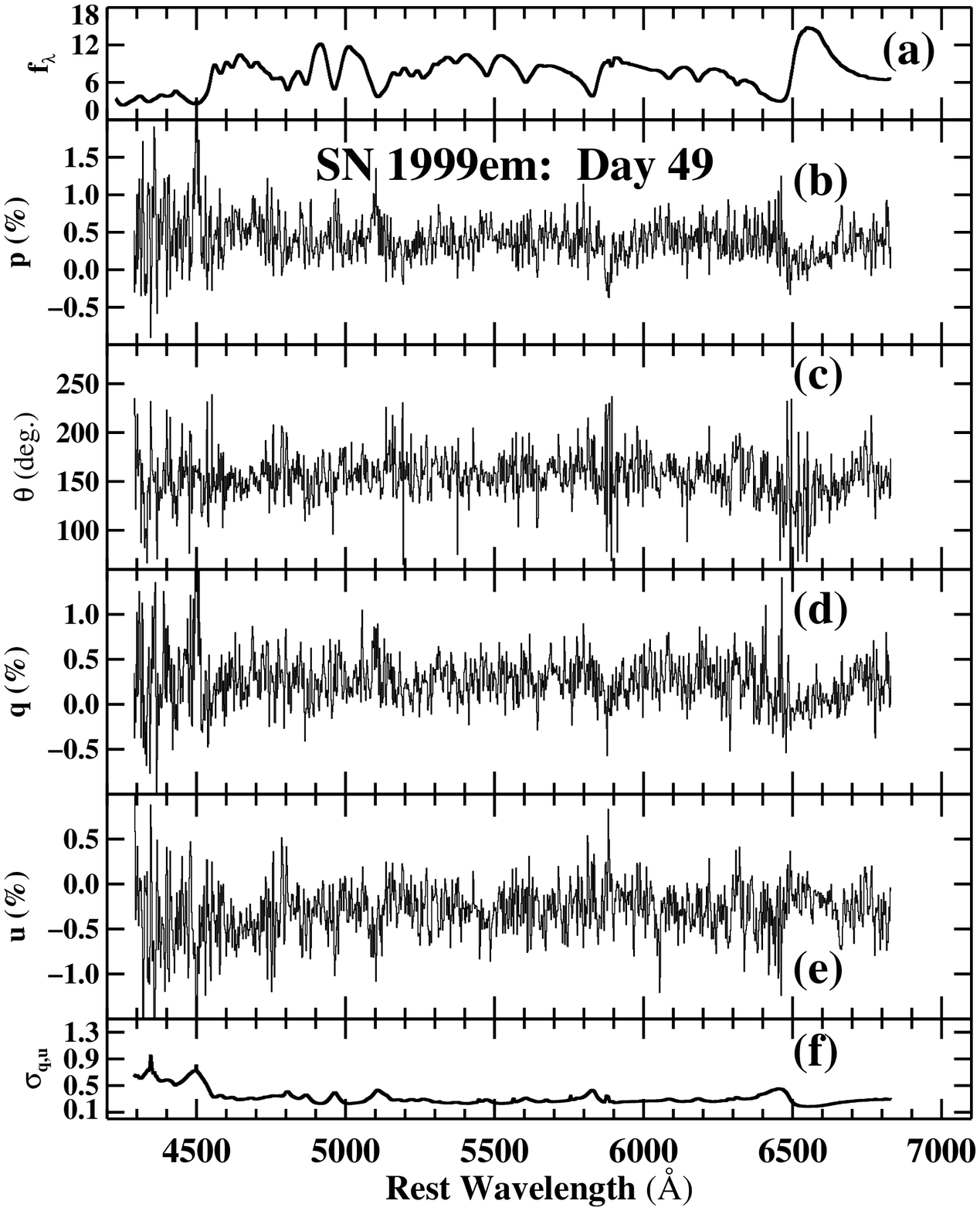}
		}
\end{center}
\caption[Polarization data for SN~1999em 49 days after discovery]
{As in Figure~\ref{fig3:2}, but for data obtained on 1999 December 17.}
\label{fig3:4}
\end{figure}

\begin{figure}
\begin{center}
 \scalebox{0.9}{
	\plotone{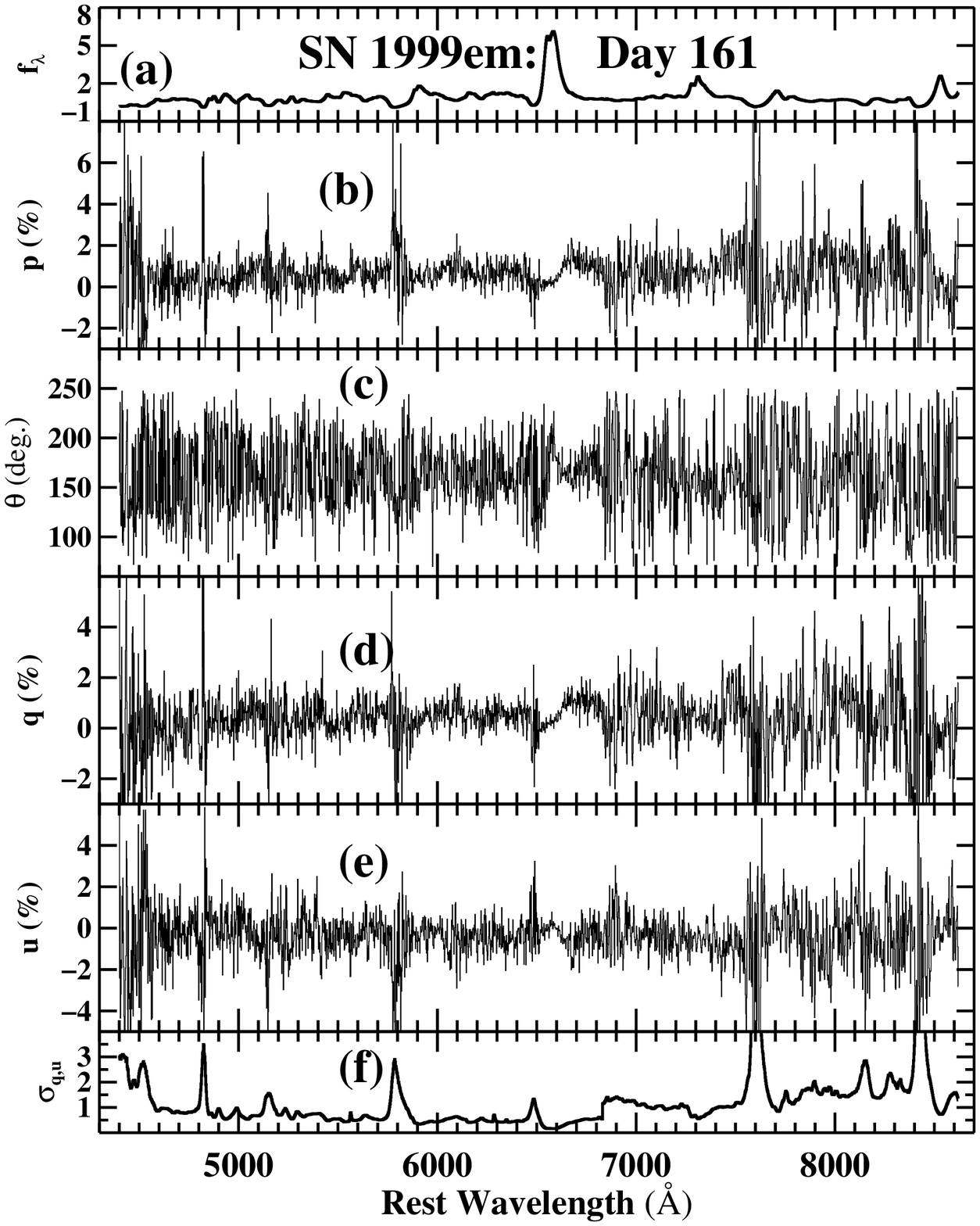}
		}
\end{center}
\caption[Polarization data for SN~1999em 161 days after discovery]
{As in Figure~\ref{fig3:2}, but for the weighted average of polarization data
obtained on 2000 April 5 and 9, 159 and 163 days after discovery,
respectively.}
\label{fig3:5}
\end{figure}

\begin{figure}
\begin{center}
 \scalebox{0.7}{
	\plotone{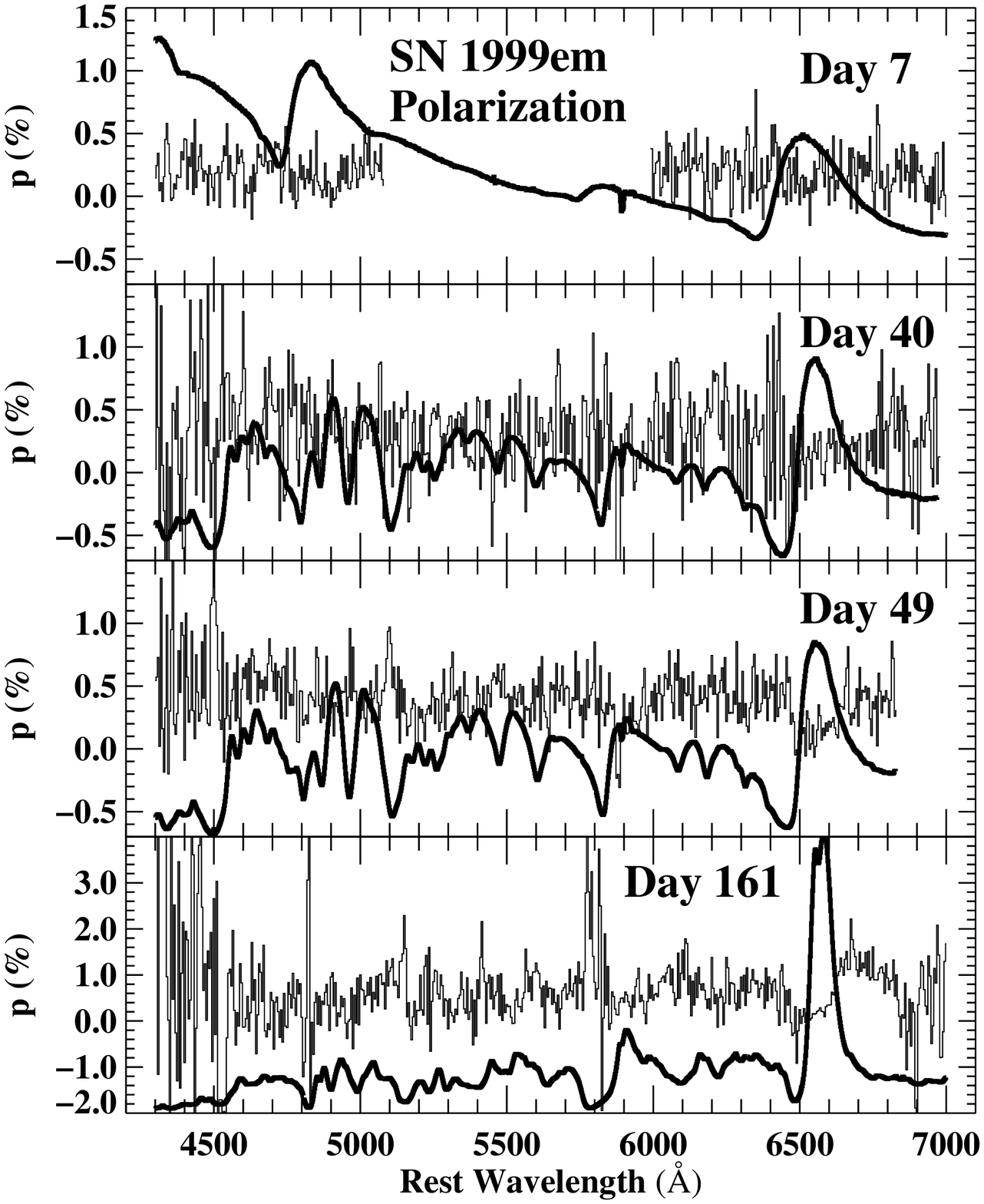}
		}
\end{center}
\caption[Time evolution of the observed optical polarization of SN~1999em]
{Montage of the observed optical polarization of SN~1999em, with arbitrarily
scaled total flux spectra ({\it thick lines}) overplotted for comparison of
features.  The polarization spectra are binned 5 \AA\ bin$^{-1}$ to improve the
S/N ratio.}
\label{fig3:10}
\end{figure}

\section{Analysis and Discussion} 
\label{sec3:4}

\subsection{Spectropolarimetry}
\label{sec3:4.1}

\subsubsection{Introduction}
\label{sec3:4.1.1}

We interpret the spectropolarimetry of SN~1999em in terms of the physical
picture shown (for a spherical atmosphere) in Figure~\ref{fig1:3}, and
described in \S~\ref{sec3:1}.  Early theoretical work on SN spectropolarimetry
was stimulated by observations of SN~1987A.  Investigators initially proposed
that the observed polarization (up to $p_V \approx 0.7\%$ during the first 100
days after explosion) was due to the polarizing effect of line scattering
(Jeffery 1989).  It was soon realized, however, that scattering by free
electrons extending out into the line-forming regions of the envelope dominates
the overall polarization level of photospheric SNe (H91), and that weak
collisions in the line-forming region of the SN atmosphere tend to destroy the
intrinsic line polarizing effect, except perhaps in the outer regions of the
atmosphere at late times (Jeffery 1991b).  The loss of geometric information
caused by the collisions generally produces unpolarized emission components for
the P-Cygni line profiles, an assumption often used to constrain the
interstellar polarization (ISP; see, e.g., Leonard et al. 2000a).  Detailed
models specific to SN~1987A suggest that the net polarization seen was likely
produced by a modest asphericity of about $10\%-20\%$ (Jeffery 1991b; H91; see
also Wang \& Wheeler 1996).

Translating observed SN continuum polarization into an estimate of the
asphericity of the electron-scattering atmosphere is quite model-dependent.
The original problem of the polarization of light emerging from a
semi-infinite, plane-parallel, pure electron scattering atmosphere with
constant net flux was solved by Chandrasekhar (1946).  The first attempt to
apply the formalism to SNe was by Shapiro \& Sutherland (1982), who solved it
for a plane-parallel, optically-thick, electron-scattering atmosphere wrapped
around an ellipsoid.  The plane-parallel model, however, requires extreme
asymmetry to produce even modest polarization: to get $p = 1\%$ demands an
asphericity of at least 60\% (the percent asphericity is defined here as
$(1-[a/b]) \times 100$, where $a$ and $b$ are the semiminor and semimajor axis
lengths, respectively) for the oblate case; prolate spheroids are unable to
produce polarization of more than 0.77\% for any axis ratio or viewing angle.
Realistic models of the extended atmospheres of SNe, however, have recently
been shown to be more polarizing than the plane-parallel approximation
predicts.

There are many factors that can influence the degree of SN continuum
polarization.  The most recent, general study of the polarization expected from
an aspherical SN atmosphere is H91, in which the continuum polarization
resulting from an axially symmetric (either oblate or prolate),
electron-scattering dominated photosphere is calculated as a function of the
photosphere's geometry and size relative to the envelope, the scattering
atmosphere's optical depth and density profile, and the axial ratio of the
envelope.  Particularly useful is the lower bound placed on the asphericity
implied by polarization measurements: by fixing all other parameters to
maximize the resulting polarization (i.e., scattering optical depth = 1,
continuum region physically small compared with envelope, object viewed
equator-on), the maximum polarization as a function of axis ratio is produced
(Figure 4 of H91).  Determination of intrinsic SN continuum polarization can
thus lead to a lower limit on the implied asphericity of the scattering
atmosphere from this model.

In addition to the information gleaned from the continuum polarization level,
sharp polarization changes across strong line features can also contain a
wealth of information about the geometry of the scattering environment.  The
influence of asphericity on the polarization properties of P-Cygni lines was
first addressed in detail by McCall (1984), who predicted that polarization
should increase in the absorption troughs.  His basic idea is that P-Cygni
absorption by gas in region 1 (Fig.~\ref{fig1:3}) selectively blocks photons
coming from the central, more forward-scattered (hence less polarized) regions,
thereby enhancing the relative contribution of the more highly polarized
photons from the limb (i.e., from regions 2 and 3 in Fig.~\ref{fig1:3}; see
also Fig.~1 of Leonard et al. 2000b).  Jeffery (1991b) extended the argument
by predicting drops in polarization at the location of emission peaks due to
the dilution of the polarized continuum light by unpolarized emission-line
photons.

The basic prediction, then,  is that a ``reverse'' P-Cygni polarization profile should
characterize strong P-Cygni line features formed above or within an aspherical
electron-scattering atmosphere.  (We note that polarization increases in
P-Cygni absorption troughs are often seen in spectropolarimetric studies of
broad absorption line QSOs, and are generally given a similar physical
interpretation [e.g., Brotherton et al. 1997; Ogle et al. 1999]).  Taking the
limiting case of blockage of only unpolarized light (or light with electric
vectors that exactly cancel) in a line trough and addition of completely
unpolarized light in a line peak, the polarization change expected across a
P-Cygni feature from this admittedly simple model is
\begin{equation}
p_{trough} \leq p_{max} = p_{cont} \times \frac{I_{cont}}{I_{trough}},
\label{eqn3:1a}
\end{equation}
\begin{equation}
p_{peak} \geq p_{min} = p_{cont} \times \frac{I_{cont}}{I_{peak}},
\label{eqn3:1b}
\end{equation}
\noindent where $p_{cont}$ is the continuum polarization and $I_{cont},
I_{peak}, {\rm\ and\ } I_{trough}$ represent the flux intensity in the
continuum, line peak, and line trough, respectively.  The continuum
polarization and the continuum flux intensity are both interpolated from the
continuum regions surrounding the line.

In general, the polarization observed across a P-Cygni profile will differ from
that predicted by equations~(\ref{eqn3:1a}) and (\ref{eqn3:1b}).  The most
important factor neglected by the simple geometric argument is the influence
that resonance scattering of continuum photons has on the polarization
properties of the line profile.  In a SN atmosphere, continuum photons that are
resonantly scattered by a line are unpolarized (Jeffery 1991b; H\"{o}flich et
al.  1996), and will diminish the polarization level throughout the entire
P-Cygni profile.  This depolarizing effect will be especially pronounced in
lines that are optically thick throughout the envelope, such as H$\alpha$.  In
addition, the strong emission component of \halpha\ will tend to fill in the
absorption trough, increasing $I_{trough}$.  Countering this is the
polarization that line photons themselves may acquire upon being scattered by
free electrons, which likely exist in the line-formation region itself (H\"{o}flich et
al.  1996 ).
A final reason that the polarization measured will differ from that predicted
is purely observational: since the P-Cygni absorption features in SNe~II-P are
quite sharp (the FWHM of the absorptions in SN~1999em due to prominent metal
lines on day 49, for example, are all in the range $35 - 45$~\AA), the typical
grating-CCD resolution of $5 - 10$~\AA\ coupled with the coarse binning needed
to improve the statistics inevitably reduces measured peak polarization
values. Indeed, in the limit of infinite bin size, $p_{trough} \rightarrow
p_{cont}$.

There are thus several mechanisms in addition to the simple geometric model
that may contribute to the polarization measured across a P-Cygni line feature
and it is not obvious which should dominate.  In this regard, it is instructive
to study the characteristics of the line polarization presented by the detailed
spectropolarimetric studies of SN~1987A (Jeffery 1991a, 1991b), SN~1993J
(Trammell et al. 1993; H\"{o}flich et al.  1996 ; Tran et al. 1997), and
SN~1998S (Leonard et al. 2000a).

For SN~1987A, many epochs of the ISP-subtracted data show large polarization
increases at the locations of P-Cygni absorptions due to H$\gamma$, H$\beta$,
\ion{Fe}{2} $\lambda 5169$, \ion{Na}{1} D $\lambda\lambda 5890, 5896$,
\ion{Ca}{2} $\lambda\lambda 8498, 8542, 8662$, and possibly \ion{Fe}{2}
$\lambda 5018$ and H$\alpha$ as well (see Jeffery 1991a and references
therein).  Polarization decreases occur across the emission peaks of H$\alpha$
and \ion{Ca}{2} $\lambda\lambda 8498, 8542, 8662$.  Since error bars are not
given and the polarization data were evidently not corrected for statistical
biases, it is somewhat difficult to quantitatively assess the polarization
changes (especially the trough increases) seen across the lines.  Taken at face
value, though, none of the trough increases appears to exceed the limit given
by equation~(\ref{eqn3:1a}).  Polarization decreases across H$\alpha$, however,
do occasionally go below the limit set by equation~(\ref{eqn3:1b}).

The three ISP-subtracted spectropolarimetric epochs of SN~1993J (Type IIb)
presented by Tran et al. (1997) also show distinct polarization increases in
the absorption troughs of H$\beta$, \ion{Fe}{2} $\lambda 5169$, and
\ion{He}{1} $\lambda 5876$ (probably blended with \ion{Na}{1} D $\lambda\lambda
5890, 5896$), and polarization decreases across the \ion{He}{1} $\lambda 5876$
and blended H$\alpha$/\ion{He}{1} $\lambda 6678$ emission lines (the reverse
P-Cygni profile across the \ion{He}{1} $\lambda 5876$ is particularly
pronounced).  Detailed modeling of a single early-time spectropolarimetric
epoch of SN~1993J presented by H\"{o}flich et
al.  1996  successfully reproduces the depolarization
across the entire H$\alpha$ profile, but does somewhat less well at fitting the
\ion{He}{1}/\ion{Na}{1} D feature: whereas the model predicts a polarization
decrease at the line's absorption trough, the data indicate a small
polarization rise.  It is possible that the increased importance selective
blocking of forward-scattered light has in lines of low optical depth
contributes to the different polarization behavior seen here between H$\alpha$
and \ion{He}{1}/\ion{Na}{1} D.  

Perhaps the clearest demonstration of the influence that unpolarized emission
lines have on SN spectropolarimetry, however, is the single spectropolarimetric
epoch of SN~1998S discussed by Leonard et al. (2000a).  At the very early epoch
studied (just 5 days after discovery), this peculiar Type IIn event displayed a
smooth, blue continuum with superposed broad, symmetric emission lines lacking
any absorption components.  The spectropolarimetry was characterized by a flat,
highly polarized ($p \approx 3\%$ after ISP removal) continuum with pronounced
polarization decreases across all the emission features.

Aspects of the predicted reverse P-Cygni polarization profile are thus found in
the spectropolarimetry of all previously studied SNe~II.  However, SN~1987A,
SN~1993J, and SN~1998S were all quite unusual core-collapse events.  SN~1993J
and SN~1998S both showed strong spectroscopic evidence for interaction with
circumstellar material.  In addition to being a {\it blue} supergiant, the
progenitor of SN~1987A may also have been spun up by a merger with a companion
star prior to explosion (Collins et al. 1999).  Such circumstances add
complexity to the interpretation of spectropolarimetry.  Since SN~1999em
appears to be a relatively normal SN~II-P, it presents a great opportunity to
study the impact of asphericity on line polarization without any added
complication by ``external'' influences.

\subsubsection{Interstellar Polarization}
\label{sec3:4.1.2}

A major source of uncertainty in SN spectropolarimetry is the degree to which
interstellar dust along the l-o-s polarizes the SN light.  An assumption often
invoked to determine the ISP to SNe is that broad, resonance-scattered and
recombination line photons (e.g., H$\alpha$) at early phases are intrinsically
unpolarized; see \S~\ref{sec3:4.1.1} for the theoretical underpinnings behind
this assumption, and Appendix~B for details of the application of this
technique (see also Jeffery 1991b; Trammell et al. 1993; H\"{o}flich et al.
1996 ; Wang et al. 1996; Tran et al. 1997).  With this assumption, the observed
polarization of the emission-line photons is then entirely due to the ISP.  Our
earliest epoch, day 7, has very low continuum polarization and no obvious line
features (see Fig.~\ref{fig3:16} for detail of the \halpha\ region).  For
H$\alpha$ emission from $6404$ to $6622$~\AA\ ($\pm\ 5000 {\rm\ km\ s}^{-1}$
from the deredshifted line peak at 6512~\AA) and background regions
$6000-6250$~\AA\ and $6900-7500$~\AA, we find $p({H\alpha}) = 0.08 \pm 0.06\%
{\rm\ at\ }\theta = 171^\circ \pm 15^\circ$.  This value is consistent with the
continuum measured in the range $6900 - 7500$~\AA\ ($p = 0.12 \pm 0.02\% {\rm \
at\ } \theta = 153 \pm 5^\circ$), and therefore allows for the possibility that
all of the polarization seen is interstellar in origin.  On the other hand, the
polarization found is also marginally consistent with zero polarization, which
would allow all of the observed polarization to be intrinsic to the SN.  Our
data therefore do not distinguish well between an ISP of 0.12\% and an ISP of
zero with this test.  In any event, though, if \halpha\ emission-line photons
are really intrinsically unpolarized at this early epoch, it does suggest very
low ISP from dust in either the host galaxy or Milky Way (MW), with a formal
$3\sigma$ upper limit of ${\rm ISP} < 0.26\%$.

There are additional indications of low ISP contamination along the l-o-s to SN
1999em.  First, since the same dust that polarizes also reddens SN light, with
the maximum polarization always found to be less than $9E(B-V)$ for the MW
(Serkowski, Mathewson, \& Ford 1975), a small ISP for SN~1999em is consistent
with its low Galactic reddening of $E(B-V)_{MW} = 0.04$ mag (Schlegel,
Finkbeiner, \& Davis 1998), as well as with the {\em total} reddening of
$E(B-V)_{tot} \approx 0.05$ mag estimated by Baron et al. (2000; see also
Leonard et al. 2001) from theoretical fits to the early-time continuum shape.
This reddening restricts the ISP to values less than $0.45\%$; we note,
however, that the upper limit of $E\BminusV = 0.15$ mag given by Baron et
al. (2000) is consistent with an ISP of up to $1.35\%$.  Second, of the 11
stars with measured polarizations within $5^\circ$ of SN~1999em (Heiles 2000),
six have zero polarization within their quoted uncertainties, and all have $p_V
< 0.2\%$, although only two (HD 29248 and HD 30112) lie sufficiently far away
($d > 300 {\rm\ pc\ }$) to fully sample the ISP through the Galactic
plane. Third, even though $p$ increases through the 4 epochs, the polarization
angle of SN~1999em remains nearly constant at $\theta \approx 160^\circ $.  If
the day 7 polarization was completely due to ISP it would suggest that
$\theta_{ISP} = \theta_{SN}$, an unlikely chance occurrence.  Furthermore,
optical imaging polarimetry of face-on spirals (Scarrott, Rolph, \& Semple
1990; Scarrott et al. 1991) generally shows polarization vectors aligned
parallel to the spiral arms (i.e., perpendicular to the line connecting a point
with the galaxy's center), which would predict $\theta_{ISP} \approx
130^{\circ}$ given the location of SN~1999em in NGC~1637, somewhat far from the
observed $\theta \approx 171^\circ$.  Finally, the nearly complete
depolarization across the H$\alpha$ line to $q\approx 0\%, u \approx 0\%$ on
days 40, 49, and 161 (see Figs. \ref{fig3:17}, \ref{fig3:18}, \ref{fig3:19})
is consistent with a near-zero ISP from physical considerations discussed in
the next section.

Therefore, while none of the preceding arguments is individually conclusive,
taken together they suggest that polarimetry of SN~1999em is contaminated by a
relatively small degree of ISP.  Although impossible to know for sure, we
suspect that the ISP is less than $0.3\%$, and cautiously proceed to analyze
our data with no correction for its effects.  We note that since ISP is a
smoothly varying function of wavelength and is constant with time, uncertainty
in its value affects neither temporal polarization changes nor modulations
across specific line features.  We caution, however, that any interpretation of
polarization increases, decreases, or directional changes in line features is
affected by the choice of ISP (see, e.g., Fig.~10 of Leonard et al. 2000a).
Other sources of systematic uncertainty may add an additional $0.1\%$
uncertainty to the accuracy of the measured polarizations (see Appendix A).

\subsubsection{SN~1999em}
\label{sec3:4.1.3}

\begin{figure}
\begin{center}
\rotatebox{0}{
 \scalebox{0.70}{
	\plotone{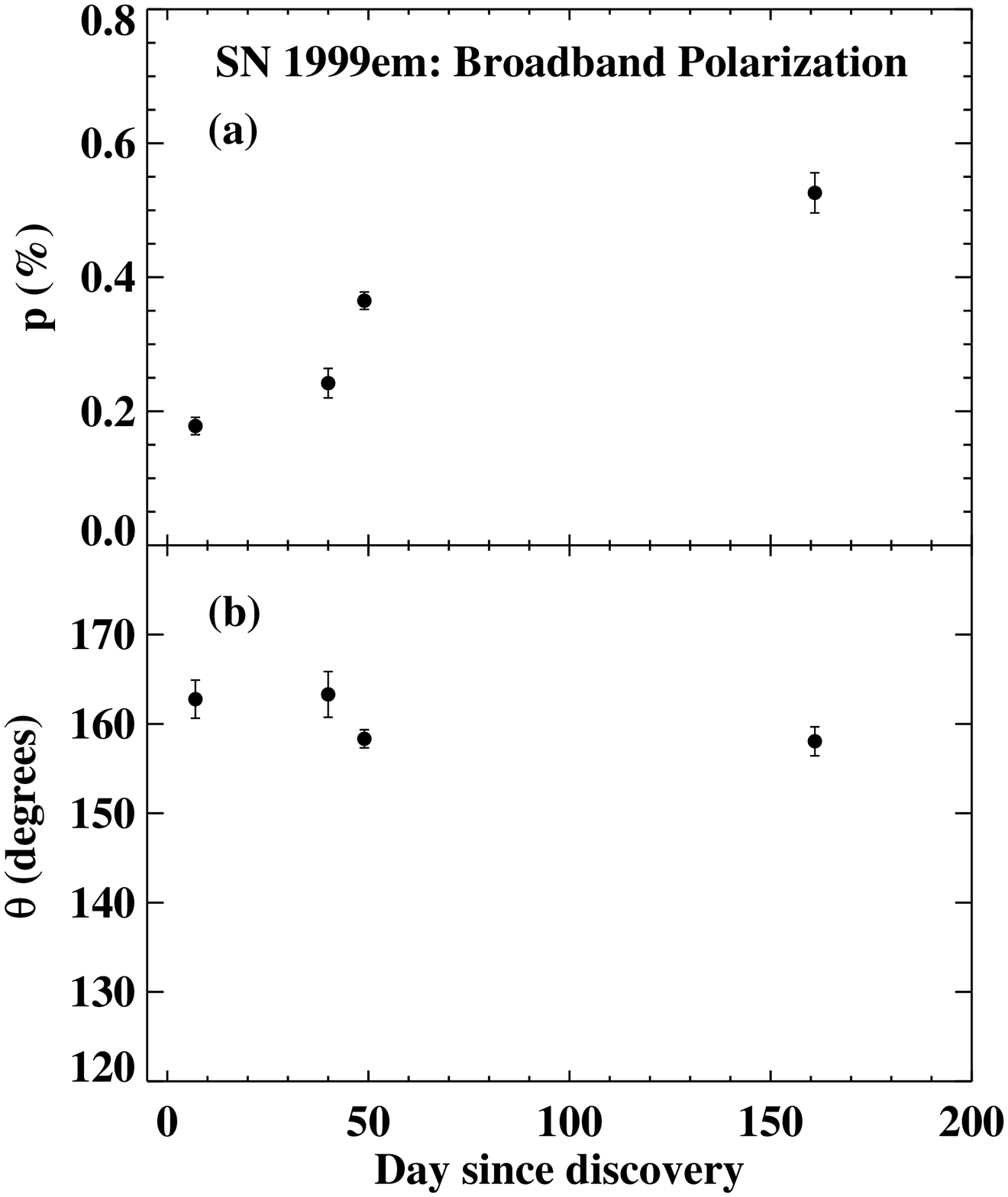}
		}
		}
\end{center}
\caption[Time evolution of the observed broadband polarization and polarization
position angle of SN~1999em] {({\it a}) Observed broadband polarization and
({\it b}) polarization position angle in the plane of the sky of SN~1999em as a
function of time.  Error bars are $1\sigma$ statistical for the observed $V$
band ($5050 - 5950$ \AA) except for day 7, which is for observed $6050 -
6950$~\AA.  Note that although the polarization more than doubles during the
first 161 days, the polarization angle remains fixed.}
\label{fig3:12}
\end{figure}

The broadband polarization of SN~1999em (generally, observed $V$) rises from
$p~=~0.18 \pm 0.01\%$ on day 7 to $p = 0.53 \pm 0.03\%$ on day 161 (Table~4 and
Fig.~\ref{fig3:12}a; quoted uncertainties are $1\sigma$ statistical only, and
do not reflect systematic effects), suggesting a modest amount of intrinsic SN
continuum polarization.  The lack of any obvious wavelength dependence of the
continuum polarization (e.g., Fig.~\ref{fig3:10}) is consistent with an
electron-scattering origin.  A remarkable feature is that although the
polarization level changes, it maintains a nearly constant polarization angle
in the plane of the sky (Fig.~\ref{fig3:12}b), similar to what was seen in
SN~1987A (Jeffery~1991a).  This strongly suggests that a preferred symmetry
axis persists as the photosphere recedes and exposes ever deeper regions of the
envelope.  The change in polarization by $\sim0.3\%$ from day 7 to day 161
cannot be produced by ISP, and is likely intrinsic to the SN itself.  We note
that the overall polarization level and characteristics of the line features
are consistent with those reported by Wang et al. (2000b) for
spectropolarimetry of SN~1999em obtained on day 72.  The continuum polarization
level implies a minimum asphericity of $\sim7\%$ during the plateau era and
$\sim 10\%$ on day 161 from the oblate, electron-scattering models of H91.

The temporal polarization increase may indicate increasing asphericity deeper
into the SN ejecta.  However, it is not clear that the models investigated by
H91 are directly applicable for either very early or very late SN epochs.  At
early times (i.e., prior to the appearance of line features due to heavy
metals; for SN~1999em this occurred on about day 9 (Leonard et al. 2001), the
electron density profile above a SN~II photosphere is believed to be very steep
(see, e.g., Jeffery 1991b; Eastman et al. 1996; Matzner \& McKee 1999).  This
type of atmosphere has much less polarizing power than one with the shallower
electron density gradients typical of SN atmospheres during the recombination
phase studied by H91.  In fact, Jeffery (1991b) suggests that the shift from a
steep to a shallow electron density gradient may explain the polarization rise
seen in SN~1987A early on (from $p_V \approx 0.2\% {\rm\ to\ } p_V \approx
0.7\%$ over the first 30 days) without the need for any increase in
asphericity.  H\"{o}flich et al.  1996 cautiously propose the same mechanism to
explain the early-time polarimetric behavior of SN~1993J as well.  It is
possible that a similar effect is at work in SN~1999em.  The characteristics of
the electron-scattering atmosphere during the transition to the nebular phase
(i.e., day 161) also differ significantly from those modeled by H91.  As
SN~1999em ages the optical depth to electron scattering decreases; at some
point in the nebular phase, in fact, it will drop well below 1, rendering
spectropolarimetry ineffective as a geometric probe.  It is possible that a
reduced electron-scattering optical depth is responsible for the relatively low
polarization measured at this late epoch.

\begin{figure}
\begin{center}
 \scalebox{0.7}{
	\plotone{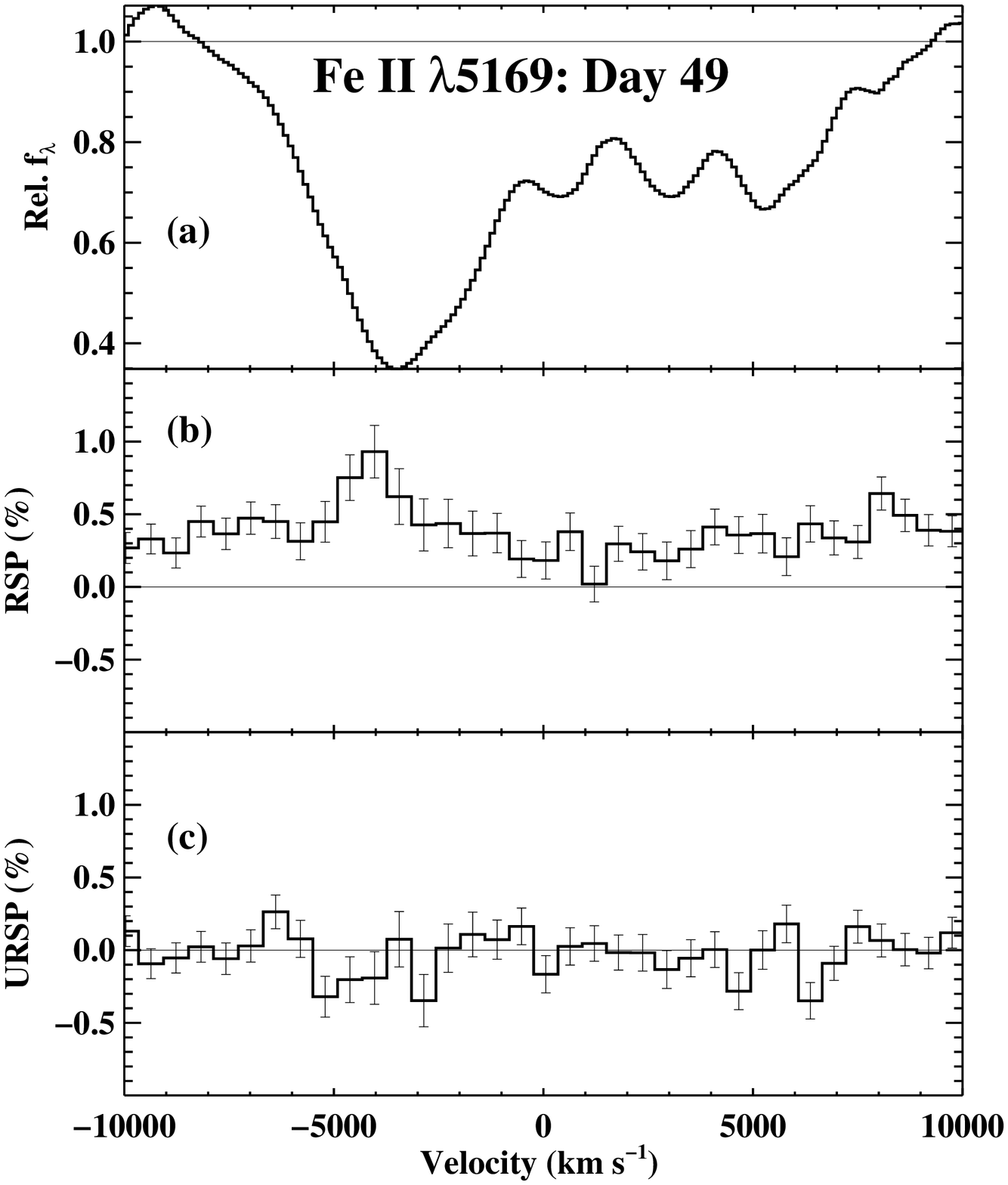}
		}
\end{center}
\caption[Polarization detail of the region around the absorption trough of \protect\ion{Fe}{2}
$\lambda 5169$ on day 49 for SN~1999em]
{Detail of the region around the absorption trough of \ion{Fe}{2}
$\lambda 5169$ on 1999 December 17.  ({\it a}) Normalized flux, displayed at 2
\AA\ bin$^{-1}$.  ({\it b}) Observed degree of polarization, measured as the
rotated $q$ parameter.  ({\it c}) Observed degree of polarization in the
rotated $u$ parameter.  Error bars in ($b$) and ($c$) are $1\sigma$ statistical
for 10~\AA\ bin$^{-1}$. In the absence of rapid P.A. change with wavelength,
the URSP should scatter about zero; see Appendix B for details.}
\label{fig3:13}
\end{figure}

\begin{figure}
\begin{center}
 \scalebox{0.8}{
	\plotone{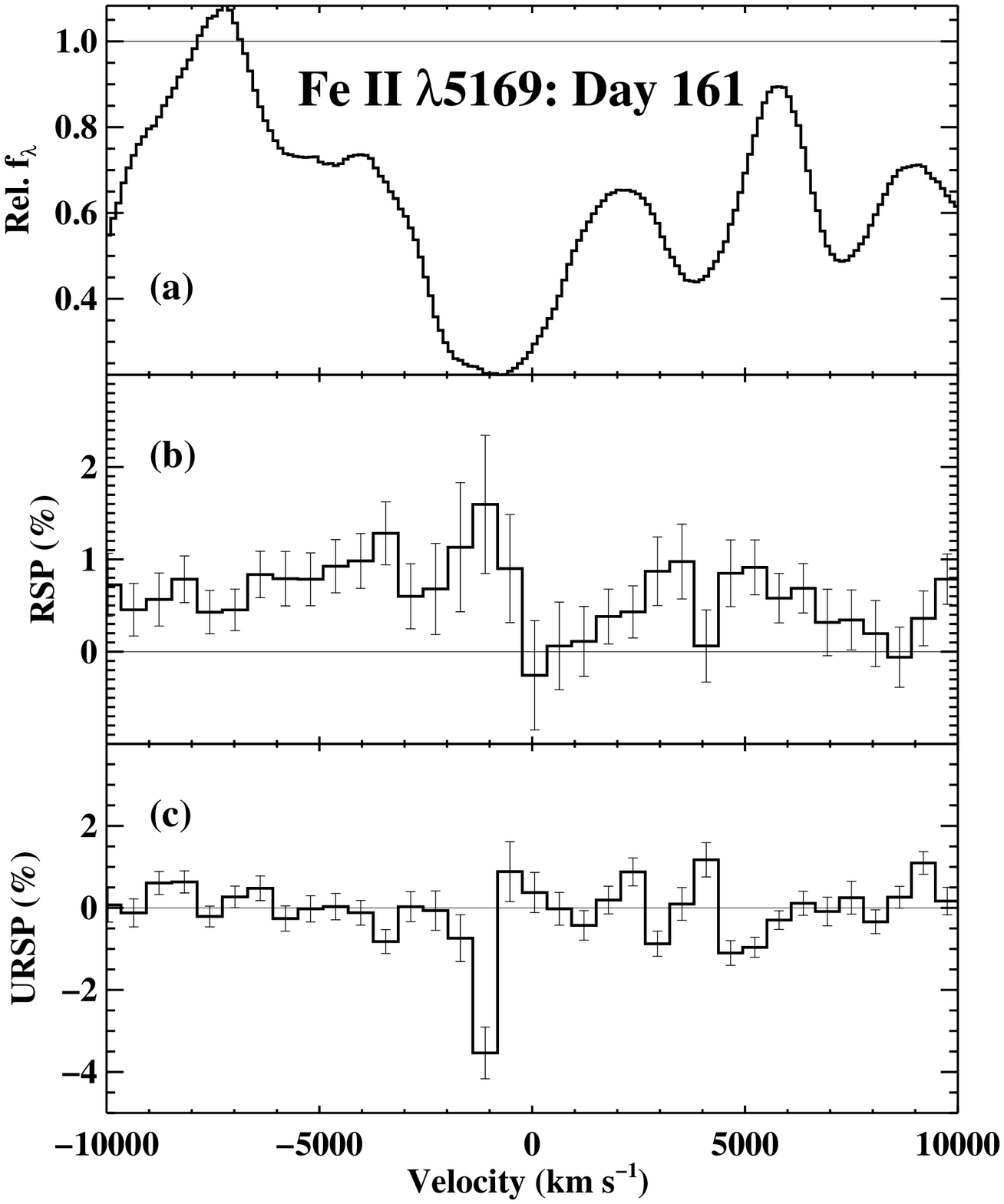}
		}
\end{center}
\caption[Polarization detail of the region around the absorption trough of \protect\ion{Fe}{2}
$\lambda 5169$ on day 161 for SN~1999em]
{As in Figure~\ref{fig3:13}, but for the weighted average of data obtained on 2000
April 5 and 9.  The sharp spike in URSP at the bottom of the line trough may
indicate that RSP underestimates the true trough polarization. }
\label{fig3:14}
\end{figure}

\begin{figure}
\begin{center}
 \scalebox{0.8}{
	\plotone{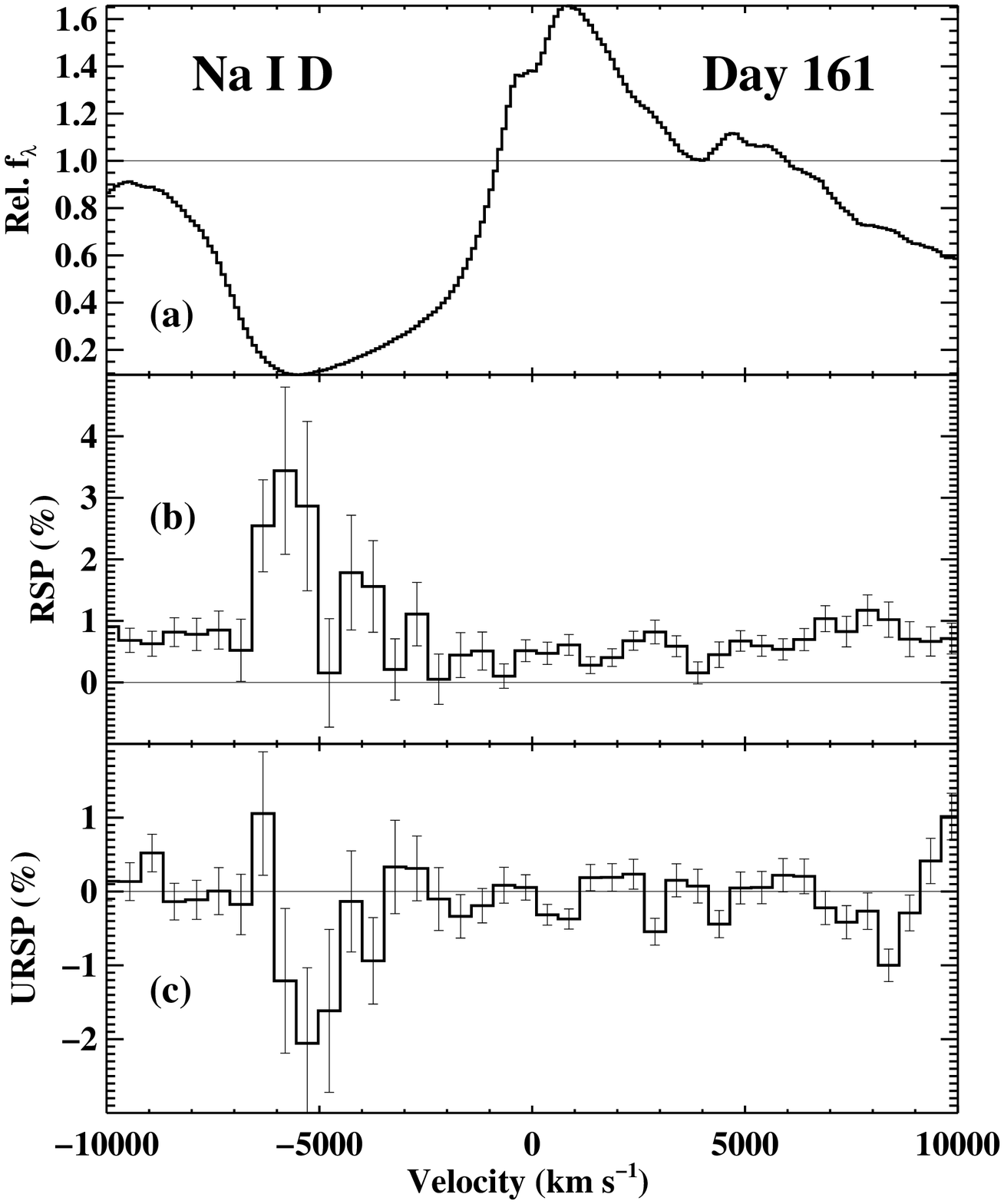}
		}
\end{center}
\caption[Polarization detail of the region around the absorption trough of
\protect\ion{Na}{1} D on day 161 for SN~1999em] {As in
Figure~\ref{fig3:13}, but for the weighted average of data obtained on 2000
April 5 and 9 near the absorption trough of \ion{Na}{1} D $\lambda\lambda 5890,
5896$ (\ion{Ba}{2} $\lambda 5853$ and \ion{He}{1} $\lambda 5876$ probably also
contribute).  Note that ${\rm URSP} \approx -1.7\%$ in many consecutive pixels
near the absorption minimum, indicating that RSP may underestimate $p$ by as
much as $0.4\%$.}
\label{fig3:15}
\end{figure}

While interpretation of the overall polarization level is hampered somewhat by
possible ISP contamination and other sources of systematic uncertainty
(Appendix A), the distinct polarization modulations across strong P-Cygni
features are not.  As anticipated (\S~\ref{sec3:4.1.1}), sharp polarization
peaks are associated in the later epochs with deep P-Cygni line troughs, most
conspicuously \ion{Fe}{2} $\lambda 5169$ and \ion{Na}{1} D $\lambda\lambda
5890, 5896$ (probably blended with \ion{Ba}{2} $\lambda 5853$ and \ion{He}{1}
$\lambda 5876$ on day 161).  We measure trough polarization in 10~\AA\ bins to
help improve the statistics in these low S/N ratio regions, and display the
results in Figures~\ref{fig3:13}, \ref{fig3:14}, \ref{fig3:15} and in Table~4.
The most dramatic increase is seen in the blended \ion{Na}{1} D line on day
161, in which the measured RSP is $p = 3.4\% \pm 1.4\%$, and the true polarization is
likely even higher given the evidence for P.A. rotation found in its URSP (see
Appendix~B).  The measured polarization increases in all troughs are in accord
with the limit set by equation~(\ref{eqn3:1a}).

\begin{figure}
\begin{center}
 \scalebox{0.7}{
	\plotone{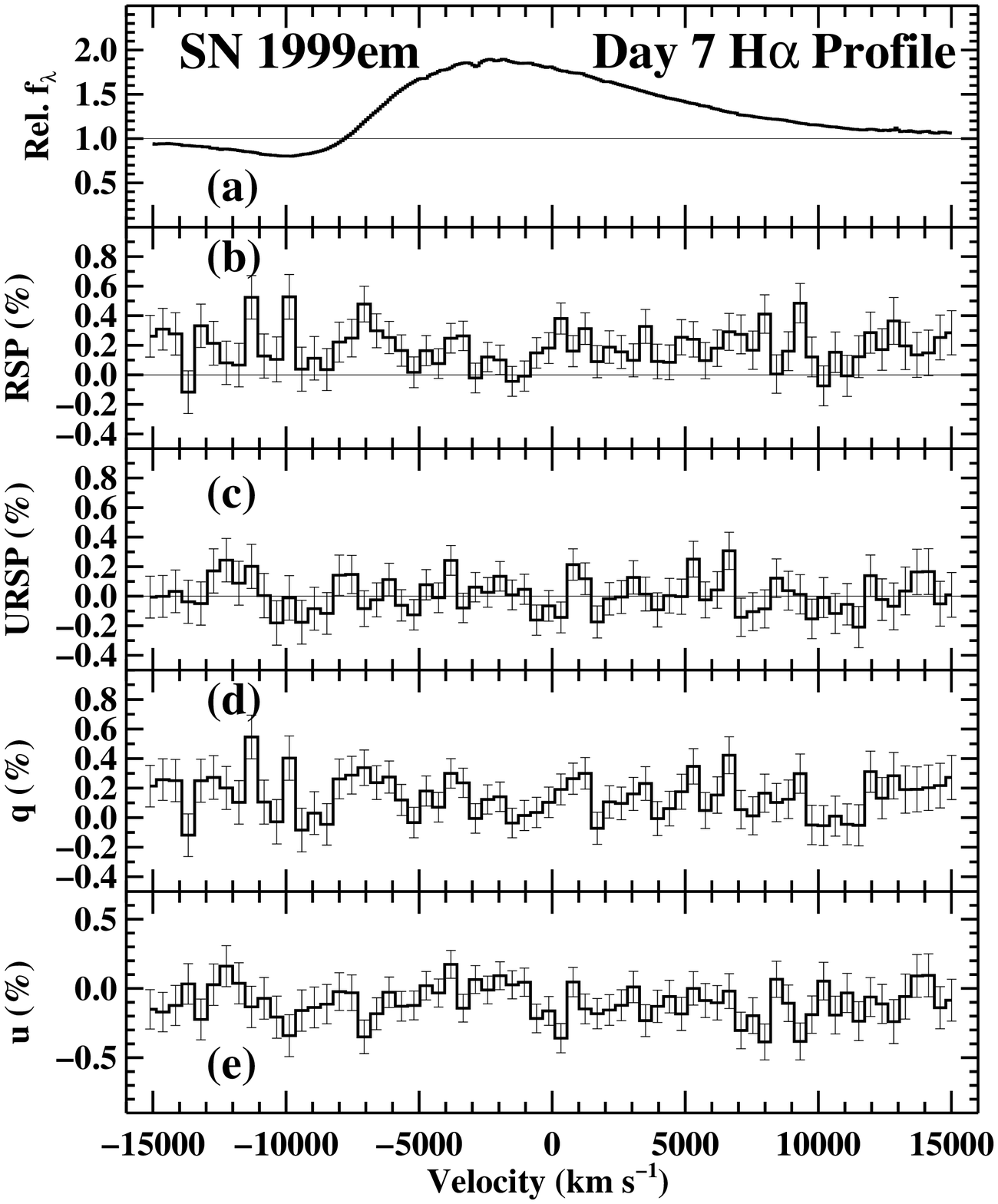}
		}
\end{center}
\caption[Spectropolarimetry of H$\alpha$ region 7 days after discovery for SN~1999em] {Region
around H$\alpha$ on 1999 November 5.  Error bars are $1\sigma$ statistical for
10~\AA\ bin$^{-1}$.  ({\it a}) Total flux, normalized by a spline fit to the
continuum, displayed at 2~\AA\ bin$^{-1}$ for better resolution.  ({\it b})
Observed degree of polarization, determined by the rotated Stokes parameter
(rotated $q$).  ({\it c}) Observed degree of polarization in the rotated $u$
parameter.  ({\it d, e}) Normalized $q$ and $u$ Stokes parameters. To
facilitate comparison, the same ordinate ranges are used in
Figures~\ref{fig3:16}, \ref{fig3:17}, and \ref{fig3:18}.}
\label{fig3:16}
\end{figure}

\begin{figure}
\begin{center}
 \scalebox{0.9}{
	\plotone{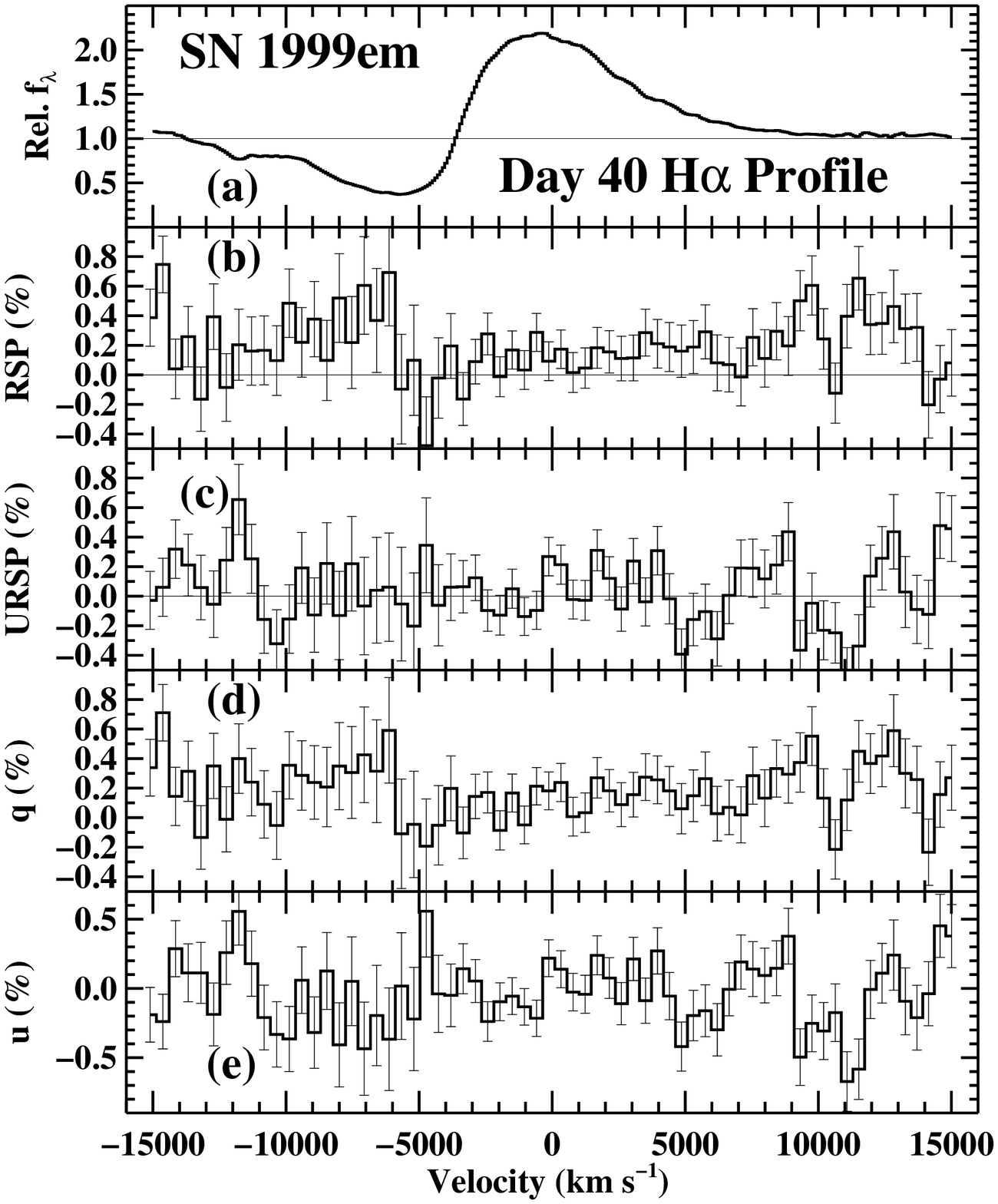}
		}
\end{center}
\caption[Spectropolarimetry of H$\alpha$ region 40 days after discovery for SN~1999em]{As in
Figure~\ref{fig3:16}, but for data obtained on 1999 December 8.  Note the
depolarization evident across the emission profile.}
\label{fig3:17}
\end{figure}

\begin{figure}
\begin{center}
 \scalebox{0.9}{
	\plotone{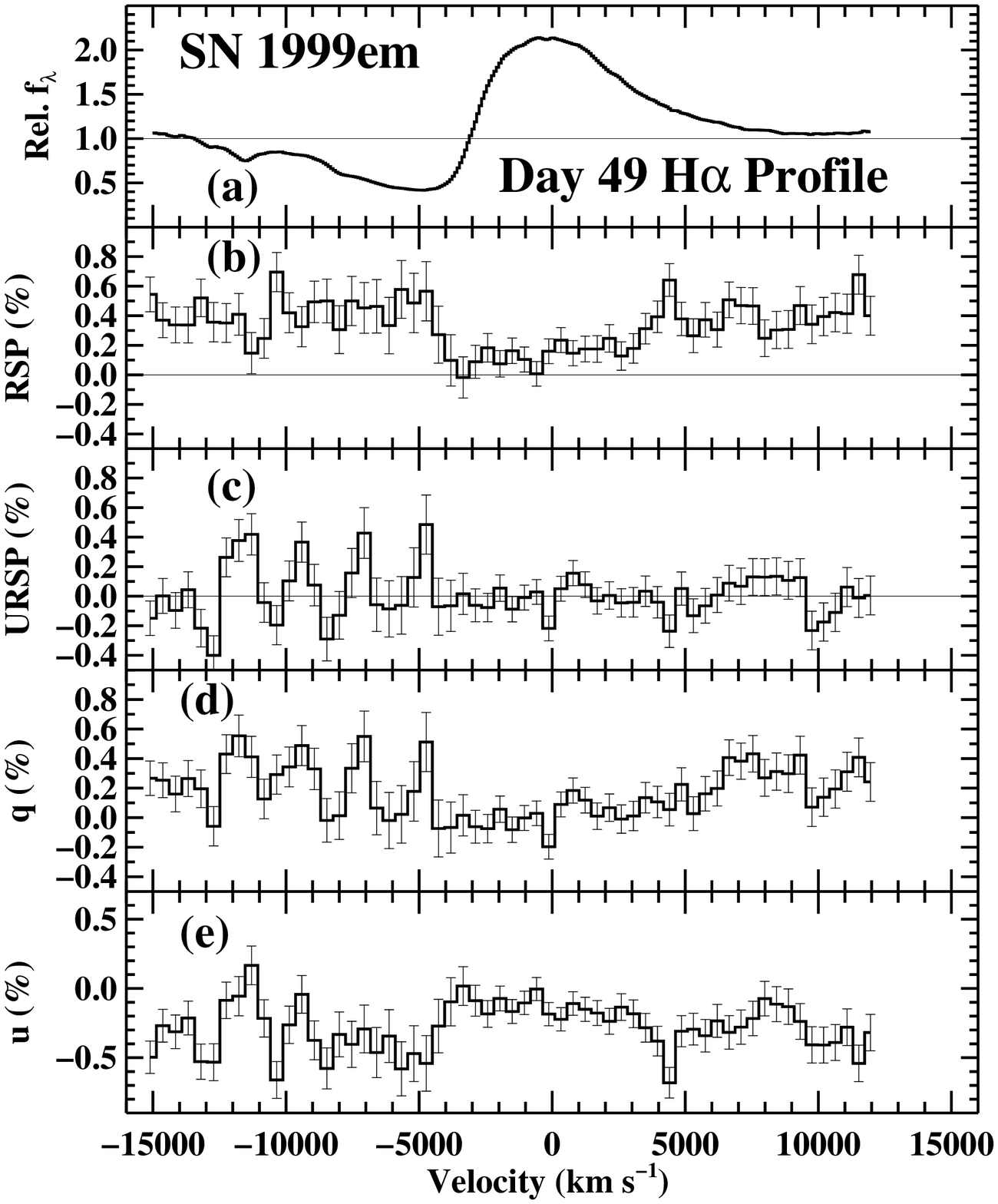}
		}
\end{center}
\caption[Spectropolarimetry of H$\alpha$ region 49 days after discovery for SN~1999em] {As in
Figure~\ref{fig3:16}, but for data obtained on 1999 December 17.  The
depolarization seen across H$\alpha$ on day 40 (Figure~\ref{fig3:17}) is
evident here as well.}
\label{fig3:18}
\end{figure}

\begin{figure}
\begin{center}
 \scalebox{0.7}{
	\plotone{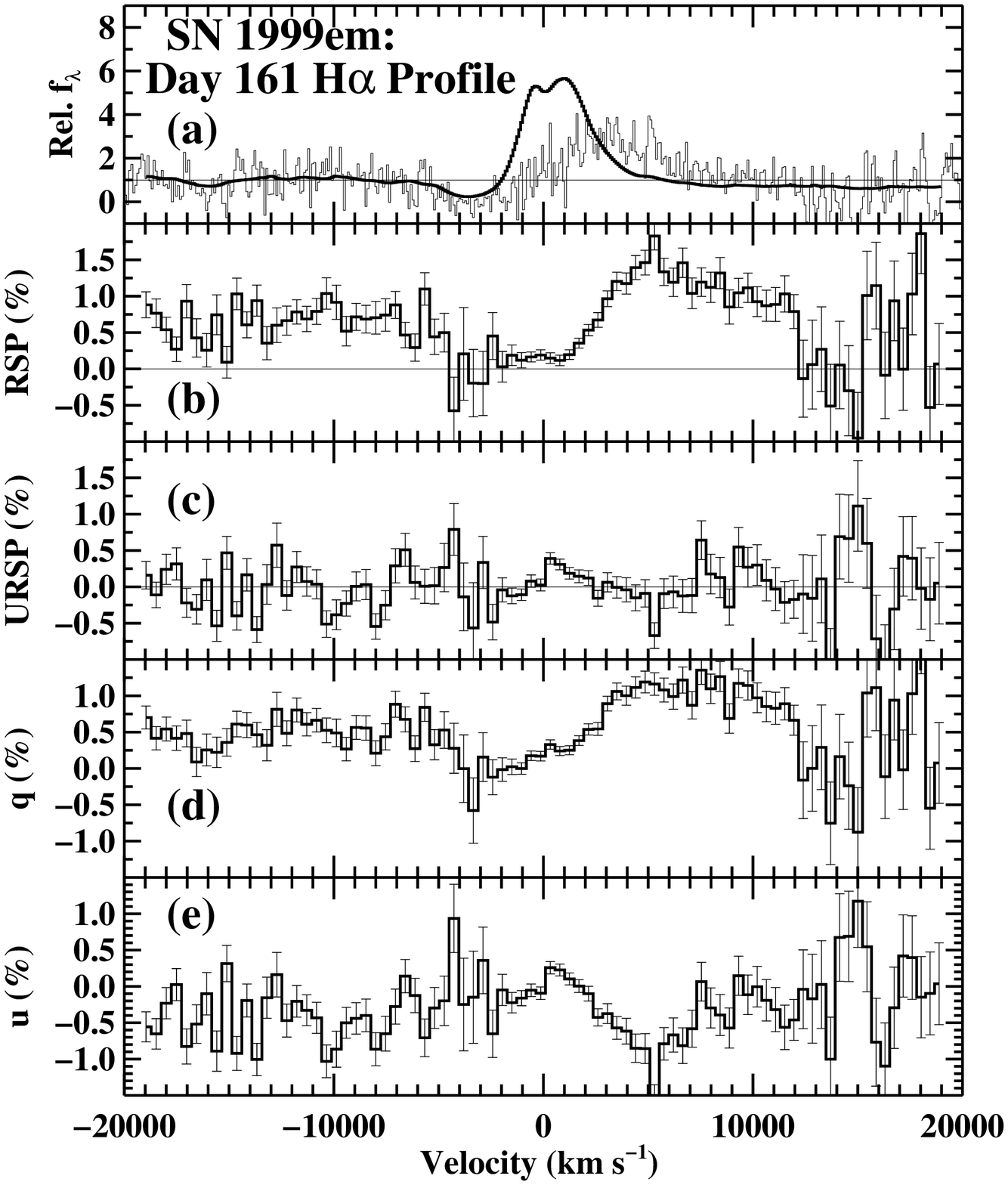}
		}
\end{center}
\caption[Spectropolarimetry of H$\alpha$ region 161 days after discovery for SN~1999em]
{As in Figure~\ref{fig3:16}, but for the weighted average of data obtained on 2000
April 5 and 9, and shown with an extended velocity range.  The normalized
polarized flux (RSP $\times f_{\lambda}$) is overplotted ({\it thin line}) in
({\it a}).  Note the increased polarization in the red wing of the line.}
\label{fig3:19}
\end{figure}

The strong depolarization across the H$\alpha$ profile also implies intrinsic
SN polarization (Figs.~\ref{fig3:16} through \ref{fig3:19}), with the change in
polarization across the line increasing by $\sim 0.6\%$ during the first 161
days (Table~4).  On days 40, 49, and 161, nearly complete depolarization exists
from slightly redward of the flux absorption minimum to the line peak near $v
\approx 0 {\rm\ km\ s^{-1}}$.  The fact that the depolarization goes well below
the value expected from the simple model of dilution of polarized continuum
light with unpolarized H$\alpha$ photons (eq.~[\ref{eqn3:1b}]) and the
lack of a polarization increase associated with the H$\alpha$ trough may be due
to a depressed underlying continuum polarization resulting from resonance
scattering of continuum photons by H$\alpha$ (\S~\ref{sec3:4.1.1}).  In effect,
since H$\alpha$ is optically thick throughout the atmosphere, it presents an
absorbing screen that ``covers'' both the limb and the central regions
(Fig.~\ref{fig1:3}), unlike the optically thin metal lines (i.e., \ion{Fe}{2}
or \ion{Na}{1} D) that predominantly block unpolarized (i.e.,
forward-scattered) light (\S~\ref{sec3:4.1.1}).  A similar H$\alpha$
depolarization was observed and theoretically modeled in SN~1993J (H\"{o}flich
et al.  1996 ), and was seen in SN~1987A as well (Jeffery 1991a; see also
\S~\ref{sec3:4.1.1}).

On day 161 the polarization increases steadily through the red wing of the
\halpha\ profile, reaching $p \approx 1.5\%$ at $v = 5,000 {\rm\ km\ s^{-1}}$,
and then drops back to the continuum value at $v \gtrsim 12,000 {\rm\ km\
s^{-1}}$.  The polarized flux (shown overplotted in Fig.~\ref{fig3:19}a) also
peaks about 3200 km s$^{-1}$ to the red of the nominal line center.  One agent
that may contribute is the redshift that line photons acquire upon being
scattered by electrons moving away from the point of origin (i.e., the
expanding universe paradigm; see, e.g., Bailey 1988; Witteborn et al. 1989).
Another possibility is that at this late epoch the density and temperature
conditions in the outer atmosphere are such that the depolarizing effect of
weak collisions is reduced, allowing the intrinsic line polarizing effect due
to resonance scattering to contribute (see Jeffery 1989, 1990 for a thorough
analysis of polarized line radiative transfer in an expanding atmosphere).  A
final, intriguing possibility is that H$\alpha$ photons on the ``far'' side of
the expanding SN (region 4 in Fig.~\ref{fig1:3}), are scattered into the l-o-s
by electrons in region 3.  Although at this late epoch the physical size of
region 4 is quite small and would contribute little flux, the photons would be
highly polarized due to the near $90^\circ$ scatter they suffer, perhaps
allowing them to significantly affect the polarization of the line's red wing.

\begin{figure}
\begin{center}
 \scalebox{0.7}{
	\plotone{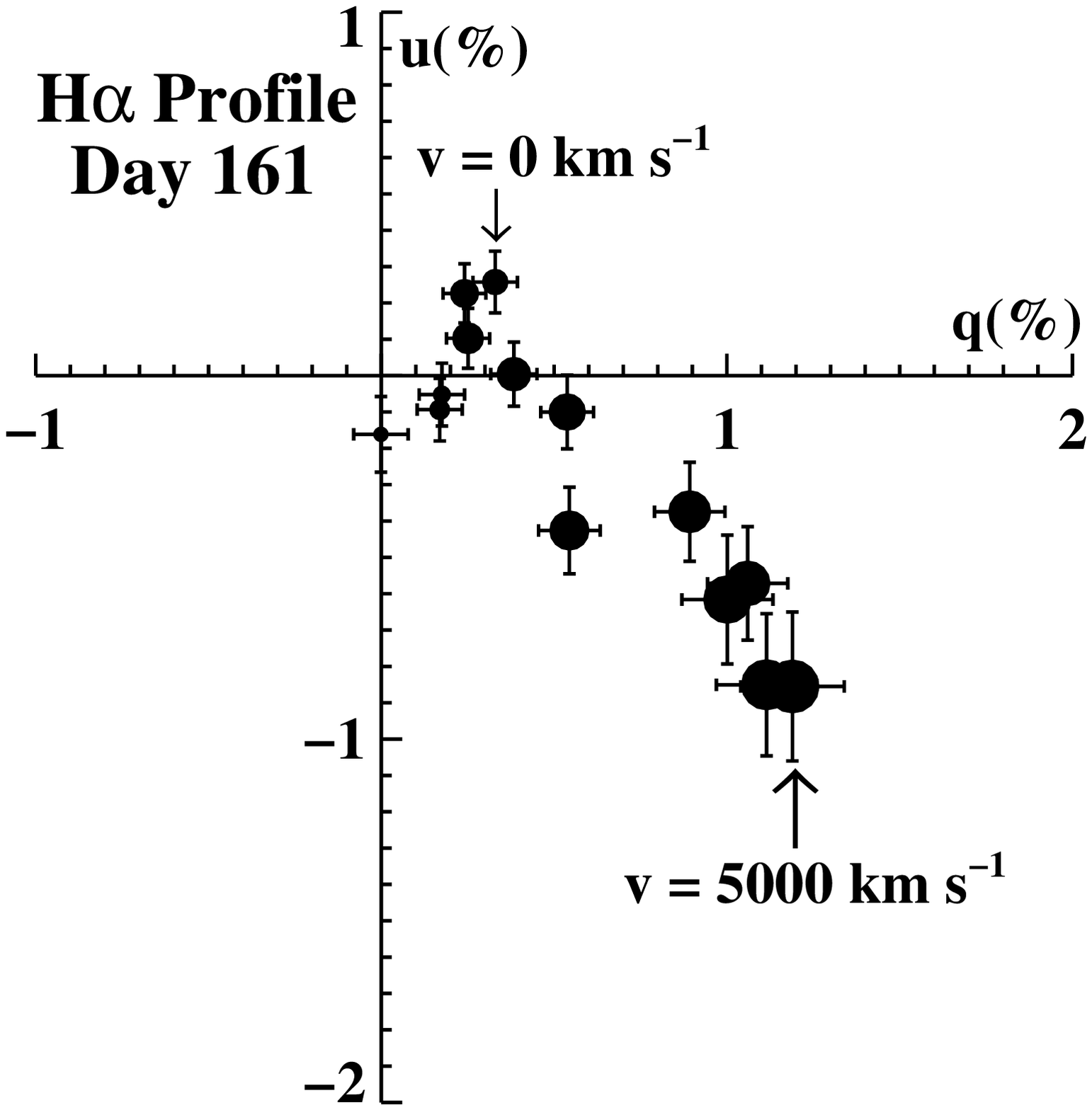}
		}
\end{center}
\caption[Polarization of the H$\alpha$ line profile in the $q$-$u$ plane, day
161] {Polarization of the H$\alpha$ line profile in the $q$-$u$ plane from $v =
-1000$ to $v = 5000$ km s$^{-1}$.  Symbol size increases with velocity. Error
bars are $1\sigma$ statistical for $10$~\AA\ bin$^{-1}$.  The linear path
traced from $v = 0$ to $v = 5000$ km s$^{-1}$ suggests an axisymmetric
scattering environment for those line photons; the different $q$-$u$ location
favored by the photons blueward of $v = 0$ \kms\ may indicate a breakdown of
axisymmetry across the line profile, but the small number of points makes this
difficult to confirm.  }
\label{fig3:20}
\end{figure}

The existence of a broad, strong, polarized emission line allows us to
investigate a key question about the line-scattering region's geometry: is it
axisymmetric?  That is, does the line profile present a uniform polarization
angle in the plane of the sky?  The best way to test this is to follow the path
traced by the data in the $q$-$u$ plane, since a common axis of symmetry for
the different scattering regions will result in a linear path.  This method is
preferred over simply looking to see if the polarization angle changes across
the line (i.e., $\theta$ in Fig.~\ref{fig3:5}), since ISP can produce
apparent P.A. rotations, but will not alter the shape of the path in $q$-$u$.
Figure~\ref{fig3:20} shows the results for H$\alpha$ on day 161, extending from
$0$ to $5000 {\rm\ km\ s^{-1}}$.  The path traced by the red side of the
profile is quite linear, and the polarization angle is consistent with that
measured for the continuum: for line photons in the range $v=0 - 5000 {\rm\ km\
s^{-1}}$, we measure $p = 0.48\% {\rm\ at\ } \theta = 161^{\circ}$.  This
measurement is complicated by the lack of a clear continuum, however, which
required us to estimate the continuum level by hand in the Stokes parameter
fluxes (see Appendix~B).  This implies that the scattering region responsible
for the polarization exhibited by the red wing of H$\alpha$ is axisymmetric and
has the same geometry as the continuum scattering region, consistent with the
idea that both the line and the continuum are scattered by the same electrons
in the line-forming region of the SN envelope.  Unfortunately, the low S/N
ratio for $v > 5000 {\rm\ km\ s^{-1}}$ prohibits an investigation out to the
highest velocities; similarly low S/N ratio in other emission lines (and
\halpha\ at other epochs) prohibits detailed analysis in them as well.  On the
blue side of the line profile, it is {\em possible} that the line prefers a
path different from that seen on the red side, but the limited number of points
preclude a definitive conclusion about the breaking of axisymmetry there.  In
all, though, the near-zero polarization of the \halpha\ emission profile seen
in Figure~\ref{fig3:20} near $v = 0$ \kms\ bolsters the belief that the ISP
may, in fact, be quite small.

\subsubsection {The Effect of Asphericity on the Expanding Photosphere Method}
\label{sec3:4.1.4}

If the geometry of a SN~II-P photosphere resembles the geometry of its
electron-scattering region, then asphericity will affect the distance derived
by any method that assumes a spherical flux source, such as EPM.  Wagoner (1991
[W91]) derives the error in EPM distances resulting from an oblate SN
photosphere with constant flux over its surface for all viewing orientations
and degrees of asphericity.  For instance, a 10\% asphericity is shown to
produce an EPM distance that overestimates the actual distance by $\sim5\%$ for
an edge-on view and underestimates it by $\sim10\%$ for a face-on l-o-s.  The
error naturally rises with increasing asphericity, reaching over 50\% for axis
ratios of 2:1 viewed face-on.  While individual distances may thus be quite
inaccurate for aspherical flux distributions, an encouraging result is that the
{\em average} distance derived to SNe~II-P with random viewing orientations is
quite robust, with $\overline{d}_{EPM} = \overline{d}_{actual}$ to within 2\%
(in the sense $\overline{d}_{EPM} < \overline{d}_{actual}$) for mean axis
ratios of up to 2:1.  The main effect of asphericity on the determination of
$H_0$ from SNe~II is to increase the {\em scatter} of EPM distances about the
true values, but not to substantially bias the overall measurement of $H_0$.
The amount of increased scatter expected for a given mean SN~II asphericity is
shown in Figure 4 of W91.

A useful characteristic of EPM is that each observational epoch provides an
{\it independent} estimate of the SN's distance once the time of explosion is
derived (see, e.g., Schmidt, Kirshner, \& Eastman 1992a).  If the asphericity
of SN~1999em really does increase with time while maintaining a constant
viewing orientation (as possibly suggested by its increasing polarization at a
constant P.A.), then the EPM-derived distance should deviate more and more from
the correct value as time goes on.  The EPM technique itself can thus be used
to answer some questions concerning the nature of a particular SN's
asphericity: EPM-derived distances that {\em increase} with time would indicate
a more edge-on viewing orientation, while those that {\em decrease} with time
would indicate a more face-on viewing orientation, with the transition from
``more edge-on'' to ``more face-on'' occurring at $\sim38^\circ$ for the simple
model of W91.  (EPM will produce a non-varying distance to either a
$38^{\circ}$ viewing orientation or a constant degree of asphericity; subtle
combinations of temporally varying viewing orientation and asphericity could
also conspire to produce this result.)

In this regard, it is useful to reexamine EPM distances previously derived for
the 16 SNe~II tabulated by S94, which include 10 SNe~II-P and 6 SNe~II-L.  None
of the SNe~II studied shows any obvious temporal trends in its derived distance
(e.g., Fig.~8 of Schmidt et al. 1992a; Fig.~6 of Schmidt et al. 1994b), with
several having data points that extend to more than 60 days after shock
breakout.  In the EPM analysis of SN~1999em (Leonard et al. 2001) the derived
distance to SN~1999em also remains quite stable with time.  Barring a
conspiracy of offsetting errors,\footnote{We note that the main source of
systematic uncertainty in EPM distances comes from the ``dilution factor'' used
to correct the observed flux due to the departure of the photospheric flux from
that expected for a blackbody radiating at the location of the optical
photosphere.  For more details, see Eastman et al. 1996, H\"{o}flich 1991b, and
Baron et al. 1995.} the temporal constancy of EPM distances implies that any
asphericity that exists must not change appreciably as a particular SN evolves
during the photospheric phase.

If SNe~II are aspherical but present a fixed degree of asphericity during the
photospheric phase, a systematic increase in the {\it scatter} in a Hubble
diagram beyond that which can be explained by statistical uncertainties alone
will result.  Figure~4 of W91 shows the anticipated degree of this systematic
scatter (given as $\Delta Q_{RMS}$, where $Q \equiv d_{A}/d_{E}$ and $d_A$ is
the ``Hubble distance'' [defined below], $d_E$ is the EPM-derived SN distance,
and $\Delta Q \equiv Q - 1.0$) as a function of mean SN asphericity.  We may
therefore gain some insight into mean SN~II asphericity during the photospheric
phase by examining the scatter of $Q$ for the 16 SNe~II reported by S94.  To do
this, we form a $\chi^2$ statistic, where
\begin{eqnarray}
\chi^2 & = & \sum_{1} \frac{(Q_i - 1.0)^2}{\sigma_{Q_{sys,i}}^2 +   
	     \sigma_{Q_{stat,i}}^2} \nonumber \\
       & = & \sum_{1} \frac{(Q - 1.0)^2}{\sigma_{Q_{sys,i}}^2 + 
	     \left(\frac{\sigma_{d_{A},i}}{d_{E,i}}\right)^2 +	
	     \left(\frac{d_{A,i} \sigma_{d_{E},i}}{d_{E,i}^2}\right)^2}
\label{eqn3:2}
\end{eqnarray}
\noindent where $\sigma_{Q_{sys}}$ is the systematic uncertainty in $Q$
(presumably due to asphericity), $\sigma_{Q_{stat}}$ is the statistical
uncertainty in $Q$, $d_E$ and $\sigma_{d_{E}}$ are the ``EPM distance'' and
statistical uncertainty, respectively, taken from column~6 of Table~6 by S94,
$d_A$ is the ``Hubble distance'' ($d_A \equiv v_{rec}/H_0$, where $v_{rec}$ is
the recession velocity of the host galaxy, taken from column~3 of Table~6 by
S94 and $H_0$ is chosen so that the weighted mean of $Q$ is 1.0), and
$\sigma_{d_{A}}$ is the uncertainty in the Hubble distance due to peculiar
velocities of the host galaxies.  Peculiar galaxy velocities are generally
thought to be in the range $100 < \sigma_{v_{rec}} < 400$ \kms\ (Lin et
al. 1996; Marzke et al. 1995).  We stress that the point of our investigation
here is {\it not} to derive $H_0$, but rather to set a uniform fiducial scale
from which to quantify the scatter of EPM-derived distances about their
``actual'' distances.  The particular value of $H_0$ required to make $Q = 1.0$
is sensitive to both $d_E$ and $d_A$ of the SN sample, as well as the errors
assigned to $d_E$, $d_A$, and $Q_{sys}$.  The main assumption made by setting
$d_A = v_{rec}/H_0$ is that the Hubble law is linear out to 180 Mpc, the
farthest EPM distance yet reported (S94; Schmidt et al. 1994b).

\begin{figure}
\begin{center}
\rotatebox{90}{
 \scalebox{0.75}{
	\plotone{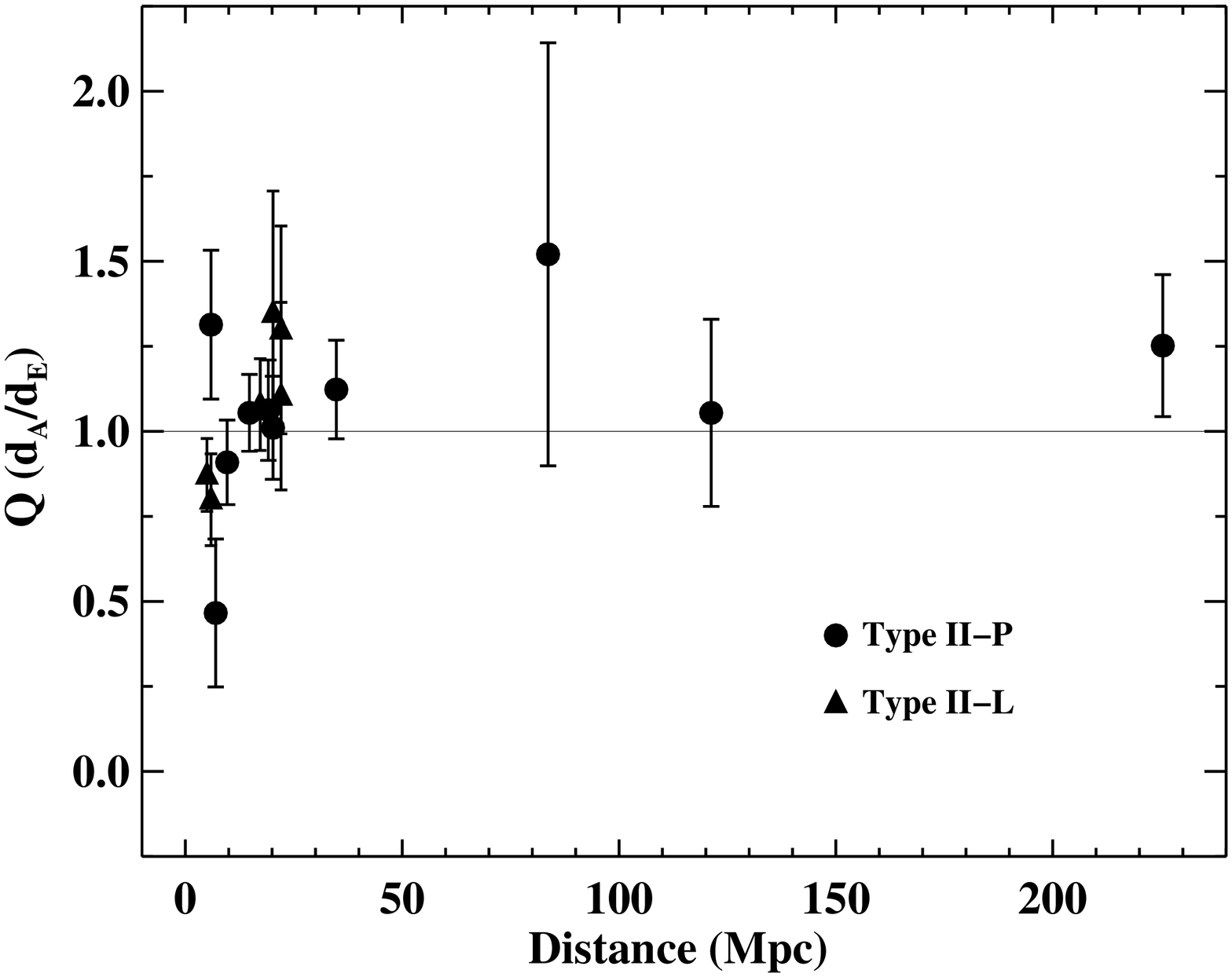}
		}
	}
\end{center}
\caption[Ratio of distance determined by assuming a linear Hubble law to the
distance determined by the Expanding Photosphere Method (EPM), vs. Hubble law
distance, for 16 SNe~II ] {Ratio ($Q$) of the distance determined by assuming a
linear Hubble law ($d_{A} \equiv v_{rec}/H_0$, where $v_{rec}$ is the host
galaxy's recession velocity) to the distance determined by EPM ($d_{E}$) for 16
SNe~II, vs. $d_{A}$.  The displayed $1\sigma$ uncertainties in $Q$ are derived
entirely from the statistical uncertainty reported for $d_{E}$.  To determine
$Q$, we chose $H_0$ so that the weighted mean of $Q$ is equal to 1; for these
16 SNe~II with the displayed $1\sigma$ uncertainties this required setting $H_0
= 64$ \kms\ Mpc$^{-1}$.  }
\label{fig3:21}
\end{figure}

The resulting scatter of $Q$ about unity is shown in Figure~\ref{fig3:21} for
all 16 SNe~II.  The $1\sigma$ errors on $Q$ and $\sigma_{Q}$ incorporate only
the statistical uncertainty reported for $d_{E}$ by S94.  For the full set of
16 SNe~II we derive a reduced $\chi^2$ of 1.20, implying that the uncertainty
resulting from $\sigma_{d_{E}}$ alone will produce at least this amount of
scatter 26\% of the time.  Including a small contribution to the uncertainty
in $Q$ from peculiar velocities of $\sigma_{v_{rec}} = 100$ \kms\ increases
this likelihood to $70\%$.  In other words, assuming that the only sources of
uncertainty in $Q$ are the reported errors on $d_E$ and a small contribution
from peculiar velocities leads to the conclusion that the observed scatter is
consistent with no directional dependency of the SN flux at the $1\sigma$
level.  If we follow S94 and assign a $300$ \kms\ uncertainty to $v_{rec}$ due
to peculiar velocities, then the likelihood that the observed degree of scatter
can be explained with no additional sources of uncertainty rises to $99\%$, and
the formal $3\sigma$ upper limit on $\Delta Q_{RMS}$ due to directionally
dependent flux is found to be $\Delta Q_{RMS} < 0.16$.  Using Figure~4 of W91,
this translates into a $3\sigma$ upper limit on mean SN asphericity during the
photospheric phase of about $30\%$.

Therefore, unless the uncertainties reported by S94 do not accurately reflect
(i.e., they overestimate) the true statistical error on $d_E$ or a conspiracy
of offsetting errors exists, significant directional dependence of SN~II
luminosity is inconsistent with the existing EPM data; the subsets of SNe~II-P
and SNe~II-L alone support a similar conclusion for each sample individually,
although the reduced sample sizes produce somewhat less robust constraints.
Although additional spectropolarimetric data are needed to create a sample
large enough to directly estimate average SN~II asphericity, the apparent
success of EPM thus already significantly limits its value.  The good agreement
among EPM, Tully-Fisher, and Cepheid distances (S94; Eastman et al. 1996) also
argues for low SN~II asphericity during the photospheric phase.

\subsection {Flux Spectra}
\label{sec3:4.2}

A nice byproduct of the long exposure time required for spectropolarimetry
(typically about 4 hours to achieve $\sigma_{q,u} \approx 0.3\%$ with 2~\AA\
bin$^{-1}$ using the Lick 3-m reflector while SN~1999em was on the plateau with
$m_V \lesssim 14$ mag) is the very high S/N ratio total flux spectrum that
results. Although the spectral evolution of SN~1999em will be discussed in more
detail elsewhere (Leonard et al. 2001), we comment briefly on two of the more
interesting features seen in the flux spectra presented here.

\begin{figure}
\begin{center}
 \scalebox{0.7}{
	\plotone{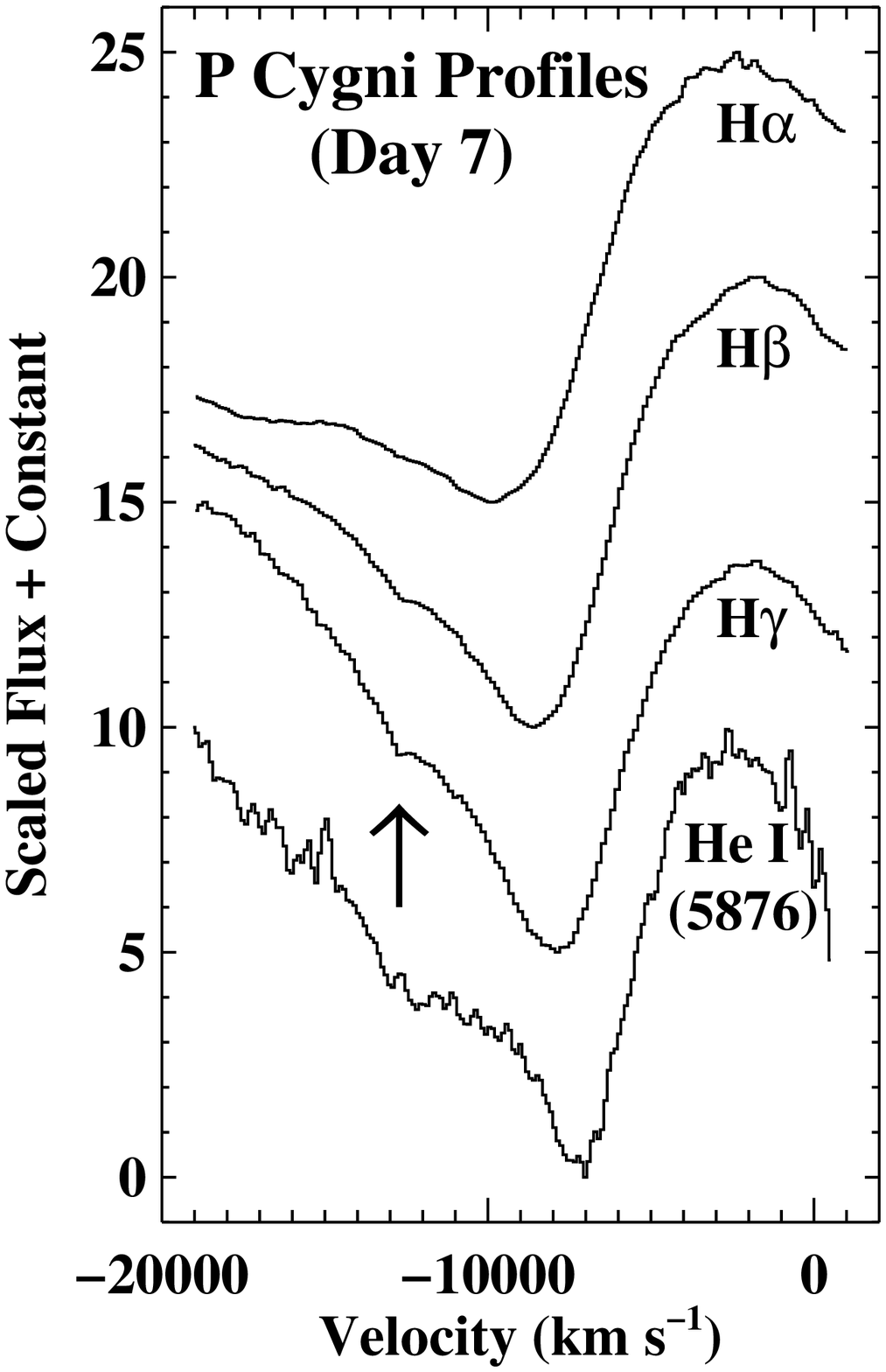}
		}
\end{center}
\caption[P-Cygni Profiles of SN~1999em, Day 7]
{P-Cygni profiles observed on 1999 November 5.  A dip near $v \approx
-12,600 {\rm\ km\ s^{-1}}$ is seen in the higher-order Balmer lines and
possibly \ion{He}{1} $\lambda 5876$ as well.  Due to telluric absorption near
6290~\AA\ (see, e.g., Matheson et al. 2000), its presence could not be verified
in H$\alpha$.}
\label{fig3:22}
\end{figure}

Figure~\ref{fig3:22} shows the absorption troughs for the \ion{He}{1} $\lambda
5876$ and hydrogen Balmer lines.  A curious ``dip'' exists at $v \approx
-12,600 {\rm\ km\ s^{-1}}$ in the higher-order Balmer lines, and possibly
\ion{He}{1} as well.  One possibility is that these dips are produced by
Chugai's (1991) proposed mechanism for the anomalous blueshifted ``absorption''
seen in SN~1987A (i.e., the ``Bochum event'').  Under this model, the features
are actually produced by a {\it lack} of absorption by gas at velocities
surrounding the ``dip,'' caused by ineffective screening of continuum photons
due to excitation stratification in the envelope.  On the other hand, the
features may be related to the ``complicated'' P-Cygni profiles reported by
Baron et al. (2000) from analysis of a spectrum of SN~1999em taken within a day
of discovery.  From detailed spectral modeling, Baron et al. (2000) conclude
that broad absorption shelves seen in the H$\beta$ and \ion{He}{1} $\lambda
5876 $ profiles at $v \approx 20,000 {\rm\ km\ s^{-1}}\ $ are due to the
combination of a shallow density gradient and high photospheric temperature at
this early epoch.  Similar features are seen in very early spectra of SN~1990E
and were identified with \ion{N}{2} $\lambda\lambda 4623, 5679$ (Schmidt et
al. 1993), but Baron et al. (2000) favor the ``complicated'' P-Cygni
explanation from their model fits.  The consistent velocity of the dips in our
spectrum and its presence in H$\gamma$ also argue against the \ion{N}{2}
identification for these features.  Whether the dips are in fact related to the
earlier features seen by Baron et al. (2000) will have to await further
modeling.

\begin{figure}
\begin{center}
 \scalebox{0.7}{
	\plotone{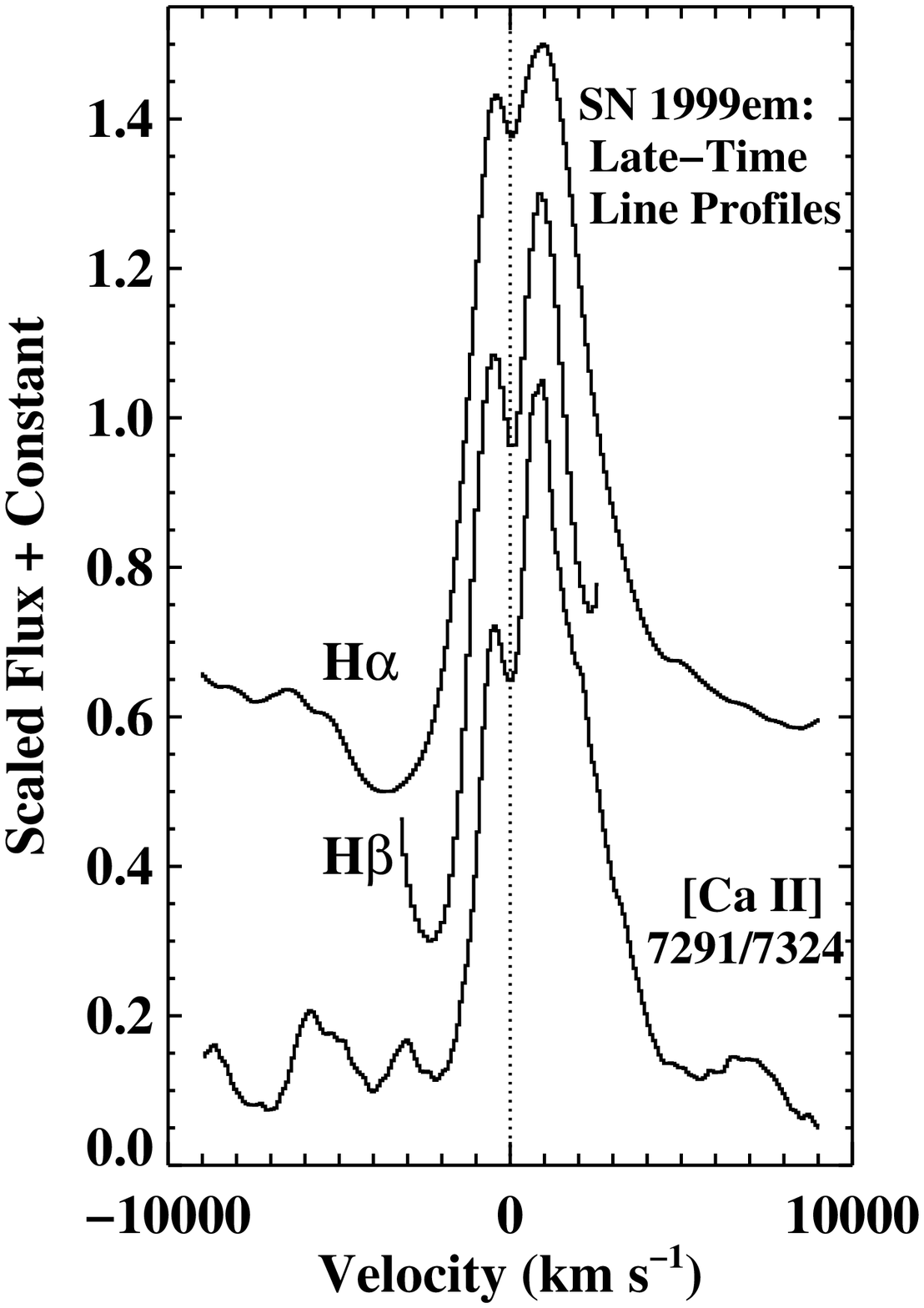}
		}
\end{center}
\caption[Emission-line profiles at late times for SN~1999em]
{Emission-line profiles at late times for SN~1999em, taken from spectra
obtained on days 159 (H$\alpha$ and H$\beta$) and 163 ([\ion{Ca}{2}]
$\lambda\lambda 7291, 7324$; $\lambda_{\circ} = 7291$).  The ``notch'' near
zero velocity first appeared between days 95 and 124 and persists through the
most recent spectrum obtained on day 333 as well.}
\label{fig3:23}
\end{figure}

Another interesting aspect is the development of a distinct ``notch'' centered
near the rest wavelengths of prominent emission lines on day 161
(Fig.~\ref{fig3:23}), creating a subtle ``double peak'' to the emission
profiles of H$\alpha$, H$\beta$, and [\ion{Ca}{2}] $\lambda\lambda7291, 7324$.
Inspection of earlier spectra indicates that this feature developed sometime
between days 95 and 124, coinciding with the time SN~1999em fell off the
plateau.  It exists in a spectrum obtained 333 days after discovery as well.  A
mechanism traditionally invoked to explain double-peaked profiles in SNe is
interaction of the ejecta with a circumstellar disk of material, perhaps
ejected by the progenitor prior to explosion (e.g., SN~1998S; see Leonard et
al. 2000a).  Since there are few other indications of circumstellar interaction,
however, we are quite skeptical of this explanation.  It is also possible that
slowly moving circumstellar material exterior to the expanding ejecta would
produce an ``absorption'' feature near $\lambda_{\circ}$.  Alternatively, the
appearance of the features at the same moment the hydrogen envelope fully
recombined (i.e., the end of the plateau) coupled with their persistence
through at least the first part of the nebular stage might suggest an origin
within the expanding ejecta itself, perhaps due to an asymmetry in the
line-emitting region.

\section{Conclusions}
\label{sec3:5}

We present spectropolarimetry of SN~1999em on days 7, 40, 49, 159, and 163 days
after discovery, with the last two epochs combined to form a single ``day 161''
epoch.  Our main results are as follows.

\begin{enumerate}

\item The observed broadband polarization rises from $p \approx 0.2\%$ on day 7
to $p \approx 0.5\%$ on day 161.

\item The continuum polarization does not exhibit any noticeable variation with
wavelength.

\item The polarization angle is $\theta \approx 160^{\circ}$ throughout
the first 161 days.

\item From several lines of evidence, the ISP along the l-o-s to SN~1999em is
thought to be quite small, although values as high as $0.3\%$ can not be ruled
out.

\item Increases in polarization of up to $0.54 \pm 0.2\%$ during the plateau
and $2.9 \pm 1.5\%$ after the plateau are associated with P-Cygni absorption
troughs of metal lines.

\item Nearly complete depolarization across the H$\alpha$ profile is seen on
days 40, 49, and 161; the depolarization is neither confirmed nor rejected in
the data from day 7.

\item On day 161, the red wing of the H$\alpha$ profile is polarized up to
$0.7\%$ higher than the surrounding continuum.  The scattering medium
responsible for the polarization of the red wing of the H$\alpha$ profile
traces a linear path in the $q$-$u$ plane, with a polarization angle consistent
with that of the continuum.

\item Analysis of EPM-derived distances to 16 SNe~II reveals little evidence
for directional dependency of SN flux.  Making reasonable assumptions about the
extent of other sources of uncertainty in EPM distances, we derive a $3\sigma$
upper limit of $30\%$ on mean SNe~II asphericity during the photospheric phase.

\item In the total flux spectra, a ``dip'' exists in the absorption profiles of
H$\beta$ and H$\gamma$ at $v \approx -12,600 {\rm\ km\ s^{-1}}$ on day 7.  A
``double peaked'' emission profile for H$\alpha$, H$\beta$, and [\ion{Ca}{2}]
$\lambda\lambda$7291, 7324 on day 161 is observed, which first appeared between
days 95 and 124 and persists through at least day 333, the latest epoch
obtained thus far.
 
\end{enumerate}

There are several mechanisms that can produce SN polarization in addition to
asphericity of the electron-scattering atmosphere, including scattering by dust
(e.g., Wang \& Wheeler 1996), asymmetrically distributed radioactive material
within the SN envelope (e.g., H\"{o}flich 1995), and aspherical ionization
produced by hard X-rays from the interaction between the SN shock front and a
nonspherical progenitor wind (Wheeler \& Filippenko 1996).  While subtle
combinations or fine tuning of models using these production mechanisms cannot
be ruled out, the primary polarization characteristics seen in SN~1999em are
exactly what one expects from an aspherical, electron-scattering, expanding
atmosphere: a wavelength-independent continuum polarization with polarization
peaks at the locations of P-Cygni troughs and depolarization across strong
emission features. From the models of H91, we infer that SN~1999em thus had a
global asphericity of $\gtrsim 7\%$ and a symmetry axis that remained fixed
throughout the first 161 days of its photospheric development. The observed
temporal increase in continuum polarization for SN~1999em may indicate greater
asphericity closer to the explosion center.  However, the temporal constancy of
its EPM-derived distance argues against significant geometric change and
suggests that the observed polarization increase may be due to the changing
density structure of the electron-scattering region.

The low polarization found for SN~1999em and the lack of significant systematic
scatter in the Hubble diagram for previous SNe~II EPM distances suggest that
significant asphericity at early times is not prevalent among this SN class.
This conclusion is encouraging for cosmological applications of SNe~II, and is
in accord with the observed trend that SN asphericity decreases with increasing
envelope mass (Wheeler 2000).  The anticipated increase in polarization as the
photosphere recedes through the hydrogen envelope is indeed observed, although
the increase is quite modest and the interpretation is hampered somewhat by the
changing density structure of the SN photosphere and envelope.  However,
imprinting even a small degree of asphericity on the outer regions of the thick
hydrogen envelope of a SN~II-P requires a very large explosion asymmetry, since
asymmetric explosions tend to turn spherical as they expand (Chevalier \& Soker
1989; H\"{o}flich, Wheeler, \& Wang 1999).  Whether the inferred asymmetry
necessitates explosion asymmetry as extreme as the ``bipolar,'' jet-induced
mechanism proposed by Khokhlov et al. (1999) will have to await detailed
modeling and acquisition of more spectropolarimetry of core-collapse events.

This study brings the total number of SNe studied in detail with
spectropolarimetry to four.  Clearly, additional spectropolarimetry of
core-collapse SNe, and SNe~II-P in particular, are needed to fully assess the
nature and ultimate cause of SN asphericity at early times.  Nevertheless, it
is not unreasonable to speculate that explosion asymmetry, or asymmetry in the
collapsing Chandrasekhar core, plays a dominant role in the explanation of
pulsar velocities (e.g., Burrows \& Hayes 1996), the mixing of radioactive
material seen far out into the ejecta of young SNe (e.g., SN~1987A; Sunyaev et
al. 1987; Lucy 1988; Tueller et al. 1991), and even gamma-ray bursts (MacFadyan
\& Woosley 1999; Wheeler 2000; Wheeler et al. 2000).  With many large
telescopes with polarimeters currently (or soon to be) on line (e.g., Keck,
VLT, Gemini), prospects for testing these connections are bright.

\section{Acknowledgments}

It is a pleasure to thank Bob Becker, Alison Coil, Aaron Barth, and Richard
White for assistance with the observations, and Tom Matheson and Weidong Li for
useful discussions.  Elinor Gates and Ryan Chornock are particularly thanked
for their assistance with this project.  We thank Peter H\"{o}flich for many
useful suggestions that improved the manuscript.  Some of the data presented
herein were obtained at the W.M. Keck Observatory, which is operated as a
scientific partnership among the California Institute of Technology, the
University of California, and the National Aeronautics and Space
Administration.  The Observatory was made possible by the generous financial
support of the W.M. Keck Foundation.  This research has made use of the
NASA/IPAC Extragalactic Database (NED), which is operated by the Jet Propulsion
Laboratory, California Institute of Technology, under contract with NASA.  We
have made use of the LEDA database\footnote{http://leda.univ-lyon1.fr.}.  Our work was
funded by NASA grants GO-7821, GO-8243, and GO-8648 from the Space Telescope
Science Institute, which is operated by AURA, Inc., under NASA contract NAS
5-26555; additional funding was provided by NASA/Chandra grant GO-0-1009C; by
NSF grants AST-9417213 and AST-9987438; and by the Sylvia and Jim Katzman
Foundation.

\clearpage

\appendix
\section{APPENDIX}
\subsection {DETAILS OF THE REDUCTIONS}

\label{seca:a}

Miller et al. (1988) discuss the reduction of data obtained using the Lick
spectropolarimeter, and similar procedures are followed for data acquired with
the Keck spectropolarimeter.  Here we focus on three issues specific to the
data acquired for SN~1999em: determining the Stokes parameters with only 3 of
the usual 4 waveplate position observations, instrumental polarization, and
sources of systematic uncertainty in SN spectropolarimetry.

\begin{figure}
\begin{center}
 \scalebox{0.80}{
	\plotone{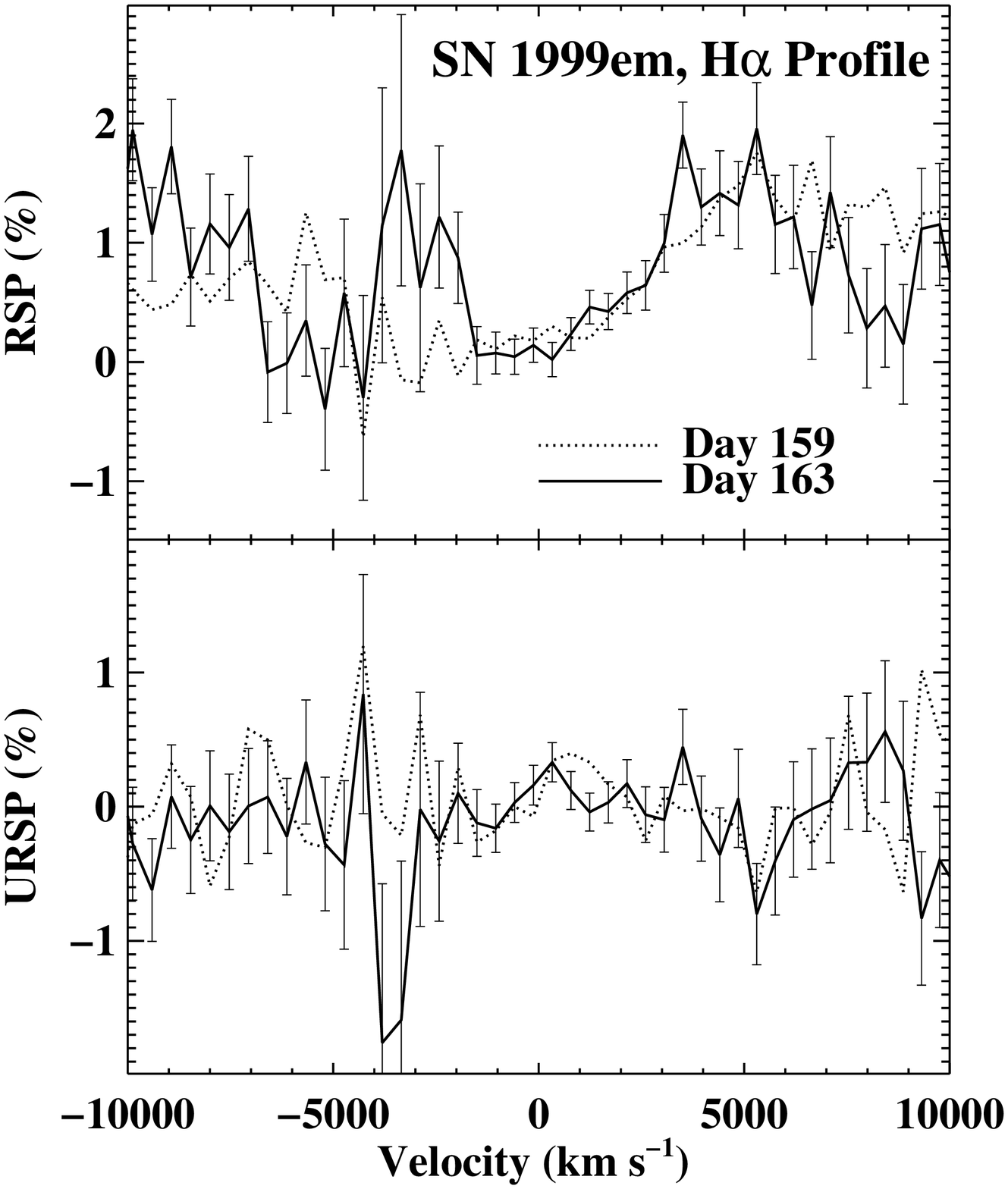}
		}
\end{center}
\caption[Comparison of polarization across the H$\alpha$ profile measured
on days 159 and 163 for SN 1999em] {Comparison of the polarization across the
H$\alpha$ profile measured on days 159 and 163.  For clarity, the $1\sigma$
statistical error bars are only shown for the day 163 data. The generally good
agreement, particularly obvious in regions of high signal-to-noise ratio,
suggests no large systematic error associated with the problematic reduction of
the day 159 data.}
\label{fig3:24}
\end{figure}

For a given observation, both the Lick and Keck spectropolarimeters place two
parallel, perpendicularly polarized beams ($B$ and $T$ for ``bottom'' and
``top'') onto the CCD chip, with the polarization angle selected by means of a
rotatable, achromatic half-wave plate.  Rotating the wave plate by an angle
$\theta$ rotates the plane of polarization by 2$\theta$.  Thus, a $45^\circ$
rotation exchanges the positions of the two parallel beams on the chip.
Generally, four observations are required (``1,'' ``2,'' ``3,'' ``4''),
corresponding to the four waveplate positions (typically observed in this
order) $0^\circ, 45^\circ, 22.5^\circ, {\rm and\ } 67.5^\circ$ (polarization
angles $0^\circ, 90^\circ, 45^\circ, {\rm and\ } 135^\circ$ in the plane of the
sky), in order to measure $q$ and $u$ with sufficient redundancy in the data
that changes in seeing, transparency, or instrumental response will not affect
the results.  The main purpose of swapping the location on the CCD of the two
beams in successive exposures is to remove differences in the instrumental
response between the $B$ and $T$ beams from the calculation of the Stokes
parameters.  It is possible to compute a gain ratio, $g_1$, which describes the
relative response of the $B$ and $T$ beams for the first two waveplate
positions:
\begin{equation}
g_1 = \sqrt \frac{B_1B_2}{T_1T_2}.
\label{eqna:a1}
\end{equation}
\noindent A gain ratio $g_2$ for the second pair of observations can be
similarly calculated.  In general, we have found $g_1 \approx g_2$.  In order
to calculate $q$ and $u$ with the day 159 data for SN~1999em that lacked the
fourth exposure (waveplate position $67.5^{\circ}$), we had to {\it assume}
that $g_1$ = $g_2$, and use $g_1$ to remove instrumental response differences
between the top and bottom beams in the third exposure.  While we were thus
able to form the complete set of Stokes parameters from this limited data set,
increased potential for systematic errors obviously existed.  Fortunately, we
obtained an additional (complete) set of observations just four days later at
Keck, which largely confirmed the earlier results (Fig.~\ref{fig3:24}).
(Unfortunately, since the zero point of the spectrograph shifted by
$\sim4^\circ$ between the two observations, we could not directly use the
fourth waveplate position from day 163 to reduce the day 159 data without
mixing the Stokes parameters; the gratings employed also had very different
resolutions.)  For the analysis of late-time SN~1999em polarimetry presented in
\S~\ref{sec3:4.1.3}, we thus took the weighted average of these two data sets
after first forming the Stokes parameters for each separately.

\begin{figure}
\begin{center}
 \scalebox{0.9}{
	\plotone{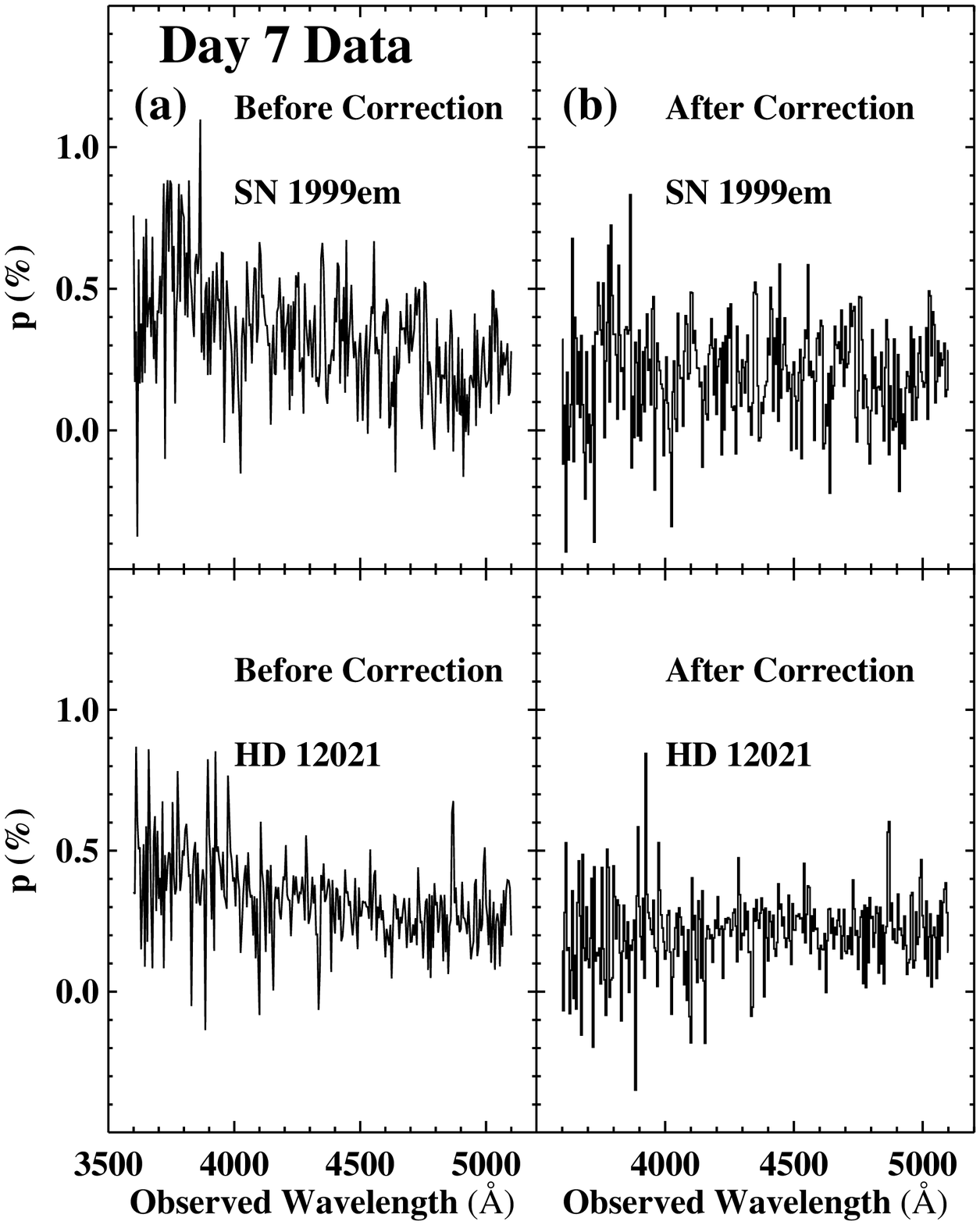}
		}
\end{center}
\caption[SN 1999em polarization 7 days after discovery compared with a null
standard] 
{Observed polarization measured on 1999 November 5 for SN 1999em and
the null standard, HD 12021, shown ({\it a}) before and ({\it b}) after removal
of instrumental polarization.}
\label{fig3:25}
\end{figure}

In general, observations of null polarization standards were null to within
$0.1\%$, and the measured polarization of polarized standards agreed well with
cataloged values, implying little instrumental polarization over the optical
passband at both Keck and Lick.  On day 7, however, there was clear evidence
for systematic problems with all of the data collected by the blue arm of the
spectrograph: both our object and null standards showed a rise of up to
$\sim0.4\%$ at blue wavelengths (Fig.~\ref{fig3:25}a), and polarization
standards showed changes in polarization angle at blue wavelengths that
disagreed with published values.  Martel (1996) found similar effects with the
Lick spectropolarimeter, and concluded that the most likely source was
instrumental polarization caused by defects in the aluminum and/or overcoat
layers of the primary and secondary mirrors.  A similar situation has been
reported at the Palomar 5-m Hale telescope (Ogle et al. 1999).  The only null
standard observed on day 7 was HD 12021, a star flagged by Schmidt et
al. (1992b) as possessing some intrinsic polarization, and we did not observe
any other null standards with this setup until after the primary was
realuminized on 2000 February 14.  Unfortunately, realuminization alters the
polarization characteristics of the mirrors (previous dates for realuminization
of the Lick 3-m primary are 1996 February 6, 1992 December 8, 1990 February 5,
and 1987 March 16; D. Severinsen 2000, private communication).  Thus, we had to
rely on HD 12021 to correct the data from this night.

As in previous studies, we assume that the telescope's intrinsic polarization
is produced by the mirrors, although polarization due to the dichroic is also
possible (see Martel 1996 for a discussion favoring the mirrors as the likely
cause).  To remove instrumental polarization, it is important to subtract it as
the first step in the polarimetric reduction, before any other corrections are
made.  Since the Lick 3-m is an equatorially mounted telescope with its primary
and secondary always maintaining the same orientation with respect to absolute
direction in the sky, it is necessary to carry out the instrumental
polarization subtraction with both the instrumental and object polarization
determined for the same position angle (P.A.) of the spectrograph.  That is,
for an object with observed (i.e., unrotated, ``raw'') Stokes parameters
($q_{\rm obs},u_{\rm obs}$), the corrected Stokes parameters ($q_{\rm
cor},u_{\rm cor}$) are given by
\begin{equation}
q_{\rm cor} = q_{\rm obs} - q_{\rm inst},\\
\label{eqna:a2}
\end{equation}
\begin{equation}
u_{\rm cor} = u_{\rm obs} - u_{\rm inst}.
\label{eqna:a3}
\end{equation}
\noindent Here ($q_{\rm inst},u_{\rm inst}$) are the instrumental polarization
rotated to the object's P.A. by
\begin{equation}
q_{\rm inst} = q_{\rm null}\cos(2\Delta PA) + u_{\rm null}\sin(2\Delta PA),\\
\label{eqna:a4}
\end{equation}
\begin{equation}
u_{\rm inst} = -q_{\rm null}\sin(2\Delta PA) + u_{\rm null}\cos(2\Delta PA),
\label{eqna:a5}
\end{equation}
\noindent where ($q_{\rm null},u_{\rm null}$) are the observed, unrotated Stokes
parameters for the null standard, and $\Delta PA \equiv PA_{\rm obj} - PA_{\rm
null}$.\footnote{Note that for a telescope like Keck-I, which has an
altitude-azimuthal mounting, the mirrors maintain their orientation relative to the
{\it horizon}.  That is, if instrumental polarization exists at Keck, two
separate observations of a null standard will produce identical ($q_{\rm
null},u_{\rm null}$) if both are made with the spectrograph at the same
orientation relative to the horizon, whereas at Lick they will be identical if
both are made with the same spectrograph orientation relative to absolute
direction in the plane of the sky.  To correct Keck data one could define
$\Delta PA$ in equations~(\ref{eqna:a4}) and ~(\ref{eqna:a5}) as $\Delta PA
\equiv ( [PA_{\rm obj} - PA_{\rm opj\_par}] - [PA_{\rm null} - PA_{\rm
null\_par}] )$, where ``par'' refers to the parallactic slit P.A. that
minimizes differential light loss (i.e., is perpendicular to the horizon; see
Filippenko 1982).}

\begin{figure}
\begin{center}
 \scalebox{0.70}{
	\plotone{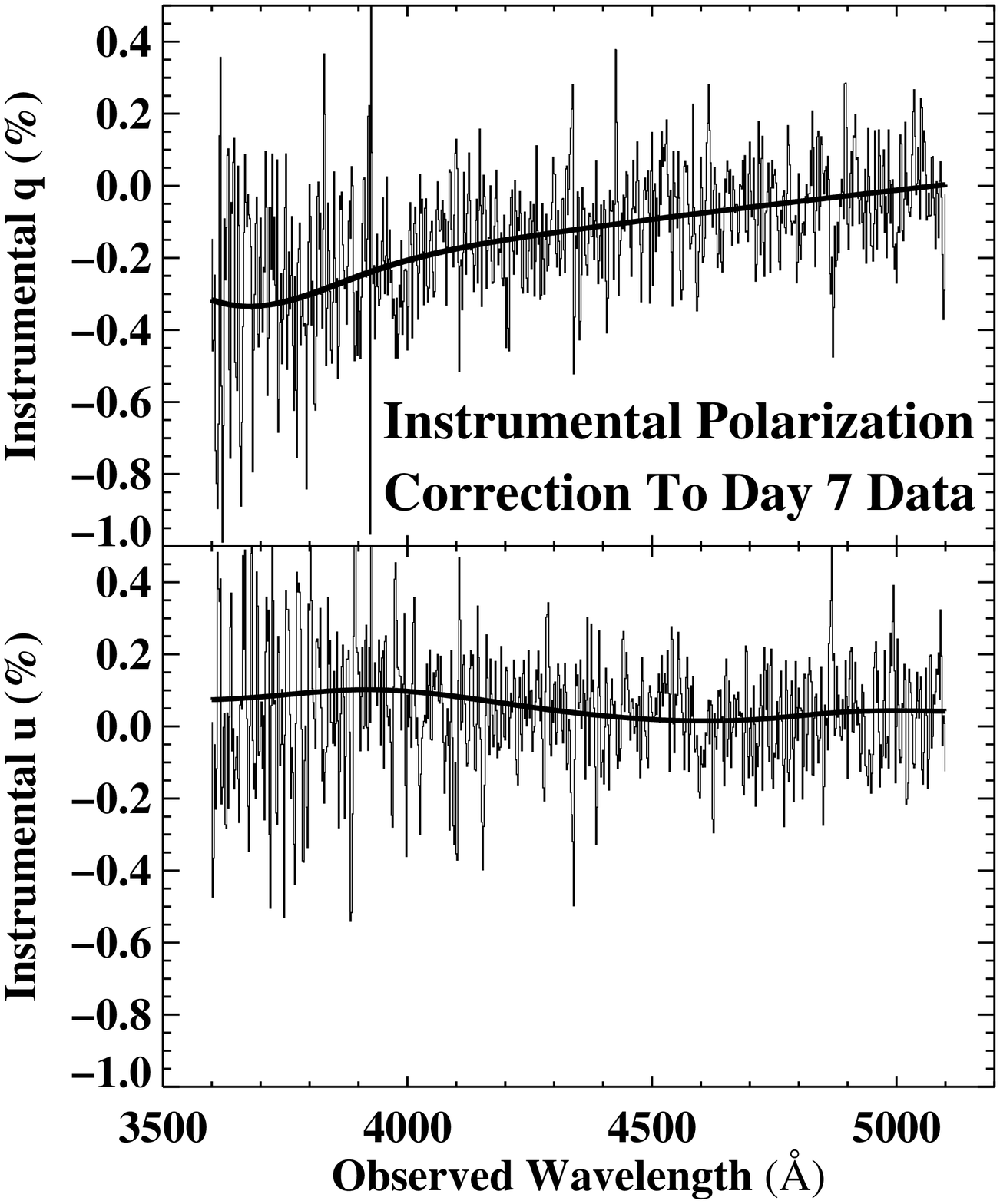}
		}
\end{center}
\caption[Instrumental polarization correction to day 7 data]
{Instrumental polarization removed from data obtained on 1999 November 5 with
the Kast double spectrograph at the Lick 3-m telescope.  Shown here is the
spline fit ({\it thick line}) to the instrumental $q$ and $u$ Stokes parameters
rotated to a telescope P.A. of 0 degrees.  The instrumental polarization was
derived using the null standard HD 12021 on this night; see text for details. }
\label{fig3:26}
\end{figure}

To derive the instrumental polarization from our observation of HD 12021, we
assumed that the star's intrinsic polarization was accurately recorded by the
red arm of the spectrograph, a belief supported by the lack of systematic
trends seen in the P.A. of the polarization standards on the red side.
Therefore, we first removed $\sim0.2\%$ from the measured polarization of HD
12021, presumably leaving us with just the ``instrumental'' portion.  We fit
low-order splines to the remaining instrumental Stokes parameters rotated to
P.A. = 0$^{\circ}$ (Fig.~\ref{fig3:26}), and then subtracted this spline as
the first step in the polarimetry reduction from all objects observed on this
night (after first rotating the object's observed polarization to a P.A. of
0$^{\circ}$).  After correction, the polarization of SN~1999em became flat at
blue wavelengths (Fig.~\ref{fig3:25}b), and the polarization angle of the
polarization standards became more stable, suggesting that we properly removed
the instrumental portion of the polarization.

One final issue is the degree systematic errors play in the determination of SN
polarization.  To test the stability of the Lick polarimeter, we took five
complete sets of observations in succession on 2000 July 28 of the polarization
standard HD 19820, without rotating the slit between observations, and computed
the observed $V$-band polarization for each set.  Although the statistical
uncertainty of each individual broadband measurement was very low
($\sigma_{q,u} < 0.008\%$), we found $\langle q_V\rangle = 3.122\% \pm 0.040\%,
\langle u_V\rangle = -3.663\% \pm 0.015\%$, where the quoted uncertainties in
$\langle q_V\rangle {\rm\ and\ } \langle u_V\rangle$ are the standard
deviations of the five individual measurements (note that the resulting $p_V =
4.81\% \pm 0.03\%$ compares well with the published value $p_V = 4.79\% \pm
0.03\%$ [Schmidt et al. 1992b]).  Clearly, systematic uncertainties dominate
over statistical ones at the $0.05\%$ level.  Note, however, that this
uncertainty was never seen to manifest itself as artificial line features;
rather, slight shifts in the {\it overall} continuum level were responsible for
the differences.

\begin{figure}
\begin{center}
 \scalebox{0.90}{
	\plotone{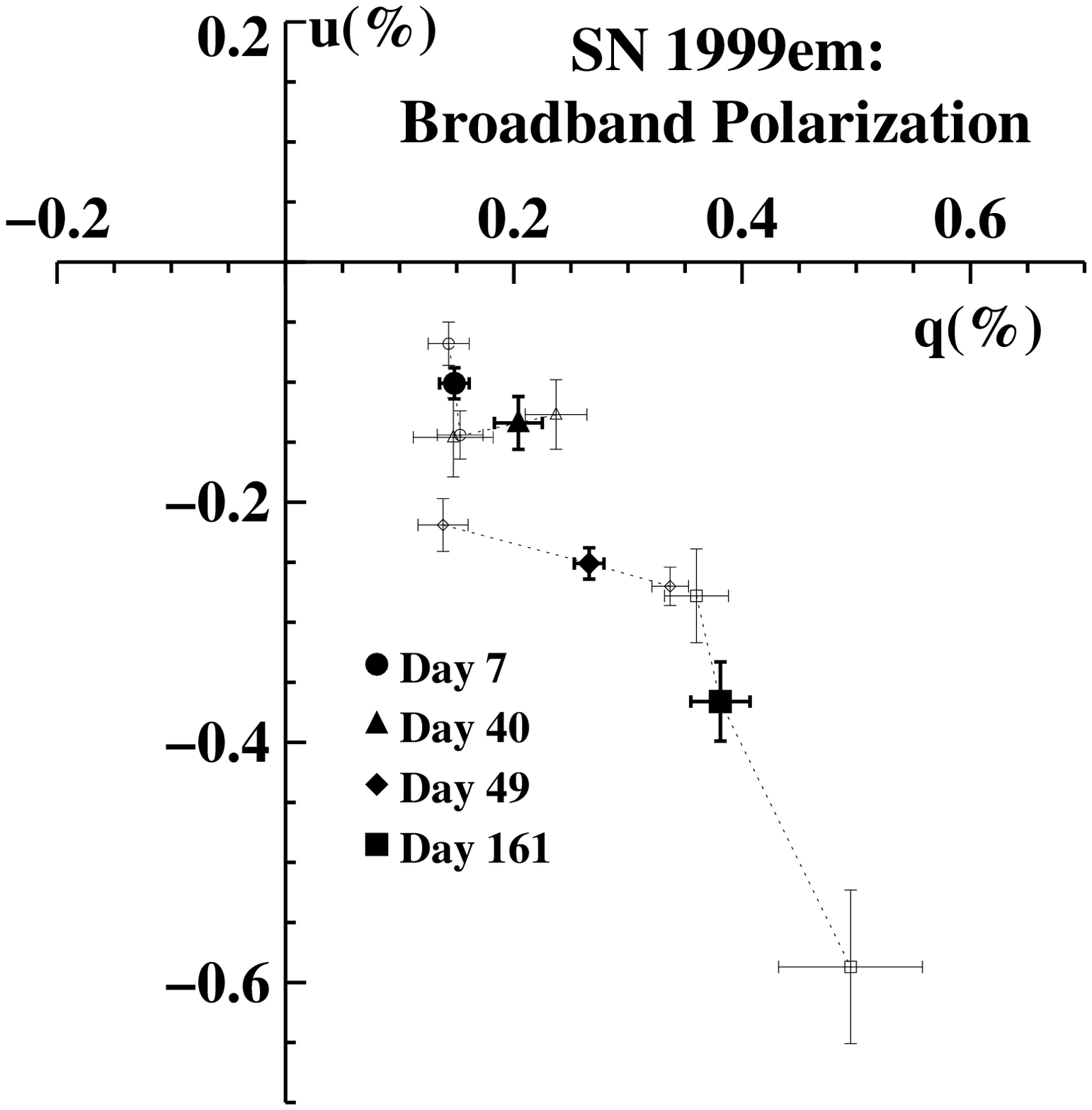}
		}
\end{center}
\caption[Evolution of the broadband polarization of SN 1999em in $q$-$u$ plane
from day 7 to day 161 after discovery] {Evolution of the broadband polarization
of SN 1999em from day 7 to day 161 after discovery, displayed as flux-weighted
averages over $5050 - 5950$ \AA\ (approximates observed $V$ band); due to its
different spectral coverage the value for day 7 was computed over $6050 -
6950$~\AA).  For each epoch, the two individual observations ({\it small open
symbols}) are connected to the combined value ({\it large filled symbols}) with
a dotted line.  Error bars represent the $1\sigma$ statistical uncertainties.
Systematic effects clearly dominate the uncertainty in the broadband
polarization measurements.  }
\label{fig3:27}
\end{figure}

\begin{figure}
\begin{center}
\rotatebox{90}{
 \scalebox{0.6}{
	\plotone{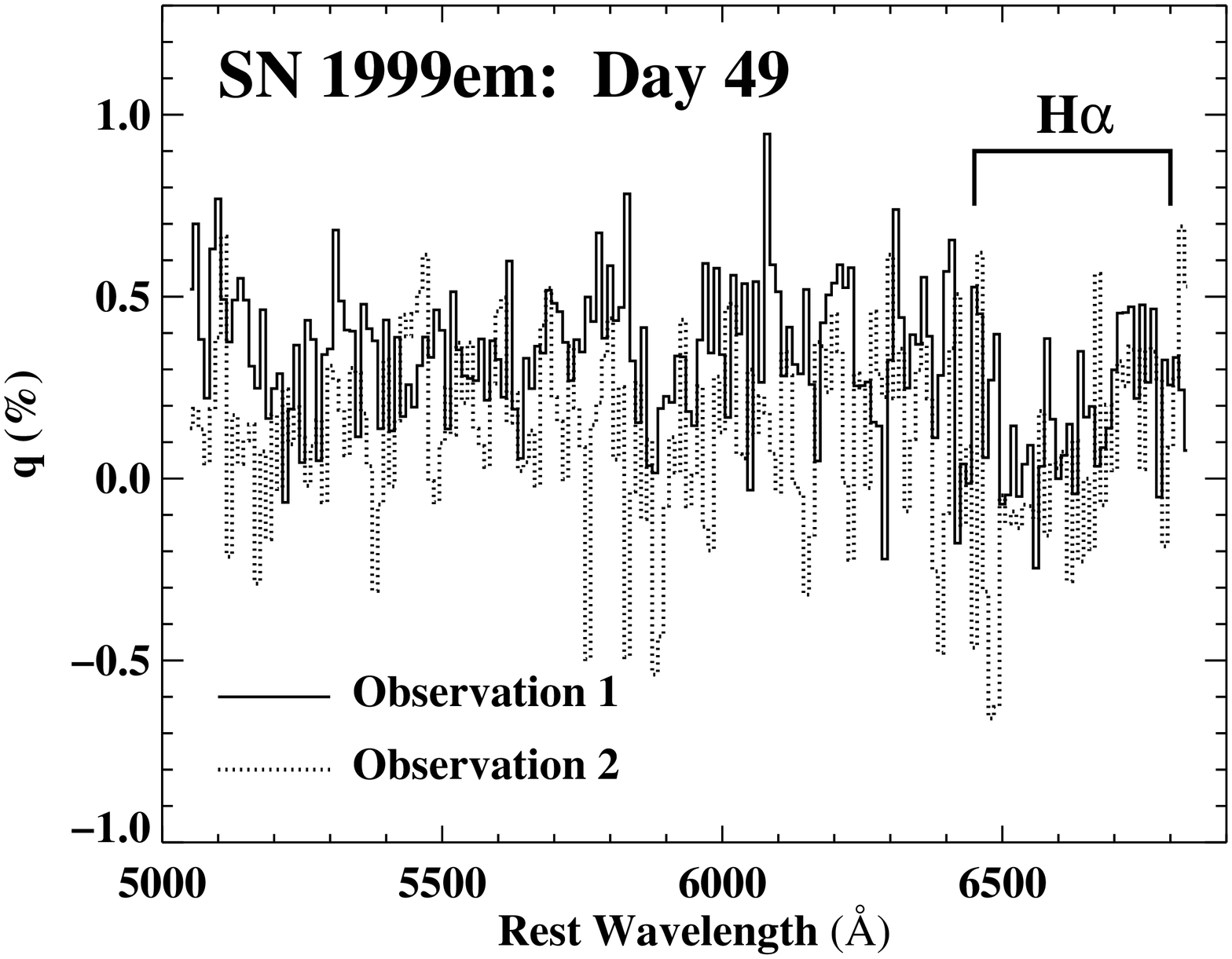}
		}
	}
\end{center}
\caption[Stokes $q$ parameter measured by successive observations of
SN 1999em 49 days after discovery] {Stokes $q$ parameter measured by successive
observations of SN 1999em on 1999 December 17.  The second observation appears
to be shifted by $\sim 0.2\%$ with respect to the first.  The offset may arise
from polarization differences in the host galaxy light in the background
regions, due to the P.A. rotation made between the two observations.
Fortunately, this systematic effect does not appear to influence the relative
strength of polarization modulations across line features, such as the dip seen
here at H$\alpha$.}
\label{fig3:28}
\end{figure}

Barring intrinsic polarization changes in SN1999em over the timescale of a few
hours, Figure~\ref{fig3:27} demonstrates that systematic errors also dominate
statistical uncertainties for SN~1999em.  In the worst case (day 49, $q$
parameter), successive exposures disagree by more than $9\sigma$
(Fig.~\ref{fig3:28}).  In addition to the systematic instrumental uncertainty
described earlier, another error source for SN spectropolarimetry is the light
contributed by the host galaxy, either in the background regions or mixed into
the SN spectrum.  Broadband imaging polarimetry of the diffuse light in nearby
spiral galaxies indicates polarizations of several percent (e.g., Scarrott,
Rolph, \& Semple 1990; Sholomitskii, Maslov, \& Vitrichenko 1999; see also
Brindle et al. [1991] in which the central $6^{\prime\prime}$ of NGC 1637, the
host of SN~1999em, is found to be polarized at 1.34\% in the $B$ band).  While
host galaxy light is typically only a few percent as bright as the SN, removal
of different background regions from a SN spectrum in successive sets of
exposures (as can happen when the P.A. is rotated to maintain the parallactic
angle) could certainly affect the overall polarization level at the $0.1\%$
level.  One possible remedy is to maintain a constant slit P.A. for all
exposures, regardless of the parallactic P.A. However, this risks allowing
polarized galaxy light to be mixed in with the SN light itself when observing
far away from the parallactic angle, which can produce a spurious wavelength
dependence to the continuum polarization (e.g., Barth et al. 1999).  Variable
seeing and poor guiding may also play a role.  Such considerations may explain
the differences seen between successive exposures of SN~1999em.  Fortunately,
while the {\it overall} polarization level may be affected by diffuse host
galaxy contamination, sharp line features are not altered (see
Figs.~\ref{fig3:28} and \ref{fig3:24}).  Nevertheless, assigning a $1\sigma$
systematic uncertainty to SN continuum polarization measurements of $\sim0.1\%$
seems warranted, especially for SNe in dusty spiral galaxies.

\subsection{Interpretation of Spectropolarimetric Data With Low Polarization} 
\label{seca:b}

Studying an object possessing nearly zero continuum polarization but sharp
polarization changes across line features demands careful attention to detail
in the polarimetric analysis.  As an aid to understanding some of the
subtleties involved, we present the following analysis of artificial data.

\begin{figure}
\begin{center}
 \scalebox{0.80}{
	\plotone{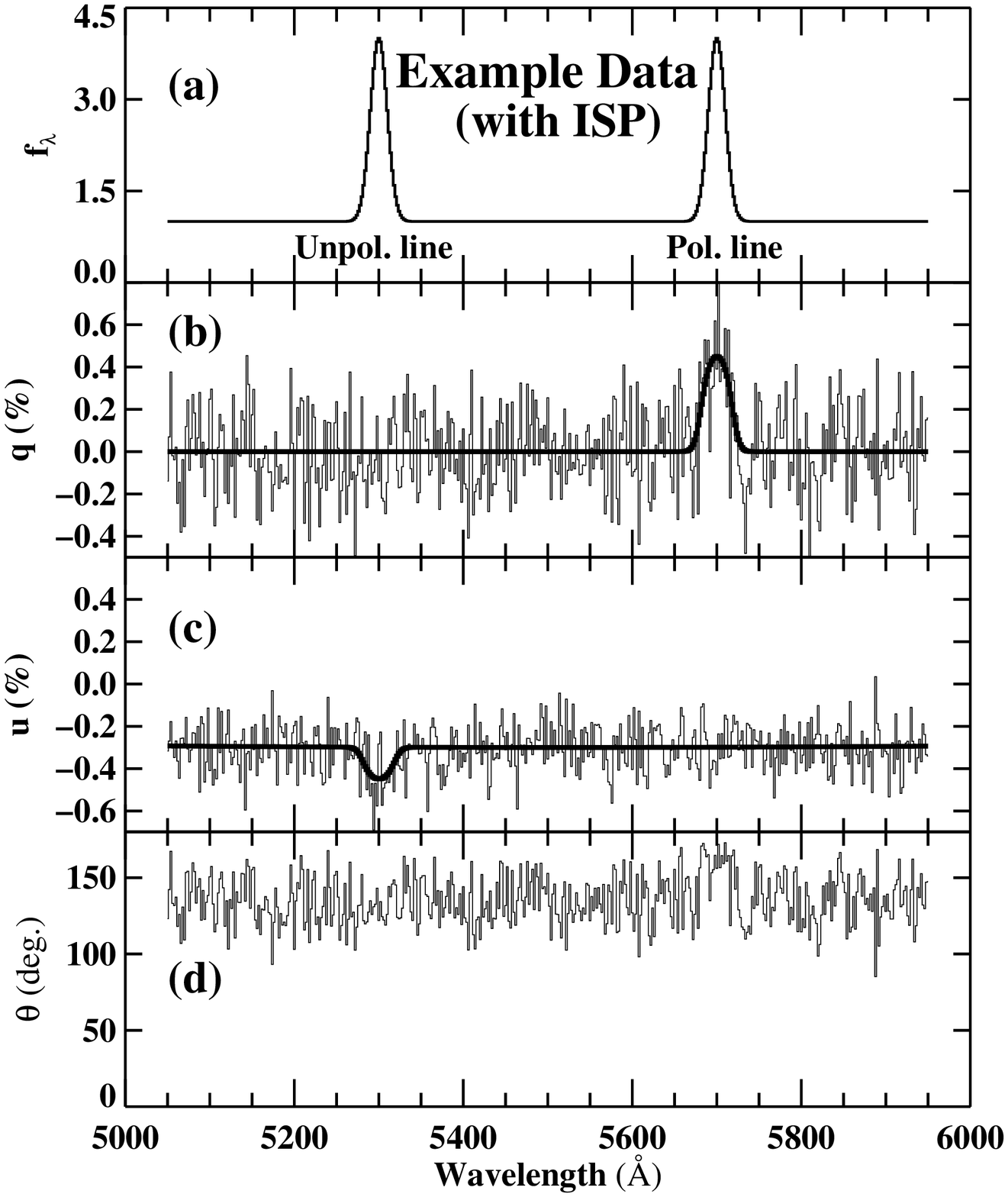}
		}
\end{center}
\caption[Spectropolarimetry of artificial data]
{Artificial data, showing ({\it a}) the ``observed'' flux, ({\it b, c})
$q$, $u$, and ({\it d}) position angle ($\theta$) in the plane of the sky,
uncorrected for the effects of ISP.  The ``noise-free'' Stokes parameters are
overplotted ({\it thick line}) in ({\it b}) and ({\it c}).}
\label{fig3:29}
\end{figure}

Figure~\ref{fig3:29} shows ``observed'' spectropolarimetry for an object with a
polarized continuum ($q_{\rm cont} = 0\%, u_{cont} = 0.2\%$), one polarized
line centered at 5700~\AA\ ($q_{pol} = 0.6\%, u_{pol} = 0.2\%$), and one
unpolarized line centered at 5300~\AA\ ($q_{unpol} = 0.0\%, u_{unpol} =
0.0\%$), to which we have added interstellar polarization ($q_{ISP} = 0.0\%,
u_{ISP} = -0.5\%$) with a ``Serkowski curve'' (Serkowski 1973; Wilking,
Lebofsky, \& Rieke 1982) peaking at 5500~\AA.  Gaussian noise of $\sigma_q =
0.2\%, \sigma_u = 0.1\%$ has also been added to the original $q$ and $u$ data.
For ease of analysis, note that all the ISP is artificially in the $u$
direction and all the excess line polarization is in the $q$ direction.

\begin{figure}
\begin{center}
\rotatebox{90}{
 \scalebox{0.65}{
	\plotone{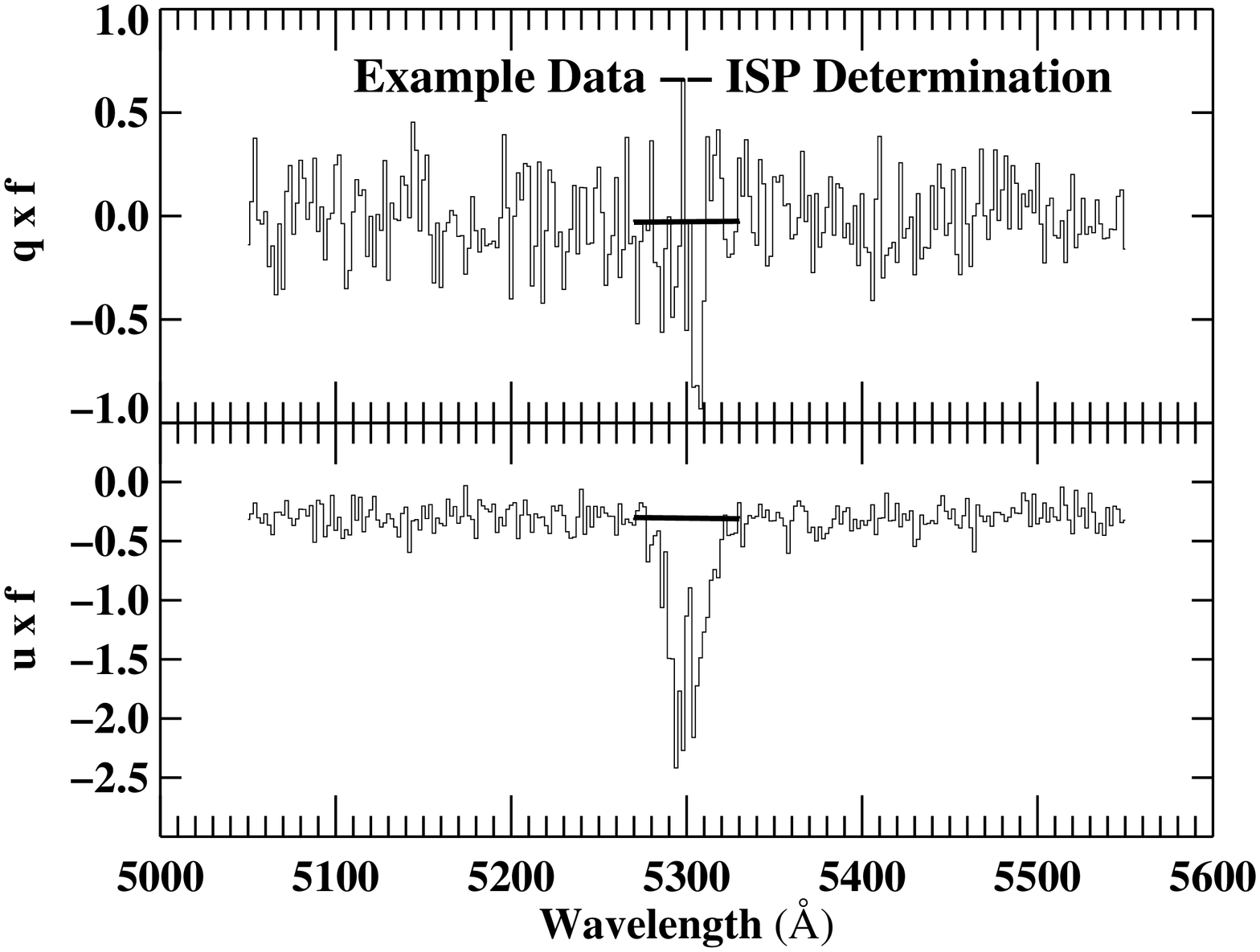}
		}
	}
\end{center}
\caption[Determining the ISP for Artificial Data]
{``Stokes parameter fluxes'' for the artificial data in the vicinity of
the unpolarized line, showing the extent of the fit to the continuum ({\it
heavy line}) used to determine the ISP.}
\label{fig3:30}
\end{figure}

To begin, the unpolarized line is used to estimate the ISP contaminating the
object's polarization (i.e., \S~\ref{sec3:4.1}).  This is done by calculating
the flux-weighted and continuum-subtracted averages of $q$ and $u$ across the
unpolarized line's profile, with $q_{ISP}$ found by measuring
\begin{equation}
q_{line} = \frac{[\sum (fq) - (fq)_{cont}]} {[\sum(f - f_{cont})]}, 
\label{eqna:b1}
\end{equation}
\noindent and setting $q_{ISP} = q_{line}$, where $(fq)_{cont}$ and $f_{cont}$
represent fits to the continuum level underlying the line in the Stokes
parameter flux and total flux, respectively.  A similar procedure is carried
out to determine $u_{ISP}$.  The fit to $(fq)_{cont}$ is preferably taken as
the interpolated value of the means found for background regions on either side
of the line profile.  If the background is complicated, however, the continuum
may be drawn by hand under the line feature in the $fq$ spectrum, although the
exact placement may be quite subjective.  The continuum fits to the Stokes
parameter fluxes for the line region $5270 - 5330$~\AA\ are shown in
Figure~\ref{fig3:30}; background regions were chosen to be $5150 - 5250$~\AA\
and $5350 - 5450$~\AA.  Using equation~(\ref{eqna:b1}), we determine $q_{ISP} =
-0.10\%\ \pm\ 0.07\% {\rm\ and\ } u_{ISP} = -0.51\%\ \pm\ 0.04\%$, in good
agreement with the actual values.  To translate this into $p$ and $\theta$\ we
use the ``debiased'' estimator of $p$ (see below) to obtain $p_{ISP} = 0.51\%\
\pm 0.04\% {\rm\ at\ }\theta = 130^\circ \pm 4^\circ$, in good agreement with
the known value.

\begin{figure}
\begin{center}
 \scalebox{0.80}{
	\plotone{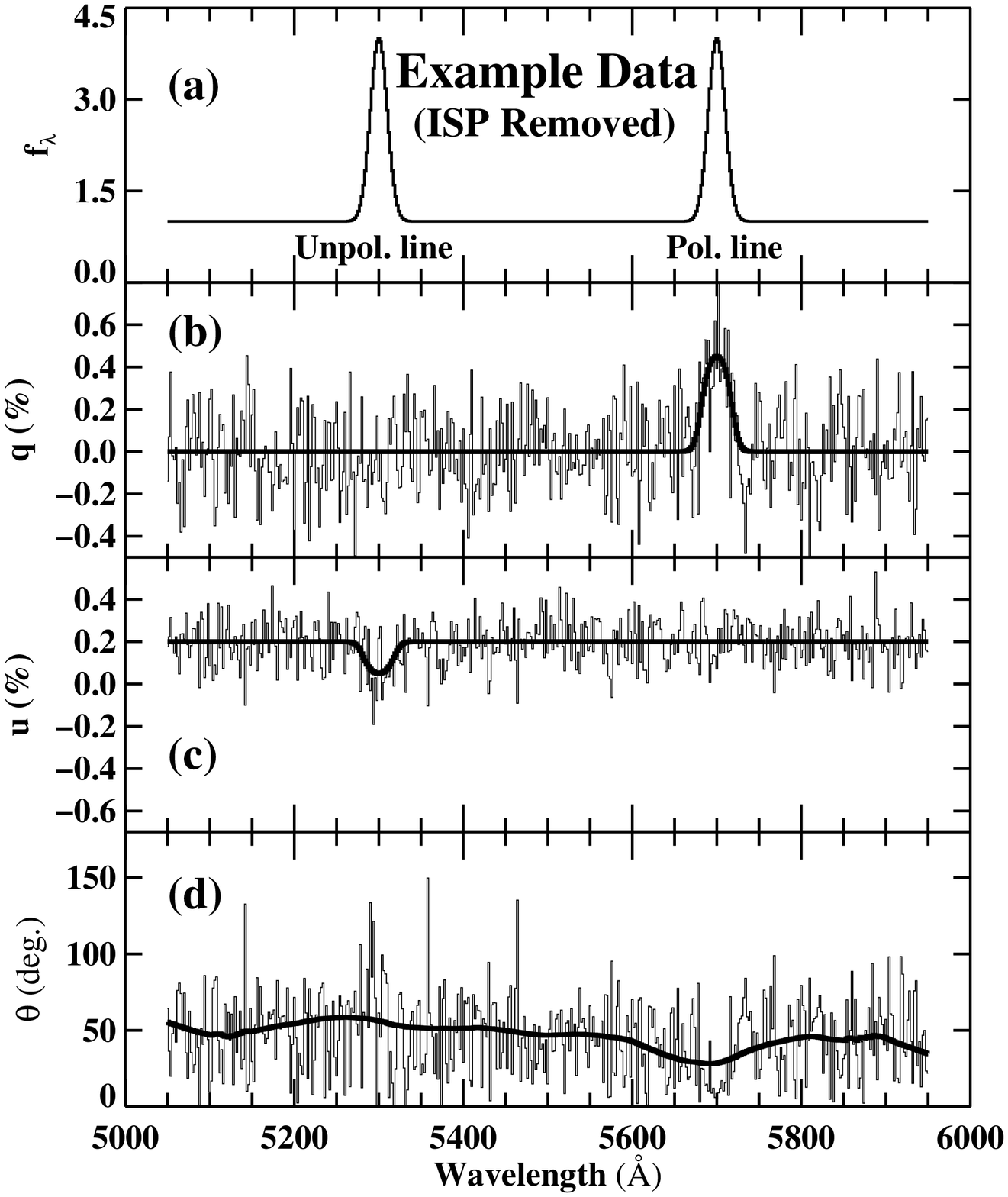}
		}
\end{center}
\caption[Spectropolarimetry of artificial data with ISP removed]
{Artificial data with ISP removed.  The ``true'' values of the Stokes
parameters are overplotted ({\it heavy line}) in ({\it b}) and ({\it c}), along
with the smoothed fit (smoothed and median filtered with a boxcar of width 100
\AA, our typical smoothing value) to the polarization angle ($\theta$) used to
determine the RSP, in ({\it d}).  Note that the sharp polarization angle change
across the polarized line at 5700~\AA\ is substantially missed by the fit.}
\label{fig3:31}
\end{figure}

The resulting ``intrinsic'' spectropolarimetry is shown in
Figure~\ref{fig3:31}; in order to simplify the following analysis by keeping
all of the extra line polarization directed parallel to the $q$ axis, we have
removed the actual, not the measured, ISP.  Translating $q$ and $u$ into $p$
can now proceed by calculating any of
\begin{equation}
^{\prime\prime} {\rm Traditional''}\ p \equiv p_{trad} = \sqrt{q^2 + u^2},
\label{eqna:b2}
\end{equation}
\begin{equation}
^{\prime\prime}{\rm Debiased''}\ p \equiv p_{deb} = \pm \sqrt{\mid q^2 + u^2 -
(\sigma^2_q + \sigma^2_u)\mid}, 
\label{eqna:b3}
\end{equation}
\begin{equation}
^{\prime\prime}{\rm Optimal}{\rm ''}\ p \equiv
p_{opt} = p_{trad} - \frac{\sigma^2(p_{trad})}{p_{trad}}, 
\label{eqna:b4}
\end{equation}
\begin{equation}
{\rm RSP} = q\cos2\theta + u\sin2\theta,
\label{eqna:b5}
\end{equation}

\noindent where $p_{deb}$ assumes the sign of ($q^2 + u^2 - [\sigma^2_q +
\sigma^2_u]$) and $p_{opt}$ is formally defined only when $p_{trad} >
\sigma(p_{trad})$.  RSP is found by rotating the $q$ and $u$ coordinate system
through an angle $\theta$ ($\theta \equiv (1/2) \arctan[u / q]$), which places
all of the polarization along the rotated $q$ parameter. A smoothed version of
$\theta$ is always used to form the RSP (Fig.~\ref{fig3:31}d).  Some of the
theory underlying $p_{deb}, p_{opt}$, and RSP is given by Stockman \& Angel
(1978; $p_{deb}$), Wang, Wheeler, \& H\"{o}flich (1997; $p_{opt}$) and Trammel,
Dinerstein, \& Goodrich (1993; RSP).  A discussion regarding the biases
introduced by using either $p_{trad}$ or $p_{deb}$ on data with low
$p/\sigma_p$ is presented by Miller et al. (1988).

\begin{figure}
\begin{center}
 \scalebox{0.70}{
	\plotone{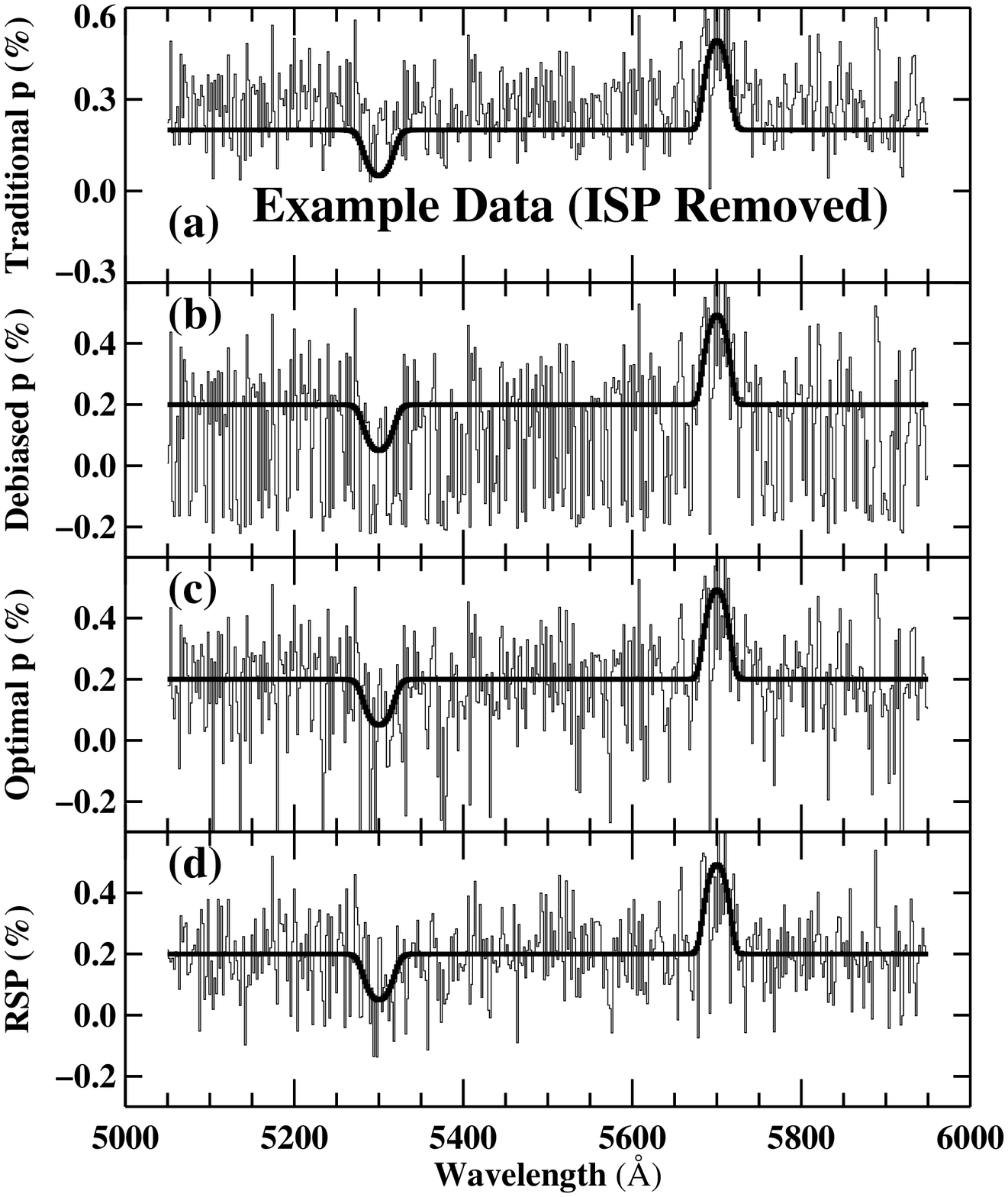}
		}
\end{center}
\caption[Four different estimates of the polarization of the artificial data]
{Four estimates of the polarization of the artificial data, with the
``true'' (noise-free) polarization overplotted ({\it thick line}).  ({\it a})
The ``traditional'' $p$ is biased high; ({\it b}) the ``debiased'' $p$ shows
many large negative values; ({\it c}) the ``optimal'' $p$ is better behaved,
but still shows a few large, negative spikes; ({\it d}) the rotated Stokes
parameter provides a good representation of the true continuum polarization,
although the polarization of the line feature at 5700~\AA\ is somewhat
underestimated (see Fig.~\ref{fig3:33}).}
\label{fig3:32}
\end{figure}

Shortcomings inherent to all four estimators of $p$ are apparent in
Figure~\ref{fig3:32}.  The ``traditional'' polarization estimator, $p_{trad}$,
is biased high, and does a poor job detecting the unpolarized line feature at
5300~\AA.  The high bias results from its positive-definite definition
(eq.~[\ref{eqna:b2}]) as well as the positive skewness exhibited by its
probability distribution (e.g., Miller et al. 1988), with the former cause
dominating for the low $p/\sigma_p$ used in this example.  Note that $p_{trad}$
becomes a progressively worse estimator of $p$ as $p/\sigma_p$ decreases,
placing a cautionary note on interpretation of increases in $p_{trad}$ seen in
spectral regions where the signal-to-noise ratio is low (e.g., line troughs).
The ``debiased'' polarization estimator, $p_{deb}$, also poorly represents the
true polarization (Fig.~\ref{fig3:32}b), since its probability distribution is
double-peaked and results in a preponderance of negative polarizations in many
pixels when $p/\sigma_p$ is low (Miller et al. 1988).  The ``optimal''
polarization estimator (Fig.~\ref{fig3:32}c) has fewer large negative data
spikes (formally, $p_{opt}$ is undefined at these locations) and overall gives
a better representation of the true polarization than either $p_{trad}$ or
$p_{deb}$.

\begin{figure}
\begin{center}
 \scalebox{0.75}{
	\plotone{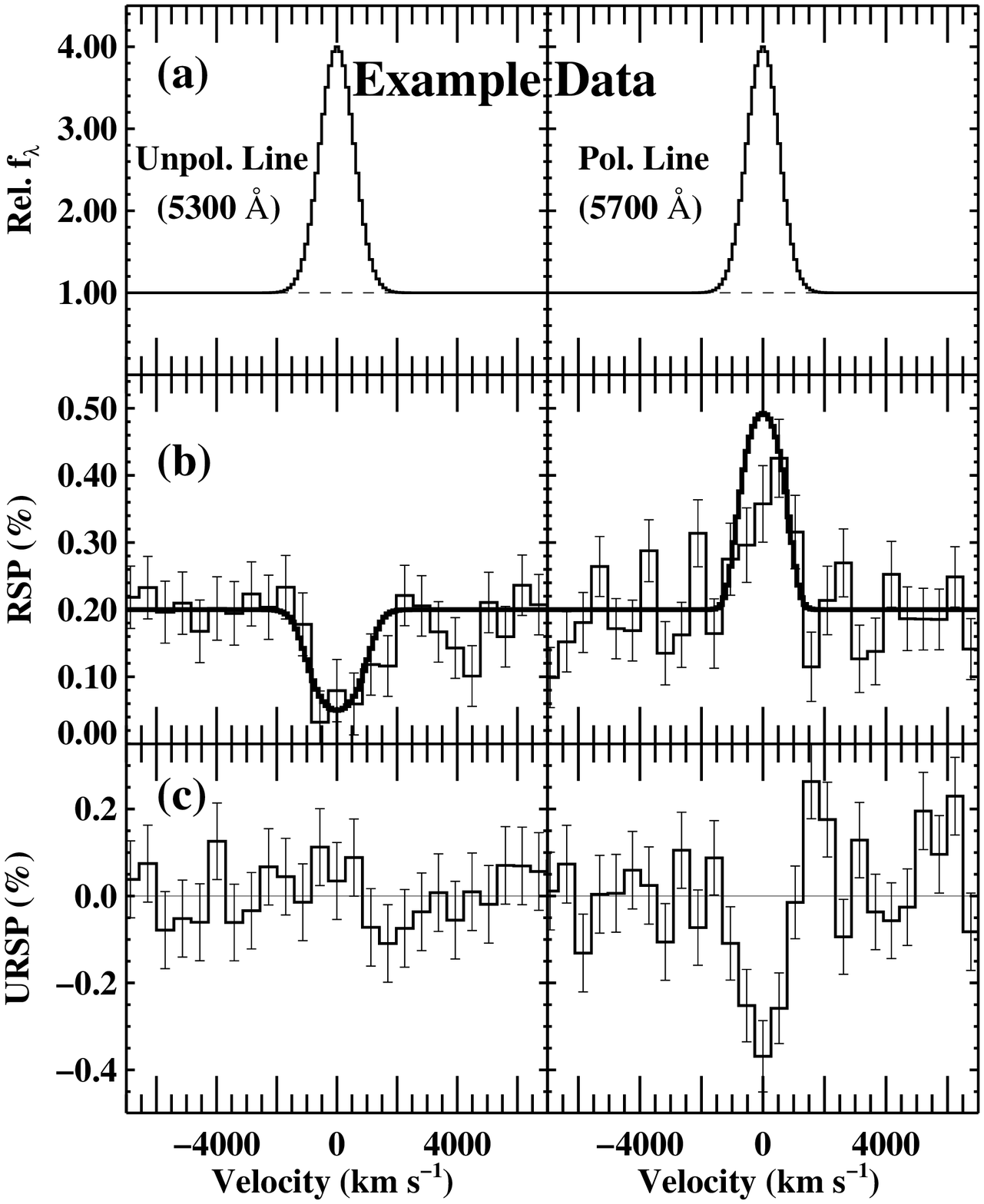}
		}
\end{center}
\caption[Line polarization in artificial data]
{({\it a}) Normalized flux, ({\it b}) RSP, and ({\it c}) URSP for the
artificial data in the vicinity of the unpolarized line feature at 5300~\AA\
and the polarized line feature at 5700~\AA.  The true polarization level is
overplotted ({\it thick line}) in ({\it b}).  The fact that the RSP
underestimates the polarization in the polarized line could be deduced by
inspection of the line's URSP, which shows large residuals with a consistent
sign in the three central pixels.}
\label{fig3:33}
\end{figure}

Figure~\ref{fig3:32}d shows that RSP yields an accurate representation of the
continuum polarization level, with a well-behaved Gaussian distribution of
values about the true continuum value.  In places where $\theta$ changes
rapidly (e.g., across the polarized line feature at 5700~\AA), however, RSP
underestimates the true polarization since $\theta_{smooth} \neq \theta$ in the
line (Fig.~\ref{fig3:31}d).  When this happens, some of the line's
polarization will be contained in the rotated $u$ parameter (URSP).  Thus, when
determining line polarization with RSP, it is often necessary to inspect both
RSP and URSP.  Figure~\ref{fig3:33} shows the data for both the unpolarized and
polarized lines, binned by 10~\AA\ to reduce noise.  Since $\theta$ does not
change across the unpolarized line, RSP accurately estimates $p_{unpol}$,
showing only random residuals about 0 in URSP.  Conversely, RSP underestimates
$p_{pol}$, with consecutive bins consistently showing large deviations (all
with the same sign) from zero in URSP.  It is generally not prudent to rectify
this situation by following the $\theta$ curve more closely (i.e., smoothing it
less), since RSP then becomes subject to the same biases that affect
$p_{trad}$.  Indeed, in the limit $\theta_{smooth} \rightarrow \theta$, RSP $
\rightarrow \sqrt{q^2 + u^2}$.

There is no perfect estimator of the true polarization for data with low $p /
\sigma_p$, and the best one to use generally depends on the polarization
properties of the type of data being analyzed.  For data with $p / \sigma_p \gg
1$ (e.g., SN~1998S), all four estimators produce essentially identical results.
When $p / \sigma_p \lesssim 3$, RSP is most robust, but only when $\theta$
varies slowly with wavelength: sharp $\theta$ changes necessitate examination
of URSP as well.  Alternatively, one can use a coarser binning to make $p /
\sigma_p > 3$ so that $p_{deb}$ and $p_{opt}$ become reliable, but care must
then be taken to ensure that polarization features are not altered by the
decreased resolution (i.e., the binning should be significantly narrower than
the scale over which polarization features appear).  Finally, when summing over
a very large wavelength band to determine an average polarization (e.g.,
$p_V$), taking the flux-weighted average of unbinned data (or data with fine
binning) and then computing either $p_{deb}$ or $p_{opt}$ is the method of
choice (see, e.g., Miller et al. 1988).  In this paper, we use $p_{deb}$ to
measure broadband polarization and RSP to characterize the spectropolarimetry.

For the artificial data, we can now estimate the line's polarization from the
\mbox{ISP-subtracted} data using equation~(\ref{eqna:b1}) and fitting the
Stokes parameter fluxes as before.  For the polarized line, we measure $q_{pol}
= 0.58\%\ \pm\ 0.07 \% {\rm\ and\ } u_{pol} = 0.21\%\ \pm\ 0.04\%$, calculated
over the line region $5670 - 5730$~\AA\ with background regions
$5550-5650$~\AA\ and $5750-5850$ \AA.  For the unpolarized line we measure
$q_{unpol} = -0.01\%\ \pm\ 0.08 \% {\rm\ and\ } u_{unpol} = -0.01\%\ \pm\
0.04\%$ using the same line and background regions described earlier for
determining the ISP.  Our analysis techniques have successfully recovered the
input values, though the uncertainty in the ISP removal would add systematic
uncertainty to the error budget.

\end{document}